\documentclass[12pt,a4paper,fleqn]{article}
\usepackage[utf8]{inputenc}
\usepackage{authblk}
\usepackage{geometry}
\usepackage{xcolor}
\usepackage[round]{natbib}
\usepackage{graphicx}
\usepackage{pdflscape}  
\usepackage{appendix}
\usepackage{threeparttable}
\usepackage{booktabs}
\usepackage{subcaption}
\usepackage{longtable}
\usepackage{threeparttable}
\usepackage{rotating}
\usepackage{multirow}

\usepackage{amssymb}
\usepackage{amsbsy}
\usepackage{bm}
\usepackage{amsmath}
\usepackage{amsthm}
\usepackage{graphicx}
\usepackage{mathtools}
\usepackage{setspace}
\usepackage[justification=centering]{caption}

\usepackage{color, colortbl}
\definecolor{ashgrey}{rgb}{0.7, 0.75, 0.71}
\definecolor{columbiablue}{rgb}{0.61, 0.87, 1.0}
\definecolor{coral}{rgb}{1.0, 0.5, 0.31}

\definecolor{colBVAR}{HTML}{bababa}
\definecolor{colBART}{HTML}{d7191c}
\definecolor{colmixBART}{HTML}{fdae61}
\definecolor{colerrorBART}{HTML}{abd9e9}
\definecolor{colfullBART}{HTML}{2c7bb6}

\definecolor{colcons}{HTML}{e31a1c}
\definecolor{colSV}{HTML}{a6cee3}
\definecolor{colhBART}{HTML}{1f78b4}

\usepackage{enumitem}
\newlist{steps}{enumerate}{1}
\setlist[steps,1]{label = Step \arabic*:}

\usepackage{adjustbox}
\usepackage{dcolumn}
\newcolumntype{d}[1]{D..{#1}} % for alignment of numbers on decimal marker

\usepackage{xcolor,colortbl}
\doublespace

\definecolor{Gray}{gray}{0.85}
\definecolor{LightCyan}{rgb}{0.88,1,1}

\newcolumntype{a}{>{\columncolor{Gray}}c}
\newcolumntype{B}{>{\columncolor{white}}c}
\usepackage{array}
\newcolumntype{H}{>{\setbox0=\hbox\bgroup}c<{\egroup}@{}}
\newcolumntype{Z}{>{\setbox0=\hbox\bgroup}c<{\egroup}@{\hspace*{-\tabcolsep}}}

\usepackage{caption}
\usepackage{subcaption}
\definecolor{nblue}{HTML}{000660}
\usepackage[colorlinks=true,urlcolor=nblue,linkcolor=nblue,citecolor=nblue]{hyperref}
\newcommand*{\myeqref}[2][Eq.~]{%
  \hyperref[{#2}]{#1(\ref*{#2})}%
}
\def\equationautorefname#1#2\null{%
  Eq.#1(#2\null)%
}

\begin{document}
\title{\onehalfspacing{}The Distributional Effects of Economic Uncertainty}
\author{ }
\date{}

\maketitle
\thispagestyle{empty}
\vspace*{-3.5em} 

%\author{Florian Huber, Massimiliano Marcellino and Tommaso Tornese\thanks{Florian Huber: University of Salzburg; e-mail: \protect\href{mailto:florian.huber@plus.ac.at}{florian.huber@plus.ac.at}.
%Massimiliano Marcellino: Bocconi University; e-mail: \protect\href{mailto:massimiliano.marcellino@unibocconi.it}{massimiliano.marcellino@unibocconi.it}.
%Tommaso Tornese: Baffi-CAREFIN Centre, Bocconi University; e-mail:
%\protect\href{mailto:tommaso.tornese@unibocconi.it}{tommaso.tornese@unibocconi.it}. \\{We would like to thank seminar participants at the ECB and Bocconi University for useful comments.} }}
%\vspace*{-0.75em}
\begin{minipage}{.49\textwidth}
  \large\centering Florian \textsc{Huber}\\[0.25em]
  \small University of Salzburg, Austria 
\end{minipage}
\begin{minipage}{.49\textwidth}
  \large\centering Massimiliano \textsc{Marcellino}\\[0.25em]
  \small Bocconi University, Italy
\end{minipage}

\vspace*{0.5em}
\begin{center}
\begin{minipage}{.49\textwidth}
  \large\centering Tommaso \textsc{Tornese}\footnotemark{}\\[0.25em]
  \small Università Cattolica del Sacro Cuore,  Italy
\end{minipage}
\end{center}
\footnotetext{
 We would like to thank Minsu Chang, Jamie Cross, Jim Hamilton, Jiaming Huang, Andrea Renzetti, Jim Stock, Allan Timmermann and participants at the 2024 NBER Summer Institute and at seminars at the European Central Bank, and Bocconi University for useful comments and suggestions. Marcellino and Tornese thank MIUR -- PRIN Bando 2017 -- prot.\ 2017TA7TYC  for financial support; Huber  gratefully acknowledges financial support from the Austrian Science Fund (FWF, grant no. ZK 35). 
 Please address correspondence to: Tommaso Tornese. Department of Economics and Finance, Università Cattolica del Sacro Cuore.  \textit{Email}: \href{mailto:tommaso.tornese@unicatt.it}{tommaso.tornese@unicatt.it}.}
\vspace*{1.5em}
%\date{April 2023}

\begin{abstract}
\begin{spacing}{2}
\noindent 
We study the distributional implications of uncertainty shocks by developing a model that links macroeconomic aggregates to the US distribution of earnings and consumption. We find that: initially, the fraction of low-earning workers decreases, while the share of households reporting low consumption increases; at longer horizons, the fraction of low-income workers increases, but the consumption distribution reverts to its pre-shock shape. While the first phase reduces income inequality and increases consumption inequality, in the second stage income inequality rises, while the effects on consumption inequality dissipate. Finally, we introduce Functional Local Projections and show that they yield similar results.

 \vspace{.5cm}

\medskip{}

\end{spacing}
\end{abstract}

\noindent \newpage{}

\section{Introduction}
\doublespacing

An extensive line of research stimulated by \cite{bloom2009impact} has studied the effects of uncertainty shocks on economic fluctuations.\footnote{See \cite{castelnuovo2019domestic} and \cite{fernandez2020uncertainty} for recent surveys of the literature.} It is generally agreed that an unexpected increase in the level of uncertainty about the future state of the economy generates a significant drop in output, employment and asset prices. Notwithstanding the importance of uncertainty shocks as drivers of the business cycle, little attention has been devoted to their distributional implications. 

In this paper, we investigate the consequences of this type of shocks on the earnings distribution among employed people and on the overall consumption distribution in the US. Focusing on both consumption and earnings is indeed crucial for understanding the propagation mechanisms of uncertainty, and whether or not inter-temporal transfers can protect consumption from shocks that affect income (see e.g. \cite{attanasio2016consumption} for a discussion). 
The main econometric specification we use is a Functional Structural Vector Autoregression (F-SVAR) model, which represents a generalization in the functional space of the popular SVAR model commonly used in the empirical macroeconomics literature. Despite having been abundantly developed in the statistical literature, models for functional data are not yet popular among econometrics practitioners. Nevertheless, many recent methodological and empirical contributions have shown the high potential such models have for economic and financial analysis. \cite{kowal2017bayesian} and \cite{chang2021heterogeneity} develop Bayesian methods for inference in functional linear models, applying their techniques to analyze the dynamics of yield curves and income distribution respectively. \cite{chang2016nonstationarity} and \cite{hu2016econometric} provide methods to analyze functional time series in a frequentist setting. Other applications of models for functional data in econometrics include \cite{diebold2006forecasting}
\cite{tsay2016some}, \cite{meeks2022heterogeneous}, \cite{inoue2021new}, \cite{chang2024effects} and \cite{bjornland2023oil}. Most of these papers, with the exception of \cite{chang2024effects}, assume that the function of interest is directly observed, while we recognize that only a sample from the distribution of interest is available but the full probability density remains unknown.

In the recent period, macroeconomics research has dedicated more and more attention to the interplay between macroeconomic phenomena and distributional developments. The aggregate effects of distributional dynamics are studied, among others, by \cite{heathcote2010macroeconomic}, \cite{athreya2017does} and \cite{rognlie2019inequality}; while the distributional implications of aggregate shocks are the focus, for instance, of \cite{anderson2016heterogeneous}, \cite{de2017business},   \cite{mumtaz2017impact}, \cite{ahn2018inequality}, \cite{kaplan2018microeconomic}, and \cite{bayer2020shocks}. By means of a functional time series model, we
contribute to the latter strand of the macroeconomics literature by showing that uncertainty shocks produce relevant and non trivial distributional effects. 

From an econometric point of view, we adopt a three-step procedure along the lines of \cite{Petersen2016FunctionalDA} and \cite{PETERSEN2022159}. In the first step, we interpolate the cross-sectional data for each time period using kernel estimators (see, e.g., \cite{silverman2018density}) to obtain period-by-period continuous earnings and conusmption distributions, in order to be able to monitor the entire densities rather than only the observed data. In the second step, we transform the  continuous earnings and consumption distributions to remove the unit-integration and non-negativity constraints as in \cite{Petersen2016FunctionalDA} and approximate the resulting curve with a set of basis functions (see e.g. \cite{ramsay2002applied}). Specifically, we use functional principal components (FPCs), along the lines of \cite{kneip2001inference} and  \cite{Petersen2016FunctionalDA}. In the final step, we jointly model the FPCs and a set of macroeconomic and financial indicators with a (Bayesian) VAR. By doing this, we are able to assess the effects of the uncertainty shocks on the FPCs, and then to reconstruct the Impulse Response Functions (IRFs) of the entire earnings and consumption distributions.

An alternative one-step approach would be to simply insert percentiles or moments of the earnings and consumption disaggregated data directly into the third step BVAR. Yet, in the case of percentiles, crossing could be an issue (see e.g. \cite{chang2021heterogeneity})  and the BVAR dimension could easily become very large, if the goal is to get a granular view of the effects of uncertainty shocks on the earnings or consumption distribution, as one would have to include a large number of percentiles into the model. A similar issue occurs in the case of moments, with the addition that higher order moments would be hardly interpretable, and their theoretical counterpart could also not exist in the presence of a heavy tailed distribution.\footnote{One may be also tempted to perform standard Principal Component Analysis (PCA) directly on disaggregated data and use them in a standard VAR as in \cite{bernanke2005measuring}. However, disaggregated data are usually collected in the form of repeated cross-section rather that in panel data form, making the use of standard PCA invalid.} 
Other empirical studies, including \cite{mumtaz2017impact}, \cite{theophilopoulou2022impact}, and \cite{choi2023impact}, only consider inequality measures, such as Gini coefficients or interquantile ranges, as endogenous variables in a VAR model. Such approach, however, conveys an inevitably partial account of the distributional dynamics, possibly providing misleading insights as discussed by \cite{heathcote2010unequal} and \cite{choi2023impact}.

Due to the three-step procedure, our methodology requires the use of "generated" regressors. Specifically, in the second step we rely on kernel-based estimated continuous distributions. In order to address  concerns related to the generated regressors problem, one could either allow for the presence of measurement error, as e.g. in \cite{chang2021heterogeneity}, or use a kernel estimator that ensures consistency also in the presence of truncated distributions, see e.g. \cite{Petersen2016FunctionalDA}. We follow the latter approach, in combination with a quite large cross-sectional dimension (about 10.000 units). Similarly, in the third step, we use the FPC in place of the estimated continuous earnings and consumption distributions. It can be shown, however, that the approximation error disappears as the cross-sectional dimension increases, when using a large enough number of FPCs (see e.g. \cite{kneip2001inference}). In related contemporaneous work, \cite{Huang2023fvar} studies the asymptotic behaviour of a similar three-steps, but fully frequentist, procedure, and shows consistency of the kernel estimators of the density and the functional principal components, under proper assumptions. \cite{Huang2023fvar} also shows that the method works well in a classical context in MC experiments, and discusses an interesting application about the distributional effects of tax cuts in the UK. 

To assess the empirical performance of our functional structural VAR in finite samples, we conduct two sets of simulation experiments. In the first one, we generate data from two different types of F-VARs, while, in the second one, we use data simulated from the log-linearized solution of the version of the \cite{krusell1998income} heterogeneous agents model considered in \cite{chang2021heterogeneity}. In all cases our approach returns estimated responses remarkably close to the true ones.

In our empirical analysis, we find that the propagation mechanism of uncertainty shocks on aggregated macroeconomic and financial data resulting from our functional VAR is similar to that reported by \cite{jurado2015measuring}. Instead, their distributional effects are novel in the literature and can be broken down in two phases. First, in the short-run phase, as employment and output drop, a larger share of low-income workers are likely laid off, while those that keep their jobs see their wage increase relative to the GDP per capita level. This is reflected in a significant reduction of the mass of employed people with an income-to-GDP per capita ratio smaller than unit, which is presumably due to a larger increase of unemployment among the less specialized workers. At the same horizons, the response of the consumption distribution shows that the proportion of households reporting a low level of consumption increases significantly after the shocks, while the mass in the middle part of the distribution decreases. The response is therefore compatible with a short-run propagation mechanism in which some of the workers with a low degree of specialization, and belonging to the middle part of the consumption distribution, become unemployed and have to cut down their consumption level due to their inability to access consumption smoothing channels. 
On the other hand, in a longer horizon phase, while the effects on unemployment and output dissipate, the mass of workers with a low relative income increases again, supposedly due to the increase of the employment rate among the unskilled workers, and the stronger decrease of their labor productivity generated by the investment foregone in the first phase. At the same time, the consumption distribution reverts to its pre-shock shape. 
While the first phase reduces the overall earnings disparity among employed people but increases the degree of inequality in the distribution of consumption
people, in the second stage the Gini coefficient of the income distribution rises considerably and that of the consumption distributions returns to its pre-shock level. 

Interestingly, this resembles the two-wave economic contraction in response to uncertainty shocks pointed out by \cite{carriero2023macro}. Just as at the aggregate level the investments foregone following an uncertainty shock generate lower economic growth in the long run, at a more granular level those missed investments may also reduce the productivity of lower-income employees and their compensation as a result.

Finally, we introduce Functional Local Projections (F-LPs), showing that they represent a viable alternative to the estimation of F-SVARs. To the best of our knowledge, we are the first to suggest the use of F-LPs in the literature. The F-LPs are obtained by regressing the FPCs on uncertainty  and a set of controls. Our FPCs summarize the dynamics of the cross-section. Treating them independent of each other thus might have adverse effects on the estimated density response and we propose modeling them jointly
in a system of equations. Since, as mentioned above, estimation uncertainty around the FPCs disappears in large cross-sections so that one can treat the estimated FPCs as observable variables, one can expect that FPC-based functional LPs and functional SVARs should lead to similar results, as standard LP and SVARs, see \cite{plagborg2021local}. Indeed, we show that for both the simulated data and the actual data, the two methods yield comparable results, which provides additional robustness to the distributional effects of uncertainty that we have described.

The remainder of the paper is organized as follows: Section \ref{sec:FSVAR} describes the functional time series model and the econometric methods employed. Section \ref{sec:Simulation} tests the ability of the model to track the propagation of structural shocks to the distribution of interest with simulated data while Section \ref{sec:JLN15} extends the empirical analysis of \cite{jurado2015measuring} to study the distributional effects of macro uncertainty shocks. Section \ref{sec:FLP} introduces and demonstrates the merits of Functional Local Projections. Section \ref{sec:Summary-and-concluding}
summarizes and concludes. A set of appendices provide additional  results.

\section{Functional Vector Autoregressions\label{sec:FSVAR}}
In this section, we introduce our econometric framework, which is closely related to \cite{chang2021heterogeneity}. The first sub-section develops a model that is capable of capturing dynamic relations between the aggregate economy and the cross-sectional distribution of earnings or consumption. Sub-section \ref{sec: fpca} then discusses the transformation and approximation techniques used to summarize cross-sectional dynamics with a few unconstrained latent factors. In Sub-section \ref{sec: inference} we summarize our Bayesian prior setup and sketch the posterior simulation algorithm.

\subsection{\normalsize{Combining cross-sectional densities with aggregate time series}\label{sec:FVAReq}}
We aim to model the joint dynamics of a cross-sectional density, labeled $p_t(\xi)$ for $\xi \in \Xi$, with $\Xi$ denoting the domain on which the probability density function (pdf) is defined, and a set of $n_v$ aggregate economic random variables  $y_{t}=\left[y_{1,t},...,y_{n_{v},t}\right]^{\prime}$. Time runs from $t= 1, \dots, T$.  The cross-sectional density, in our case, will alternately be the distribution of earnings or consumption in the US. We assume that we do not observe $p_t(\xi)$ directly but we are given a set of identically and independently distributed (iid) draws from $p_t(\xi)$, $\xi_{it}$ for $i=1, \dots, N$.  Let $f_{t} = g(p_t)$ denote an invertible transformation of the density describing the distribution of interest. Additionally, let $\bar{f}=\frac{1}{T} \sum_{t=1}^{T}f_{t}$ be the sample mean of $f_t$, and $\widetilde{f_{t}} = f_{t}-\bar{f}$  be the deviation of  $f_t$ form this mean.

The F-VAR with $p$ lags then consists of two blocks:
 \begin{align}
y_{t} &=c_{y}+\sum_{l=1}^{p}B_{l,yy}y_{t-l}+\sum_{l=1}^{p}\int B_{l,yf}\left(\acute{x}\right)\widetilde{f}_{t-l}\left(\acute{x}\right)d\acute{x}+u_{y,t},\label{eq:1} \\
\widetilde{f}_{t}\left(x\right) &=c_{f}\left(x\right)+\sum_{l=1}^{p}B_{l,fy}\left(x\right)y_{t-l}+\sum_{l=1}^{p}\int B_{l,ff}\left(x,\acute{x}\right)\widetilde{f}_{t-l}\left(\acute{x}\right)d\acute{x}+u_{f,t}\left(x\right),\label{eq:2}
\end{align}
where $x$ and $\acute{x}$ are in the domain of the transformed distribution $\widetilde{f_{t}}$, and $u_{y,t}$ and $u_{f,t}(x)$ are innovations with zero mean and variance given by:
\begin{equation*}
 \Omega(x,\acute{x}) = \left[\begin{array}{cc}
\Omega_{yy} & \Omega_{yf}\left(\acute{x}\right)\\
\Omega_{fy}\left(x\right) & \Omega_{ff}\left(x,\acute{x}\right)
\end{array}\right].
\end{equation*}
Notice that the innovations are correlated across the aggregate macro series in $y_t$ but also across the two blocks. This implies that structural economic shocks (such as the uncertainty shock we consider in this paper) can impact $y_t$ but also have a direct effect on $\widetilde{f_t}(x)$. It is also worth stressing that dynamic inter-dependencies are captured through the inclusion of lagged $y_t's$ in (\ref{eq:2}) but also by including lags of the log-densities in (\ref{eq:1}). 

Since the functions $\widetilde{f_t}(\cdot)$ and $u_{f,t}(\cdot)$ are continuous, the dimension of the model is infinite.  In order to be able to conduct inference about the parameters of the infinite dimensional model, we resort to the Karhunen-Loève theorem (see e.g. \cite{bosq2000linear}, Theorem 1.5) and write the function $\widetilde{f_{t}}(\cdot)$ as an infinite order expansion:
\begin{equation}
    \widetilde{f_{t}}\left(x\right)=\sum_{k=1}^{\infty}\zeta_{k}\left(x\right) \times \alpha_{k,t}, \label{eq: additive_repr}
\end{equation}
with $\zeta_{k}(\cdot)$ and $\alpha_{k,t}$  denoting time-invariant orthogonal functional bases and time-varying factors, respectively. Conditional on choosing an appropriate truncation level $K$, we can approximate (\ref{eq: additive_repr}) through:
\begin{equation}
    \widetilde{f_{t}}\left(x\right)\approx \sum_{k=1}^{K}\zeta_{k}\left(x\right) \times \alpha_{k,t} = \boldsymbol\zeta(x)' \alpha_t, \label{eq: finite_additive_repr}
\end{equation}
where $\alpha_t = [\alpha_{1,t}, \dots, \alpha_{K,t}]'$ and $\zeta(x) = [\zeta_{1}(x), \dots, \zeta_{K}(x)]'$. 
Likewise, the other components of equations (\ref{eq:1}) and (\ref{eq:2}) can be approximated in a similar way:
\begin{align}
    u_{f,t}\left(x\right) \approx \boldsymbol\zeta(x)' \tilde{u}_{f,t},~ c_{f}\left(x\right) \approx \boldsymbol\zeta(x)'\tilde{c}_{f},~  B_{l,yf} \left(\acute{x}\right)&\approx\tilde{B}_{l,yf}\boldsymbol\delta\left(\acute{x}\right), ~B_{l,fy}\left(x\right)\approx\boldsymbol\zeta(x)'\tilde{B}_{l,fy}
    \label{expansion_operators}
\end{align}
and $B_{l,ff}\left(x,\acute{x}\right)\approx\boldsymbol\zeta(x)'\tilde{B}_{l,ff}\boldsymbol\delta\left(\acute{x}\right)$.  Here, $\boldsymbol\zeta\left(x\right)$ and $\boldsymbol\delta\left(\acute{x}\right)$ are $K$-dimensional vectors of time-invariant orthogonal functional bases, $\tilde{u}_{f,t}$ is a vector of time-varying factors, and $\tilde{c}_{f}$, $\tilde{B}_{yf}$, $\tilde{B}_{fy}$, and $\tilde{B}_{ff}$ are matrices of scalar coefficients. 

Using these finite approximations, the F-VAR in (\ref{eq:1}) and (\ref{eq:2}) can then be written as a finite dimensional VAR: 
\begin{equation}
\left[\begin{array}{c}
y_{t}\\
\alpha_{t}
\end{array}\right]=\left[\begin{array}{c}
c_{y}\\
\tilde{c}_{f}
\end{array}\right]+\sum_{l=1}^{p}\left[\begin{array}{cc}
B_{l,yy} & \tilde{B}_{l,yf}C_{\alpha}\\
\tilde{B}_{l,fy} & \tilde{B}_{l,ff}C_{\alpha}
\end{array}\right]\left[\begin{array}{c}
y_{t-l}\\
\alpha_{t-l}
\end{array}\right]+\left[\begin{array}{c}
u_{y,t}\\
\tilde{u}_{f,t}
\end{array}\right],\label{eq:VAR}
\end{equation}
\noindent where $C_{\alpha}=\int\delta\left(\acute{x}\right)\boldsymbol{\zeta}^{\prime}\left(\acute{x}\right)d\acute{x}$,
and $u_{t}=\left[u_{y,t}^{\prime},\tilde{u}_{f,t}^{\prime}\right]^{\prime}$ has zero mean and variance $\Omega$. Conditional on knowing $\alpha_t$, the model in (\ref{eq:VAR}) is a standard VAR model and inference can be performed applying conventional frequentist or Bayesian techniques. If $K$ is large (i.e., the time variation in cross-sectional densities is complex and multi-dimensional and requires many modes of variation to approximate their behaviour), frequentist estimation of the model might suffer from overfitting issues. Hence, our approach will be Bayesian and we describe it in more detail in Sub-section \ref{sec: inference} below.

It is worth stressing that the  covariance matrix $\Omega$ is a full matrix. Hence, to attribute a structural interpretation to the model and to recover the structural shocks, we need to back out the structural representation of the model.  This can be achieved by introducing suitable restrictions on the matrix $A_0^{-1}$, such that $\Omega=A_0^{-1}\left(A_0^{-1}\right)^{\prime}$, which defines the impact matrix, and then map back the impulse responses (IRFs) of $\alpha_t$ to the original functional space to which $\widetilde{f_{t}}\left(\cdot\right)$ belongs. For instance, the structural form of the VAR is exactly identified if we assume that the system (\ref{eq:VAR}) is driven by $n=n_{v}+K$  iid structural shocks with unit variance, $\varepsilon_{t}$, and that the responses of the variables in the system to the shocks are such that we can write: 
\begin{equation}
A_{0}u_{t}=\varepsilon_{t},\label{eq:6}
\end{equation}
\noindent with $A_{0}$ an invertible lower-triangular matrix.  In this case, the matrix $A_{0}$ can be simply found by inverting the lower-triangular Cholesky factor of the estimate obtained for $\Omega$. This type of identifying restrictions are popular in the empirical macroeconomics literature and will be used in subsequent sections to identify uncertainty shocks. 
Noticeably, nothing prevents the use of other, more sophisticated, identification strategies in other applications of this class of models. As in standard SVARs, the impact matrix  $A_{0}^{-1}$ could be identified through a variety of second- or higher-moments assumptions (see e.g. \cite{kilian2017structural}) . We focus on the "Cholesky" identification scheme because it is the one adopted by the important contribution of \cite{jurado2015measuring} to the literature on uncertainty shocks, which represents the foundation of the model we employ in Section \ref{sec:JLN15} to study distributional responses.   

\subsection{Modeling cross-sectional dynamics using functional principal components}\label{sec: fpca}
Up to this point, we have not specified the kernel density estimation method we employ,  what the transformation $g\left(\cdot\right)$ does to the density $p_t$, and what kind of bases are contained in $\boldsymbol\zeta\left(\cdot\right)$ and $\boldsymbol\delta\left(\cdot\right)$.  Hence, in this sub-section we describe how we estimate the continuous earnings and consumption distributions starting from an observed sample, we define the transformation $g\left(\cdot\right)$ applied to $p_t$ to enforce the necessary non-negativity and unit-integration constraints, and illustrate the method we use to determine the functional basis that best approximates $\widetilde{f_t}\left(\cdot\right)$ for any given truncation point $K$. We should mention that in these choices we differ from \cite{chang2024effects}, for the reasons discussed below.

\subsubsection{Kernel density estimation\label{sec:Kernel}}
Since we want to model the dynamics of the entire continuous distributions of earnings and consumption, but only observe a sample from it, a density estimator has to be applied. To this end, we use a Gaussian kernel estimator specifying the bandwidth size following Silverman's rule of thumb, which has been shown to be a reliable choice in much of the applied statistics literature (see e.g. \cite{kokoszka2019forecasting}).\footnote{In our simulations and application, we have experimented with different choices of kernel functions and bandwidth sizes, which did not produce relevant differences in the results.}  Because in our application (and in most practical cases) the support of the distributions of earnings $\Xi$ and consumption are bounded due to the non-negativity of such variables and the censoring adopted by statistical agencies when conducting surveys, we apply a boundary correction to the kernel density estimator by augmenting data with their reflection near the boundaries.
More precisely, we adopt the following density estimator:
\begin{equation}
    \hat{p}_{t}\left(\xi\right)=\frac{1}{Nh}\sum_{i=1}^{N}\left\{ k\left(\frac{\xi-\xi_{it}}{h}\right)+ k\left(\frac{\xi-\xi^{L}_{it}}{h}\right)+ k\left(\frac{\xi-\xi^{U}_{it}}{h}\right)\right\},
\end{equation}
where $\xi_{it}^{L}=2L-\xi_{it}$ and $\xi_{it}^{U}=2U-\xi_{it}$, with $L$ and $U$ being respectively the lower and upper bound of the support $\Xi$.

\subsubsection{Log Quantile Density transformation \label{LQD sec}}
An important challenge when modeling the dynamics of a distribution is to ensure that it always satisfies the non-negativity and unit-integration constraints intrinsic in the space of distributions. For example, one must always ensure that the predicted distribution \textit{h} periods after a shock (i.e. the distribution IRF at horizon \textit{h}) is non-negative at every point in the domain and integrates to one. 
One could simply set  $g(p_t)=p_t$ and apply the approximation (\ref{eq: finite_additive_repr}) to the deviations of the distribution  $\widetilde{f_{t}}=p_t-\frac{1}{T} \sum_{t=1}^{T}p_{t}$. As the sample mean $\frac{1}{T} \sum_{t=1}^{T}p_{t}$ has integral equal to 1 by construction, the unit integration constraint can be enforced by ensuring that each of the $K$  basis functions integrate to 0 (for example by using FPC as basis, which have zero integral by construction). As discussed by \cite{Petersen2016FunctionalDA}, however, the fact that the basis have zero integral also implies that they must have some negative parts, which can generate approximations of the distribution taking inadmissible negative values in some regions of the support. This problem is particularly important in the context of a F-VAR, in which the structural shocks move the coefficients $\alpha_t$ that multiply the functional basis. In such a context, if a large shock is estimated to have a large effect on some of the $\alpha_t$, the negative part of the associated basis can become so important that it leads the predicted distribution to have negative regions after the shock. 

To avoid this issue, most of the existing literature has applied the approximation in equation (\ref{eq: finite_additive_repr}) to the logarithm of the density function, which naturally enforces the non-negativity constraint. This however is not sufficient to enforce the unit-integration constraint, and therefore leads again the functional IRFs to leave the space of probability functions. Some of the literature, then, brings the functional IRFs back to the space of distributions by re-normalizing them \textit{ad hoc} to have unit integral.  

Although this strategy may seem appealing, \cite{Petersen2016FunctionalDA} warn about the important potential flaws of such practice. The problem originates from the fact that standard functional data methods (e.g. FPCA) are designed to model the behaviour of random functions in the Hilbert space $L^2$, while the space of densities is only a subspace of $L^2$. A sensible modeling strategy would therefore need either to use functional data methods that restrict the analysis to the space of densities, or to transform the density in an unconstrained function in the Hilbert space $L^2$. \cite{Petersen2016FunctionalDA} propose a coherent and straightforward strategy to do the latter. In particular,  they suggest to apply standard functional data analysis to the Log Quantile Density (LQD) function associated with the distribution of interest, in place of working directly with the distribution itself. Even if there exists a one-to-one mapping between the LQD and the density function defined on a given support, the LQD has the remarkable advantage of being an unconstrained function in $L^2$, and hence no special care is needed when applying (\ref{eq: finite_additive_repr}) and (\ref{eq:VAR}) to its de-meaned version $\widetilde{f_t}$. As the mapping between the LQD and the density function is one-to-one, in fact, a simple inverse transformation can be applied to retrieve the IRFs of the distribution of interest from those inferred for the LQD. The resulting distributional impulse responses will then satisfy, by construction, the required non-negativity and unit-integration constraints. 

Given the clear advantage of the approach proposed by \cite{Petersen2016FunctionalDA} and the success of its application in \cite{kokoszka2019forecasting} and \cite{PETERSEN2022159}, in what follows we define the function $f_t\left(\cdot\right)$ to be the LQD associated to the earnings or consumption distribution; the transformation $g\left(\cdot\right)$ is therefore the mapping that defines the LQD function associated with $p_t(\xi)$.
The LQD associated with a probability distribution $p_t(\xi)$ is the logarithm of the first derivative of the quantile function $Q(z)=F^{-1}(z)$ (also known as inverse cumulated density function, icdf): 
\begin{equation}
f_t\left(x\right)=g\left\{p_t(\cdot)\right\}=\log\left\{\left.\frac{d}{dz}Q(z)\right|_{z=x}\right\},\label{eq:LQD}
\end{equation}
where $x\in[0,1]$. The resulting LQD function is therefore defined on $[0,1]$ and looses information about the support $\Xi$. Its great advantage, however, is that the LQD is unrestricted and lies in the $L^{2}$ space, and thus can be modeled without difficulty. 

As the support  $\Xi$ is known, the distribution of interest $p_t(\cdot)$ can be easily derived back from $f_t(x)$ using the inverse transformation. It simply amounts to computing the quantile function $Q(x)=\theta\int_{0}^{x} \exp[f_t(z)]dz$, where $\theta=\sup_{\xi\in\Xi}\xi\times\left\{\int_{0}^{1} \exp[f_t(z)]dz\right\}^{-1}$, and then computing the first derivative of the associated cumulative distribution function $\frac{d}{d\xi}F(\xi)=\frac{d}{d\xi}Q^{-1}(\xi)$.

\begin{figure}[t!]
\caption{\label{fig:LQD}\textbf{LQD Transformation}}
\begin{center}

\centering \subfloat[PDF]{\includegraphics[clip, trim= 0 0 0 0, scale=0.5]{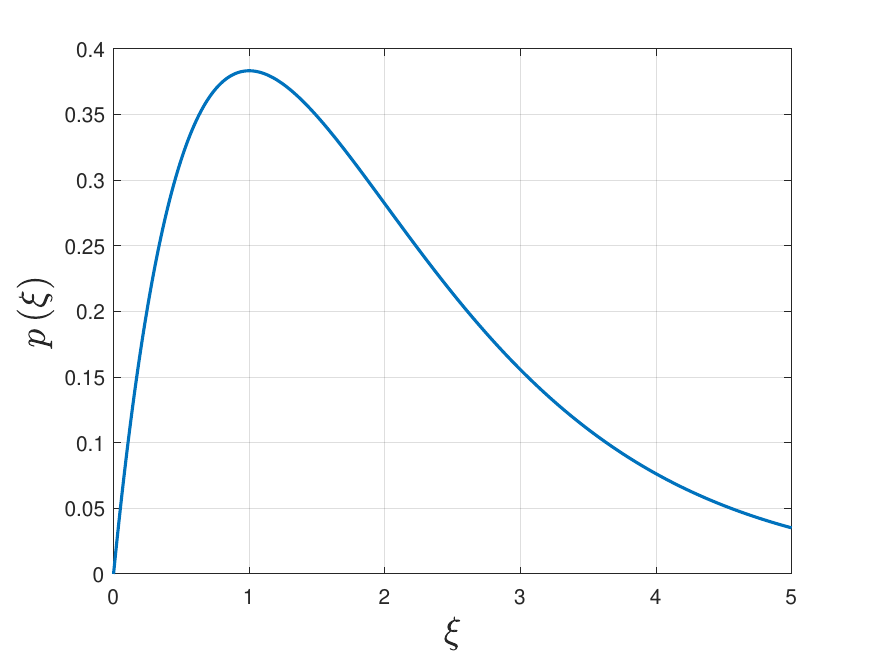}} 
\centering \subfloat[CDF]{\includegraphics[clip, trim= 0 0 0 0, scale=0.5]{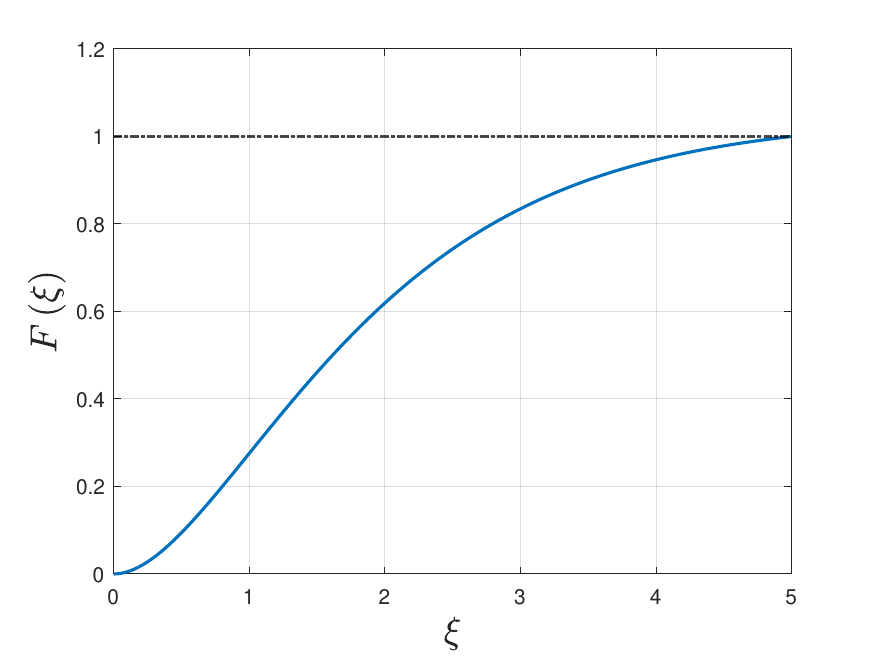}}\\ 
\centering \subfloat[Quantile]{\includegraphics[clip, trim= 0 0 0 0, scale=0.5]{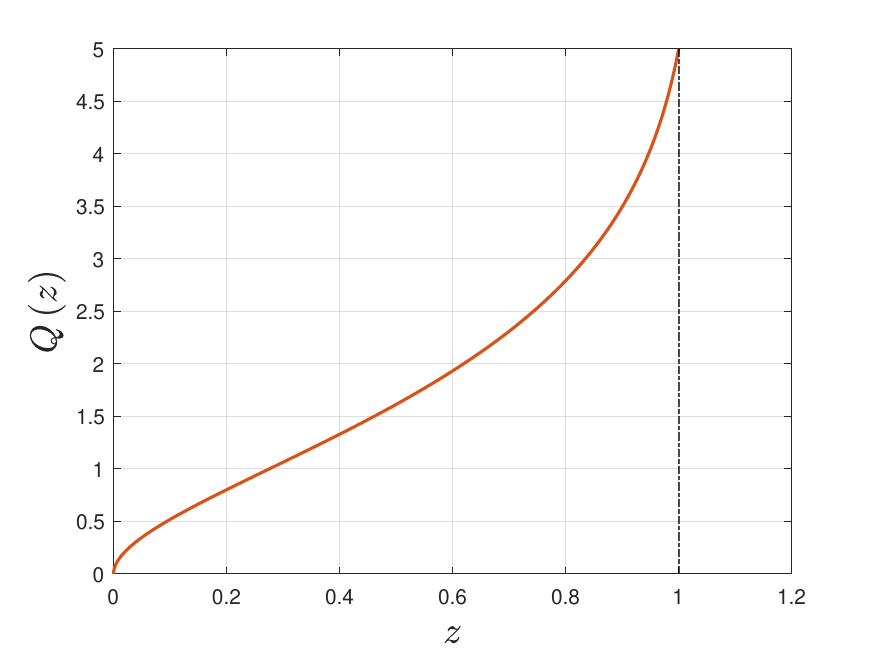}} 
\centering \subfloat[LQD]{\includegraphics[clip, trim= 0 0 0 0, scale=0.5]{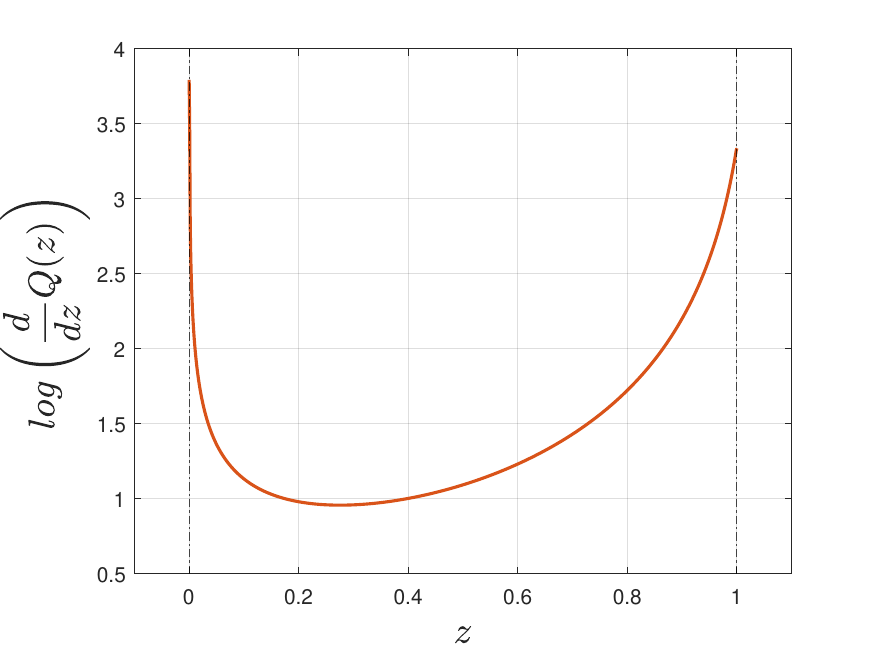}} 
\end{center}
\end{figure}

Figure \ref{fig:LQD} helps to visualize the LQD transformation and to highlight why it is convenient. The figure refers to a \textit{Gamma(2,1)} distribution censored to have support  $\Xi=[0,5]$. The four panels represent the four steps that the mapping $g(\cdot)$ applies to $p_{t}(\cdot)$ (panel (a)) to find its LQD (panel (d)). 
In the first three panels, the transformation runs through three alternative ways of describing the same distribution, namely the \textit{pdf}, the \textit{cdf}, and the quantile function. These three objects are familiar, contain exactly the same amount of information, and are all subject to some constraint, that is, the \textit{pdf} must be non-negative and integrate to 1, the \textit{cdf} must be non-negative, non-decreasing, and take values from 0 to 1, the quantile function must be non-decreasing and take values in $\Xi$. In practice, once the kernel estimate of the \textit{pdf} is available, the \textit{cdf} (and hence its inverse, the quantile function) can be easily obtained by numerical integration. The only unfamiliar function in the figure is the LQD, which is simply the logarithm of the first derivative of the quantile function in panel (c), which can also  be easily computed numerically. However, the LQD does not obey to any constraint and therefore represents an ideal modeling device. 

As can be seen from the figure, the LQD looses the information about the support, but, since $\Xi$ is known to the researcher, it is easy to retrieve any of the functions in panels (a),(b), or (c), once the LQD is available. To recover the $Q(z)$ once the LQD is known, for example, it is sufficient to integrate numerically $\exp\left[LQD\right]$ over $[0,z]$, and to re-normalize it so that $Q(1)=\sup_{\xi\in\Xi}\xi=5$.     

\subsubsection{Functional Principal Component Analysis}
As regards the functional basis used in the approximations of equation \ref{eq: finite_additive_repr} and \ref{expansion_operators}, the  choice of $\boldsymbol\delta\left(\cdot\right)$ is of no practical relevance as long as its inner product with $\boldsymbol\zeta\left(\cdot\right)$, $C_{\alpha}$, is finite; while the choice of the basis $\boldsymbol\zeta\left(\cdot\right)$ used to expand the function of interest, $\widetilde{f_{t}}(\cdot)$, represents a central theme in the literature on functional data analysis (for a recent survey, see \cite{wang2016functional}). 

Various alternative methods are available. For instance, \cite{chang2021heterogeneity} consider splines to approximate the cross-sectional distribution. While their approach is flexible and non-parametric, the truncation point $K$ (i.e. number of spline knots) needed to approximate sophisticated distributions can easily become very large, which in turn undermines inference for the associated VAR in equation (\ref{eq:VAR}). In this paper, we rely on the computation of Functional Principal Component Analysis (FPCA) described in \cite{tsay2016some}. We adopt FPCA due to its ease of use and the fact that it can be flexibly applied to summarize the bulk of the dynamic variation observed in $\widetilde{f_{t}}\left(\cdot\right)$
through a limited number of factors $\mathbf{\alpha_{t}}$.  

To set the stage, let $Z$ denote a $T \times N_{Z}$ matrix with $(j, t)^{th}$ element   $z_{jt} = \widetilde{f}_t(x_{jt})$.
We let $N_{Z}$ denote the number of grid points on which we evaluate the LQD. Typically we set this to a large value (such as $1000$). The matrix $Z$ can be decomposed using a truncated singular value decomposition (SVD):
\begin{equation}
    Z = S V D' + E
\end{equation}
where $S$ is a $T \times K$ matrix containing the first $K$ left eigenvectors, $V$ is a $K \times K$-dimensional diagonal matrix with the first $K$ largest eigenvalues on its main diagonal and $D$ is a $ N_Z \times K$ matrix of right eigenvectors. The term $E$ measures the approximation error committed by just considering the first $K$ eigenvectors. In what follows, we are going to use the first $K$ principal components, computed as $A = SV$ with $\alpha_t = A'_{t \cdot}$ denoting the $t^{th}$ row of $A$.  The associated orthogonal basis are therefore the $K$  columns of $D$.

The key intuition behind FPCA is that we aim to summarize cross-sectional information using a small number of factors. For example, if the density we wish to approximate is a time-varying Gaussian, the first two principal components of $Z$ will be closely related to the time-varying mean and standard deviation. In this case, two factors will be sufficient to adequately approximate the associated $\widetilde{f_t}$. If the cross-sectional density is non-Gaussian and exhibits more complex variation, the number of factors necessary to approximate  $\widetilde{f_t}$ becomes larger. In practice, several strategies can be used to determine the truncation point $K$. As with standard factor models, one can rely on information criteria as in \cite{bai2002determining}, perform cross-validation as suggested by \cite{tsay2016some}, or stop the truncation when the share of variation explained in the observed cross-section is above a desired threshold. Alternatively, one can try different $K$ to check how robust the analysis is to different choices. In this paper, we follow the latter strategy and repeat each application with different plausible $K$s, showing that the differences between the resulting functional IRFs of interest are mostly negligible. We also experimented with very small $K$ and found that, although the model was still able to replicate the crucial features of the responses, increasing the truncation point to $K=3$ for earnings and $K=5$ for consumption improved the performance of the model, while increasing it further did not generate any relevant differences. 

One advantage of this approach is that we introduce relatively little parametric restrictions. We start out by using a kernel density based estimate of the cross-sectional densities and transform it to become a simple function in $L^{2}$. By doing so we can decide on the grid of points at which the LQD is evaluated. Since the number of grid points determines the size of the cross-section, we can use results such as the ones in \cite{stock2016dynamic} and show that the principal components are consistent estimators of the true factors. What is more important, however, is that estimation uncertainty surrounding the factors declines appreciably. This implies that we do not have to estimate a full state space system but can simply plug in the PCA-based estimates of the latent factors. %For more details, see \cite{baumeister2023}.

Before discussing the estimation of the SVAR, it is worth stressing that the resulting functional VAR resembles a factor-augmented VAR similar to the one proposed in \cite{bernanke2005measuring}. This implies that the choice of how we capture dynamics in the cross-sectional densities impacts the intrinsic nature of the reduced form  innovations $u_{\alpha,t}$, but it does not affect the labeling of the first $n_{v}$ structural shocks in $\varepsilon_{t}$, $\varepsilon_{y,t}$, in a Cholesky identification scheme. As long as the orthogonal decomposition is able to reflect the time dependence between the aggregate variables and the cross-sectional distribution of interest, the performance of identification schemes used to disentangle the aggregate structural shocks of interest is not influenced by the chosen functional basis.

\subsection{Bayesian inference}\label{sec: inference}
Once the scores $\alpha_{t}$ are obtained via FPCA, the VAR in (\ref{eq:6})
is estimated with Bayesian methods. We specify a natural conjugate Gaussian-Inverse Wishart prior distribution for the reduced form parameters, see, e.g., \cite{sims1998bayesian, koop2010bayesian, giannone2015prior}. 

To simplify prior implementation, we rewrite the model more compactly as: 
\begin{equation}
z_{t}=\Pi x_{t}+u_{t},\label{eq:7}
\end{equation}
where $z_{t}=\left[y_{t}^{\prime},\alpha_{t}^{\prime}\right]^{\prime}$,
$x_{t}=\left[1,z_{t-1}^{\prime},\ldots,z_{t-p}^{\prime}\right]^{\prime}$
and $\Pi=\left[\Pi_{0},\Pi_{1},\ldots,\Pi_{p}\right]$, with $\Pi_{0}=\left[c_{y}^{\prime},\tilde{c}_{f}^{\prime}\right]^{\prime}$
and $\Pi_{l}=\left[\begin{array}{cc}
B_{l,yy} & \tilde{B}_{l,yf}C_{\alpha}\\
\tilde{B}_{l,fy} & \tilde{B}_{l,ff}C_{\alpha}
\end{array}\right]$ for $l=1,\ldots,p$.  This is a standard multivariate regression model which explains the observed aggregate macro series and the functional PCs using only lagged values of $z_t$.

The natural conjugate Gaussian-Inverse Wishart prior distribution for the reduced form parameters is factorized as
\begin{equation*}
 p\left(vec\left(\Pi^{\prime}\right),\Omega\right)=p\left(\Omega\right)\times p\left(vec\left(\Pi^{\prime}\right)\mid\Omega\right),
\end{equation*}
where $p\left(\Omega\right)$ is Inverse Wishart with $\nu$ degrees
of freedom and scale matrix $\Phi$, and $p\left(vec\left(\Pi^{\prime}\right)\mid\Omega\right)$
is Gaussian with mean $vec\left(\Psi\right)$ and variance $\Omega\otimes\Gamma$.
In specifying the parameters of these priors, we follow the Minnesota
tradition of \cite{doan1984forecasting} and set $\Psi$ to be a $n\times m$
($m=np+1$) matrix of zeros, except for the $(i,i+1)^{th}$ element
that is set to $1$ if the $i$-th variable of the system is known to
be persistent. The component of the prior variance $\Gamma$ is a
diagonal $m\times m$ matrix with the $(i,i)^{th}$ element equal to $10^{3}$
if $i=1$, and $\frac{\lambda_{1}^{2}}{\sigma_{j}l^{\lambda_{2}}}$
otherwise, where $j$ and $l$ are the index and lag of the right
hand variable to which the $j$-th column of $\Pi$ refers, and $\sigma_{j}$
is the error variance of an AR(1) model estimated by OLS for the $j$-th
variable. Finally, the parameters of the Inverse Wishart prior $p\left(\Omega\right)$
are set to $\nu=n+2$ and $\Phi=\text{diag}[\sigma_{1},\ldots,\sigma_{2}]$.
Throughout the paper, the hyperparameters $\lambda_{1}$ and $\lambda_{2}$
are set to $0.2$ and $2$ respectively, values commonly used in
the VAR literature.

Given these priors, the posterior distribution is also Gaussian-Inverse
Wishart 
\begin{equation*}
    p\left(vec\left(\Pi^{\prime}\right),\Omega\mid Y\right)=p\left(\Omega\mid Y\right)\times p\left(vec\left(\Pi^{\prime}\right)\mid Y,\Omega\right).
\end{equation*}
Specifically, $p\left(vec\left(\Pi^{\prime}\right)\mid Y,\Omega\right)$
is a Gaussian distribution 
 \begin{equation*}
     p\left(vec\left(\Pi^{\prime}\right)\mid Y,\Omega\right) = \mathcal{N}(\bar{\Psi}, \overline{\Omega})
 \end{equation*}
with variance and mean given by, respectively:
\begin{align*}
    \overline{\Omega} &=  \Omega\otimes\bar{\Gamma}, \quad \bar{\Gamma}=\left(\Gamma^{-1}+X^{\prime}X\right)^{-1}, \\
    \bar{\Psi}&=\bar{\Gamma}\left(\Gamma^{-1}\Psi+X^{\prime}Y\right),
\end{align*}
with $X=\left[x_{1}^{\prime},\ldots,x_{T}^{\prime}\right]^{\prime}$ and $Y=\left[z_{1}^{\prime},\ldots,z{}_{T}^{\prime}\right]^{\prime}$ denoting stacked data matrices.

The posterior of the covariance matrix follows an inverse Wishart distribution:
\begin{equation*}
    p\left(\Omega\mid Y\right) = \mathcal{W}^{-1}(\bar{\nu}, \bar{\Phi}).
\end{equation*}
The posterior degrees of freedom are $\bar{\nu}= \nu + T$ and the scaling matrix is $\bar{\Phi}=\Phi+Y^{\prime}Y+\Psi^{\prime}\Gamma^{-1}\Psi-\bar{\Psi}^{\prime}\bar{\Gamma}^{-1}\bar{\Psi}$.

The posterior just described is convenient because  draws from the stationary distribution 
can be easily obtained by direct Monte Carlo, sampling  from $p\left(\Omega\mid Y\right)$ first, and then drawing from
$p\left(vec\left(\Pi^{\prime}\right)\mid Y,\Omega\right)$. This completes the description of our functional VAR model. In Section \ref{sec:Simulation}, we will illustrate that it works well in terms of recovering the true impulse responses using simulated data. 

\subsection{Construction of distributional IRFs}\label{sec: FIRFs}
Before focusing on the application of the described procedure to artificial and real-world data, it is worth describing in more detail how we obtain posterior draws of the distributional IRFs. Following \cite{chang2021heterogeneity}, we assume that the system is at its steady state when the shock hits. For a given posterior draw of the model parameters, the LQD function before the shock is therefore $f_{ss}(\cdot)= \boldsymbol\zeta(\cdot)' \alpha_{ss}+\bar{f}$, where $\alpha_{ss}$ is the unconditional mean implied by the model parameters. The steady state distribution to which it corresponds, $p_{ss}(\cdot)$, can then be easily found by applying the inverse transformation, $p_{ss}(\cdot)=g^{-1}(f_{ss}(\cdot))$, described in Section \ref{LQD sec}.

The response of the functional PC, $\alpha_{t}$, to a structural shock of interest can be computed, for every posterior draw of the SVAR parameters, by standard methods, assuming that only the shock of interest at $h=0$ differs from zero. The resulting standard IRFs are therefore defined as $IRF_{\alpha,j,d,h}=E\left\{\alpha_{t+h}|I_{t-1},\varepsilon_{t,j}=d,\varepsilon_{t,-j}=0\right\} - E\left\{\alpha_{t+h}|I_{t-1},\varepsilon_{t}=0\right\}$. These IRFs can then be used to find the expected value of the LQD function $h$ periods after the shock, which we define as: $f_{ss+h}(\cdot)= \boldsymbol\zeta(\cdot)'(\alpha_{ss}+IRF_{\alpha,j,d,h})+\bar{f}$, and to the corresponding distribution of interest, $p_{ss+h}(\cdot)=g^{-1}(f_{ss+h}(\cdot))$, using the inverse transformation described in Section \ref{LQD sec}. The distributional IRFs that we report throughout the paper are then simply computed as the difference between $p_{ss+h}(x)$ and $p_{ss}(x)$, for every $x$ in a thin grid belonging to the support $\Xi$.

An important feature to take into consideration when constructing distributional IRFs is that, although the SVAR is linear, the map from $p(\cdot)$ to $f(\cdot)$ is not, which implies that both the initial level of the distribution, $p_{ss}(\cdot)$, and the size of the shock, $d$, are relevant and must be chosen carefully. In our applications, $p_{ss}(\cdot)$ is chosen to be the steady state implied by the SVAR parameters, while the shock size is set in terms of standard deviations. 

It is also important to notice that, unlike most of the existing literature, we do not need to re-normalize $p_{ss+h}(x)$ to ensure that it is positive and integrates to $1$ , since the use of the LQD ensures it by construction. In light of the discussion of \cite{Petersen2016FunctionalDA}, we believe that this represents a crucial advantage of our econometric procedure, since the re-normalization step operated by all the existing contributions to our knowledge can generate spurious shapes in the differences between $p_{ss+h}(x)$ and $p_{ss}(x)$. 

\section{Artificial Data Experiments\label{sec:Simulation}}
In this section we assess the ability of the F-SVAR, estimated through
FPCA and the Bayesian routine just described, to capture the propagation
of the aggregate structural shocks $\varepsilon_{y,t}$ to the distribution $p_{t}$. We do this by running
three experiments with data simulated from known Data Generating Processes
(DGPs). In the first experiment, the DGP is a simple F-SVAR of the form in (\ref{eq:6}), in which
the endogenous function is the deviation of the LQD from its sample mean, $\widetilde{f_{t}}$, which has an exact finite factor structure, as in (\ref{eq: finite_additive_repr}). In the second experiment the endogenous function of the F-SVAR is instead the logarithm of the distribution $p_{t}$ itself, which is also assumed to have an exact finite factor structure. In both cases, at each period a sample is drawn from the generated distribution and assumed to be observed, together with the realizations of the endogenous variables $y_{t}$ . In the final simulation experiment, the DGP is taken from Section 5 of \cite{chang2021heterogeneity}, which is based on a log-linearized solution of the \cite{krusell1998income} HANK model. In that case the endogenous function in the DGP represents the distribution of assets among the employed population.
Finally, in Section \ref{subsec:Alternative transformations} we use the three DGPs to examine how the choice of the transformation $g(p_t(\cdot))$ affects the goodness of fit of the approximation delivered by FPCA.

\subsection{F-SVAR: DGP 1\label{subsec:F-SVAR-DGP 1}}

The data used for the first experiment are generated by simulating
forward the model in (\ref{eq:7}),\footnote{The exact model parameters used to simulate the data are reported
in Appendix A.} setting $p=4$, $n_{v}=2$, and $K_{true}=3$. The realizations of the $K_{true}$-dimensional vector time series $\alpha_{t}$ are then transformed in LQD functions, using as basis the FPC computed from the LQDs of $50$
realizations of a mixture of Gamma distribution, where the mixing distribution is a $Beta(a,b)$, with $a$ and $b$ uniformly distributed in $[0, 3]$. Doing so, $f_{t}$ is obtained for $t=1,\ldots,500$, and then transformed into a distribution on $\Xi=[0, 6]$ through applying the inverse mapping $g^{-1}$. From the resulting distribution, a sample of size  $N=8000$ is drawn at every period and assumed to be observed by the econometrician.

\begin{figure}[t!]
\caption{\label{fig:ModesofVariation_dgp1}\textbf{Modes of variation: DGP 1}}
\begin{centering}
\includegraphics[bb=420bp 0bp 0bp 110bp,scale=1]{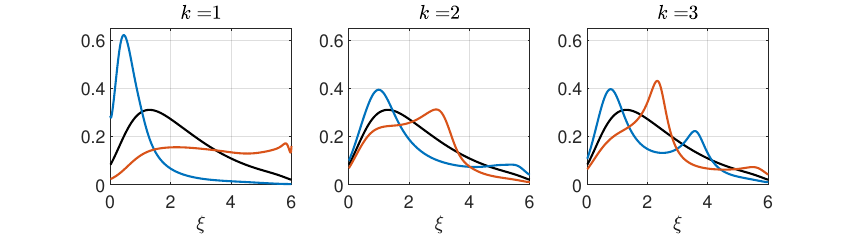}
\par\end{centering}
\centering{}%
\begin{minipage}[t]{0.9\columnwidth}%
\begin{spacing}{0.7}
{\footnotesize{}Notes: The black line depicts the sample mean of the distributions generated by the DGP. Red (blue) lines show the change implied by an increase (decrease) in $\alpha_{k,t}$ of 2 standard deviations.}
\end{spacing}
\end{minipage}
\end{figure}

To get an idea of the distributional dynamics generated by the DGP, in Figure \ref{fig:ModesofVariation_dgp1} we report the modes of variation implied by the DGP, that is the effect of a two standard deviations increase/decrease of each element of $\alpha_t$ on the endogenous distribution. 

Once the data have been simulated, we estimate the distribution $\hat{p_{t}}$ using the kernel estimator described in Sub-section \ref{sec:Kernel}, and then transform it to the LQD $\hat{f_{t}}=g\left(\hat{p_{t}}\right)$. We apply FPCA to the deviation from the mean of this functional data, setting $K=7$,\footnote{Results with different truncation points $K$ are reported in the appendix.}
and conduct inference about the VAR parameters employing the Bayesian methods described in the previous section.

\begin{figure}[t!]
\caption{\label{fig:Functional-IRFs-FSVAR_dgp}\textbf{Functional impulse responses arising from the F-SVAR: DGP 1}}
\begin{centering}
\includegraphics[bb=1000bp 0bp 0bp 450bp,scale=0.5]{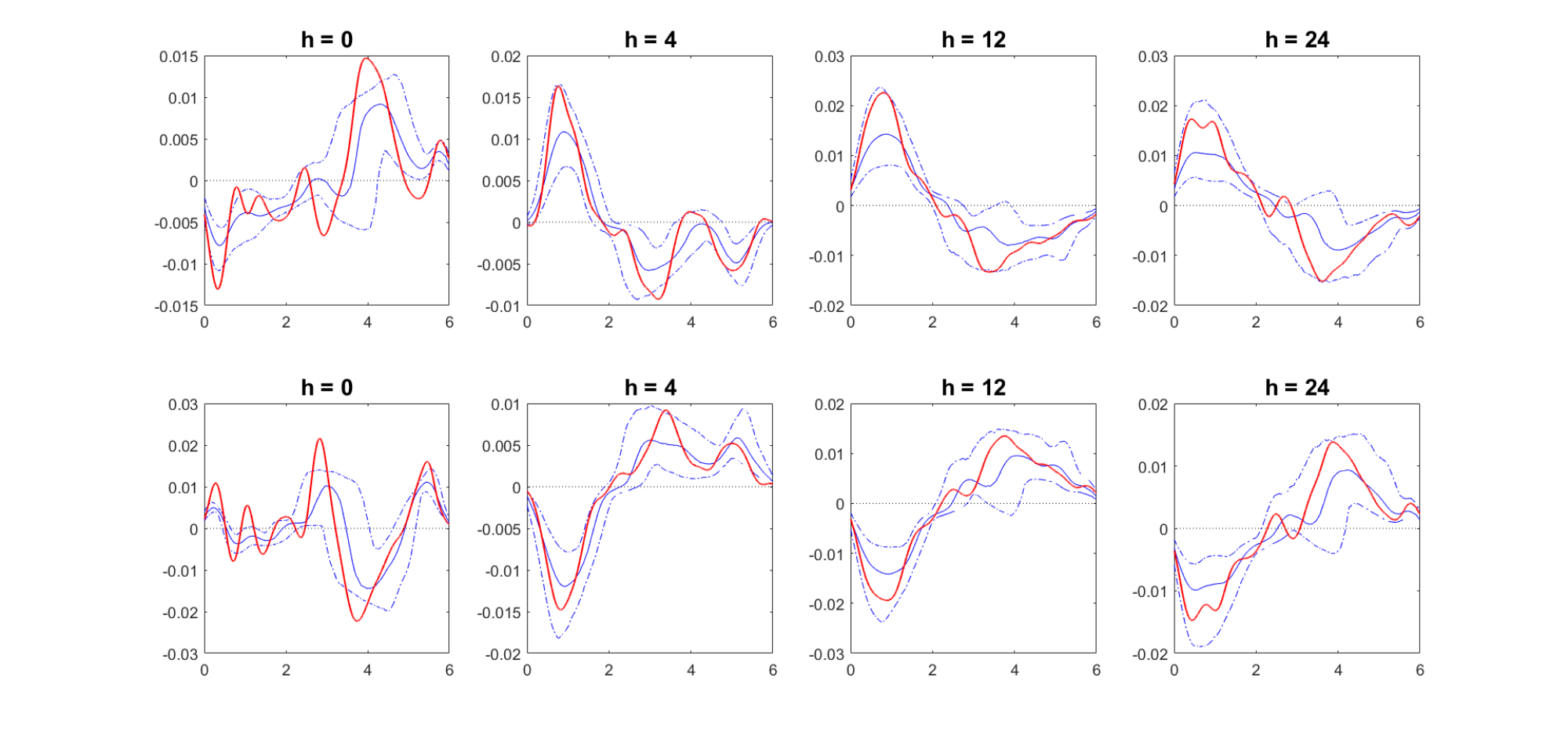}
\par\end{centering}
\centering{}%
\begin{minipage}[t]{0.9\columnwidth}%
\begin{spacing}{0.7}
{\footnotesize{}Notes: Red lines show the true responses of $p_{t}\left(\xi\right)$
to one standard deviation shocks to $\varepsilon_{1}$ (upper panels) and $\varepsilon_{2}$ (lower panels).
The solid blue lines represent the posterior median response, while
dashed blue lines delimit the 90\% credible bands. $h$ denotes the
horizon at which the response is measured. }
\end{spacing}
\end{minipage}
\end{figure}

Figure \ref{fig:Functional-IRFs-FSVAR_dgp} shows the $h$-step-ahead IRFs of $p_{t}$
to the first $n_{v}=2$ structural shocks ($\varepsilon_{y}$) implied by the DGP (red), together with the posterior median (solid blue) and the 90\% credible bands (dashed blue) obtained for the model-based functional IRFs.  We consider $h \in \{0, 4, 12, 24\}$. The two rows refer to the two structural shocks, while each panel in every row focuses on one of the horizons considered. As explained in Section \ref{sec: FIRFs}, the distributional IRFs are computed as the difference between the distribution $h$ periods after the shock and that corresponding to the steady-state. Hence, the horizontal axis reports the support $\Xi$ of the distribution.
The figure shows that the methodology described in Section \ref{sec:FSVAR} does a remarkably good job in tracking the propagation of the two  shocks
to the distribution of interest at all horizons. In particular, the model-based IRFs track the horizon-specific IRFs along the cross-section well. Only in some rare cases, the actual IRFs are outside the credible intervals. 
It is also interesting to note how the functional IRFs change as the number of principal components $K$ included in the estimated model changes. The responses produced by the F-SVAR with $K=1,2,3,15$ are reported in the appendix and show that, as the number of principal components increases from $1$ to $3$, the accuracy of the inferred IRFs improves steadily, while for $K \geq 3$  the responses produced by the model are virtually the same. In particular, the model including only one principal component, although able to capture important features of the IRFs, fails to replicate the impact responses adequately. On the other hand, when at least three principal components are included, the model is able to track most of the distributional variation with credible bands that contain  the true response most of the time. As seen from Figure \ref{fig:Functional-IRFs-FSVAR_dgpk15}, increasing the number of factors included to a moderately large number does not translate into substantially larger  credible bands. We take this as evidence that Bayesian prior shrinkage does a good job in limiting the efficiency loss resulting from estimating a larger model.

\begin{figure}[t!]
\caption{\label{fig:Corr MC 1}\textbf{\small{Average correlation between median and true functional IRF: DGP 1}}}
\begin{centering}
\includegraphics[bb=1600bp 50bp 10bp 750bp,scale=0.40]{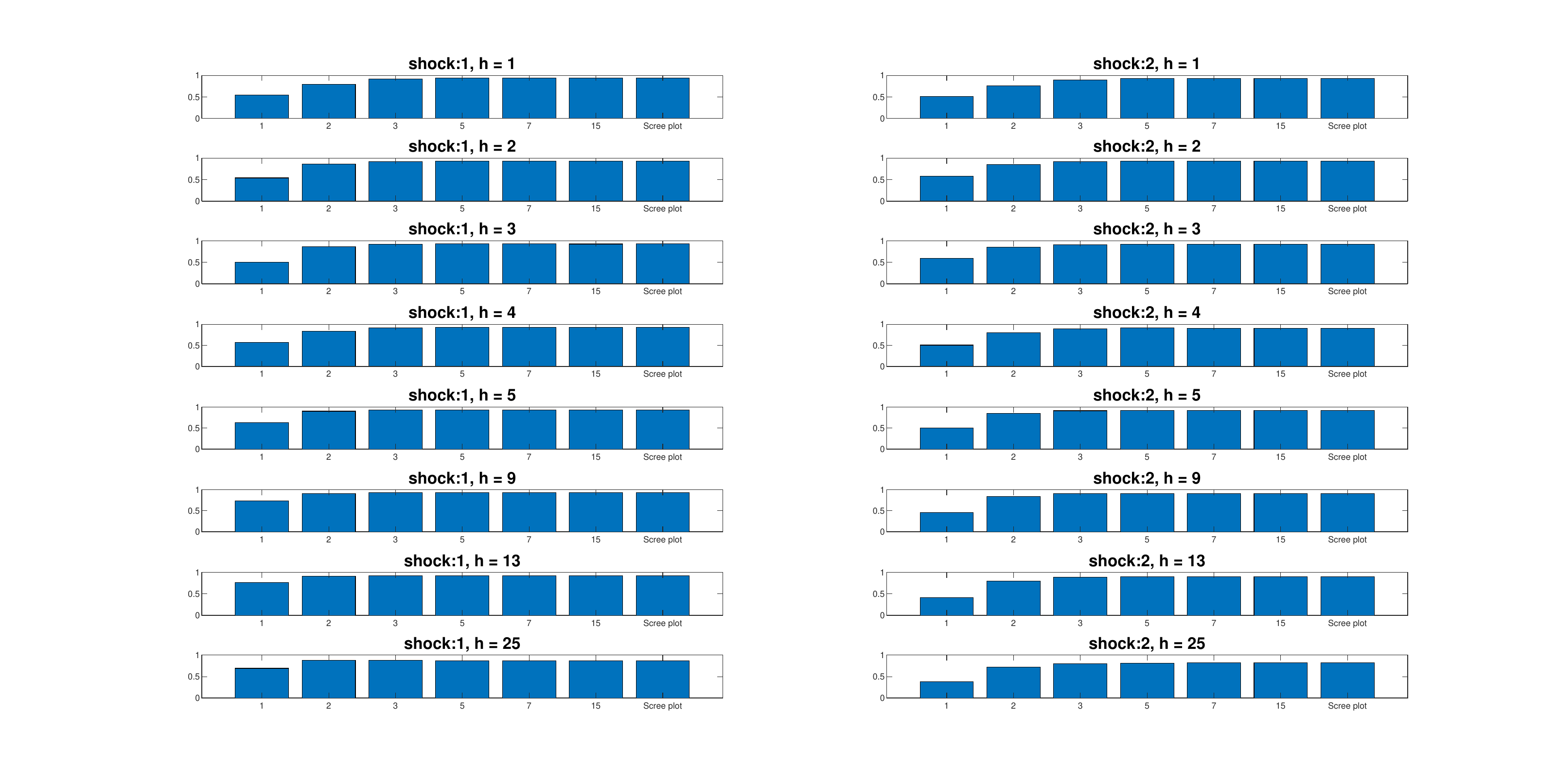}
\par\end{centering}
\centering{}%
\begin{minipage}[t]{0.9\columnwidth}%
\begin{spacing}{0.7}
{\footnotesize{}Notes: The coefficient is computed as the average across Monte Carlo repetitions and refers to the correlation between the point-wise median of the posterior distribution of functional IRFs and the true responses implied by the DGP. }
\end{spacing}
\end{minipage}
\end{figure}

To check that the desirable properties we have described so far for a specific realization from the DGP are maintained in repeated samples, we generate 200 additional samples from the same DGP and apply the inference procedure described. For every replication and different values of K, we compute the correlation between the true functional responses to the $n_{v}$ aggregated shocks $\varepsilon_{y}$ at various horizons and the posterior median functional IRF inferred by the estimated F-SVAR. In particular, we consider functional responses at horizons $h=0, 1, 2, 3, 4, 8, 12, 24$, and repeat the inference procedure for $K=1, 2, 3, 5, 7, 15$. In addition, we consider the case in which $K$ is set to be the smallest values for which the variance explained by the FPC is at least 90\% (we refer to this method as 'scree plot' method using the terminology of Tsay, 2016). The bar charts in Figure \ref{fig:Corr MC 1} show the average correlation coefficients attained across replications. As in the single replication reported above, for $K$ sufficiently large, the responses inferred by the F-SVAR are able to replicate the bulk of the distributional changes generated by the structural shocks in the true model. Moreover, as discussed earlier, increasing the number of principal components from 1 to 3 generates a substantial gain in the accuracy of the functional IRFs, but setting $K$ above 5, or selecting it through the 'scree plot' method, does not produce any additional benefit.      

\subsection{F-SVAR: DGP 2\label{subsec:F-SVAR-DGP 2}}

The data used in the first experiment were generated from a model that differed only slightly from the model used to conduct inference.\footnote{The difference between the DGP and the estimated model is in the functional basis used and in the true number of factors $\alpha_t$, however the endogenous function in the VAR DGP is the LQD, as postulated by the estimated model.} We now want to assess the reliability of our inference procedure in a more challenging setting.

\begin{figure}[t!]
\caption{\label{fig:ModesofVariation_dgp2}\textbf{Modes of variation: DGP 2}}
\begin{centering}
\includegraphics[bb=420bp 0bp 0bp 110bp,scale=1]{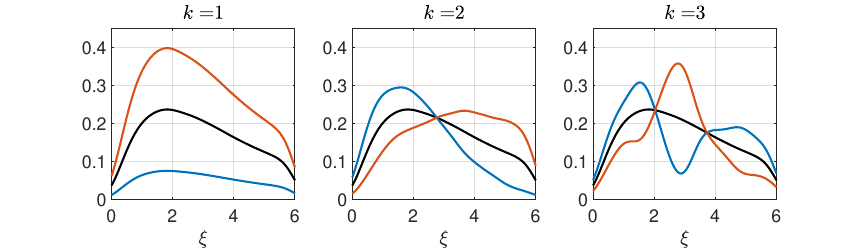}
\par\end{centering}
\centering{}%
\begin{minipage}[t]{0.9\columnwidth}%
\begin{spacing}{0.7}
{\footnotesize{}Notes: The black line depicts the sample mean of the distributions generated by the DGP. Red (blue) lines show the change implied by an increase (decrease) in $\alpha_{k,t}$ of 2 standard deviations.}
\end{spacing}
\end{minipage}
\end{figure}

In the second experiment, we increase the degree of mis-specification of the estimated model by assuming that the endogenous function in the DGP is the logarithm of the distribution $p_{t}$, rather than the LQD as postulated by the estimated model. Since the transformation $g(\cdot)$ from $p_{t}$ to the LQD  $f_{t}$ is highly non-linear, the true dynamics of the $f_t$'s are also highly non-linear, while our procedure assumes they follow a linear VAR. This second experiment therefore represents  a notably more demanding challenge for the estimated model.
As in the previous sub-section, we generate $500$ observations of the vector time series $\left[y'_{t}, \alpha'_{t}\right]^{\prime}$ by simulating
 the model in (\ref{eq:7}) forward. We  set $p=4$, $n_{v}=2$, and $K_{true}=3$.\footnote{Again, the exact model parameters used to simulate the data are reported
in Appendix A.} Differently from the first experiment, however, the obtained $\alpha_t$'s  are directly used to generate realizations of  $\log{p_{t}}$ (with support $\Xi=\left[0, 6\right]$)  using as basis the first $K_{true}$ FPC of the logarithm of the mixture of Gamma distributions described in the previous sub-section. At every time period in the sample we normalize the obtained $p_{t}$ to have unit integral and draw a sample of  $N=8000$ observations from it.
Figure \ref{fig:ModesofVariation_dgp2} depicts the modes of variation implied by the DGP.

\begin{figure}[t!]
\caption{\label{fig:Functional-IRFs-FSVAR_dgp 2}\textbf{Functional impulse responses arising from the F-SVAR: DGP 2}}
\begin{centering}
\includegraphics[bb=30bp 20bp 1030bp 450bp,scale=0.5]{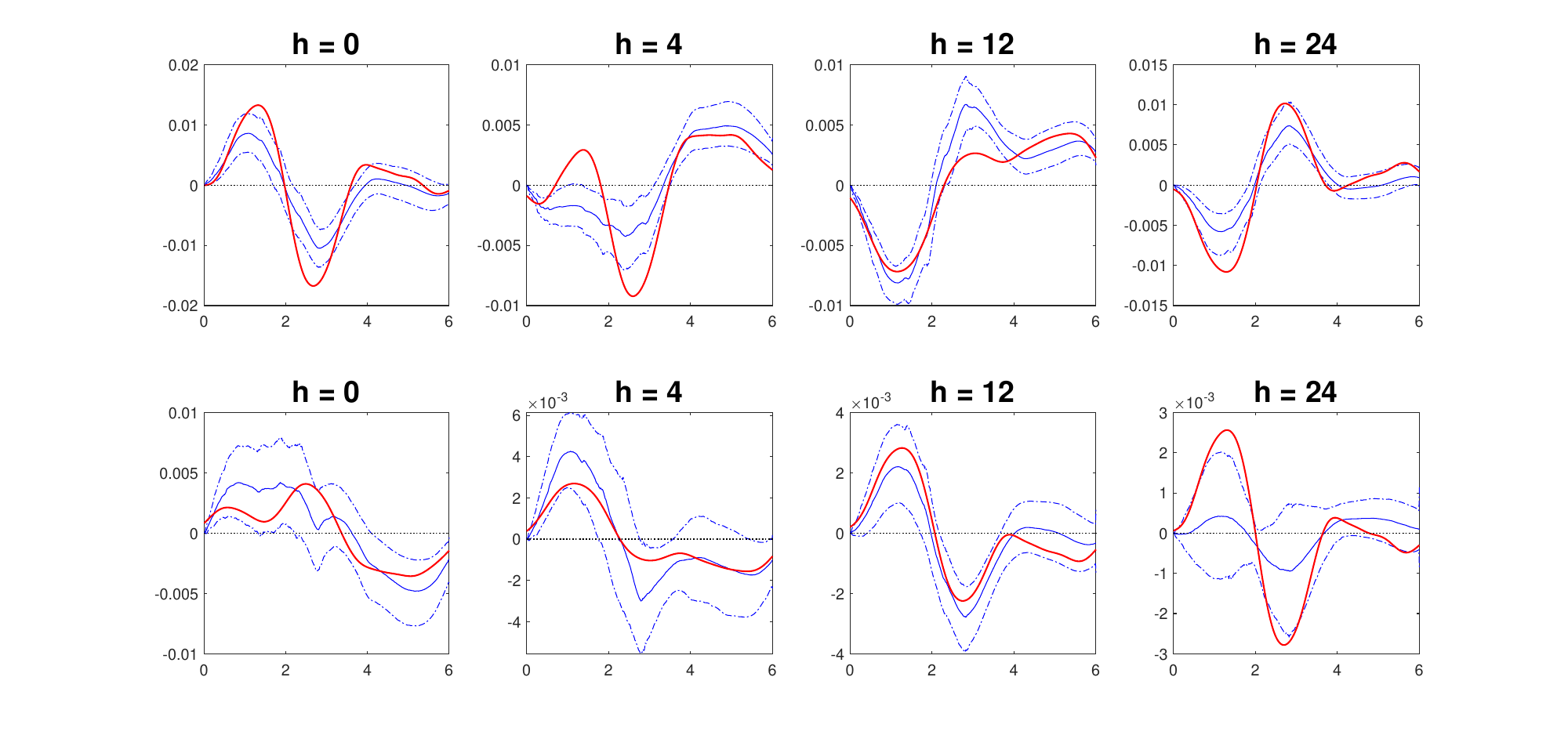}
\par\end{centering}
\centering{}%
\begin{minipage}[t]{0.9\columnwidth}%
\begin{spacing}{0.7}
{\footnotesize{}Notes: Red lines show the true responses of $p_{t}\left(\xi\right)$
to one standard deviation $\varepsilon_{1}$ (upper panels) and $\varepsilon_{2}$ (lower panels).
The solid blue lines represent the posterior median response, while
dashed blue lines delimit the 90\% credible bands. $h$ denotes the
horizon at which the response is measured. }
\end{spacing}
\end{minipage}
\end{figure}

Figure \ref{fig:Functional-IRFs-FSVAR_dgp 2} compares the IRFs of $p_{t}$
to the $\varepsilon_y$ shocks implied by the DGP (red), with the posterior median (solid blue)
and the 90\% credible bands (dashed blue) obtained by the estimated model setting $K=7$.\footnote{Results with different truncation points $K$ are very similar and are reported in the appendix.} Despite the higher degree of mis-specification of the estimated model, our procedure is able, also in this case, to replicate satisfactorily the salient features of the responses of the distribution of interest to the first $n_{v}$ structural shocks, especially at shorter and medium horizons.

\begin{figure}[t!]
\caption{\label{fig:Corr MC 2}\textbf{\small{Average correlation between median and true functional IRF: DGP 2}}}
\begin{centering}
\includegraphics[bb=1600bp 50bp 10bp 750bp,scale=0.40]{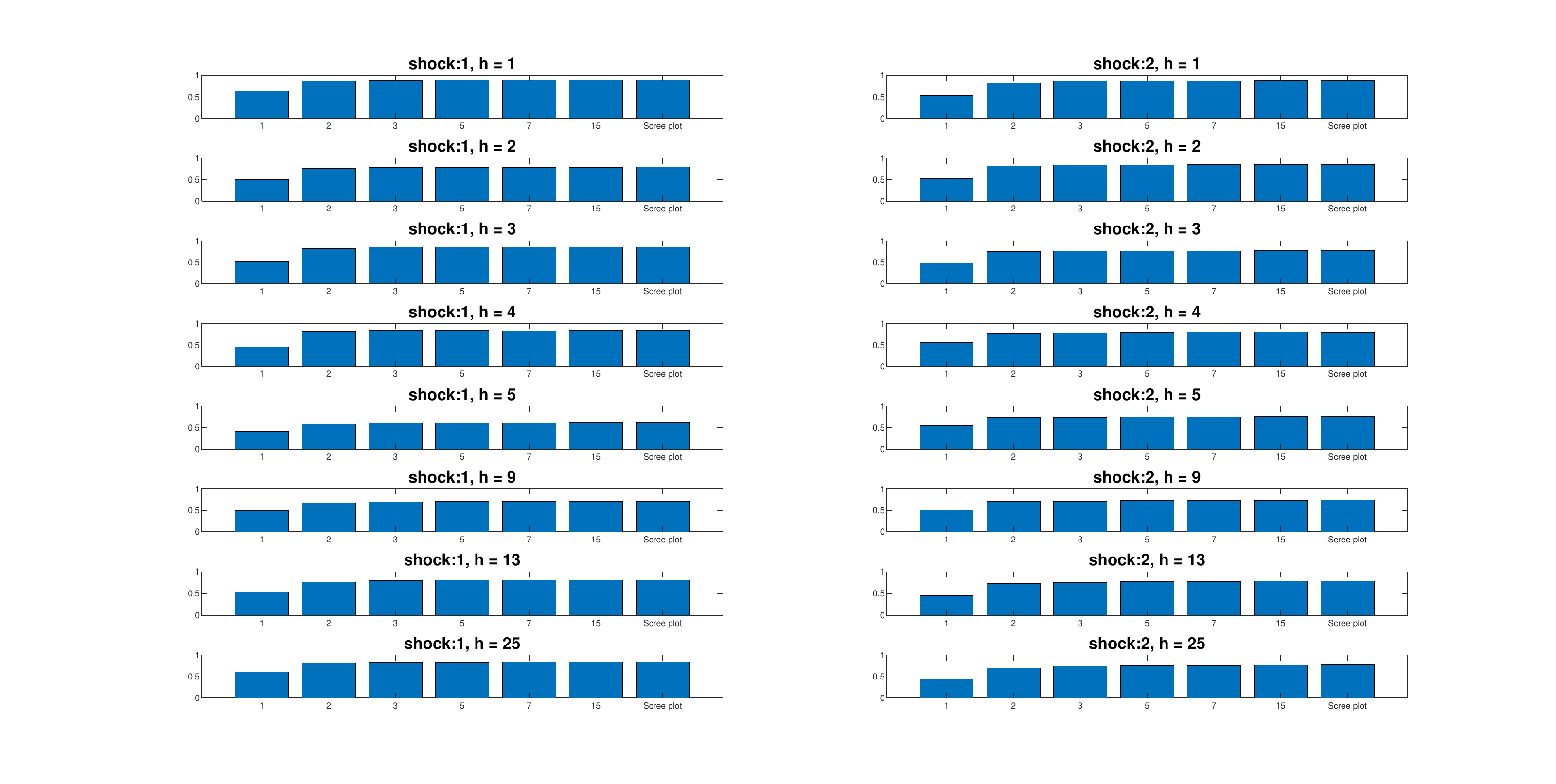}
\par\end{centering}
\centering{}%
\begin{minipage}[t]{0.9\columnwidth}%
\begin{spacing}{0.7}
{\footnotesize{}Notes: The coefficient is computed as the average across Monte Carlo repetitions and refers to the correlation between the point-wise median of the posterior distribution of functional IRFs and the true responses implied by the DGP. }
\end{spacing}
\end{minipage}
\end{figure}

Next, as we did in the previous sub-section, we produce 200 additional replications of the experiment to check that the good performance just described is not limited to the particular sample obtained. The average correlation coefficients between the median of the posterior of functional IRFs and the ones implied by the DGPs are again large and are shown in Figure \ref{fig:Corr MC 2}. Also in this case the correlation values are generally large, increasing the number of FPCs from 1 to 3 improves  the accuracy of the inferred functional responses considerably, but the improvement of setting $K$ to larger values is negligible on average.

\subsection{Krusell and Smith (1998) DGP\label{subsec:F-SVAR-DGP KS}}

\begin{figure}[t!]
\caption{\textbf{\label{fig:Functional-IRFs--KS98}Functional impulse responses arising from the F-SVAR: Krusell
and Smith (1998) DGP}}
\includegraphics[bb=210bp 0bp 250bp 750bp,scale=0.4]{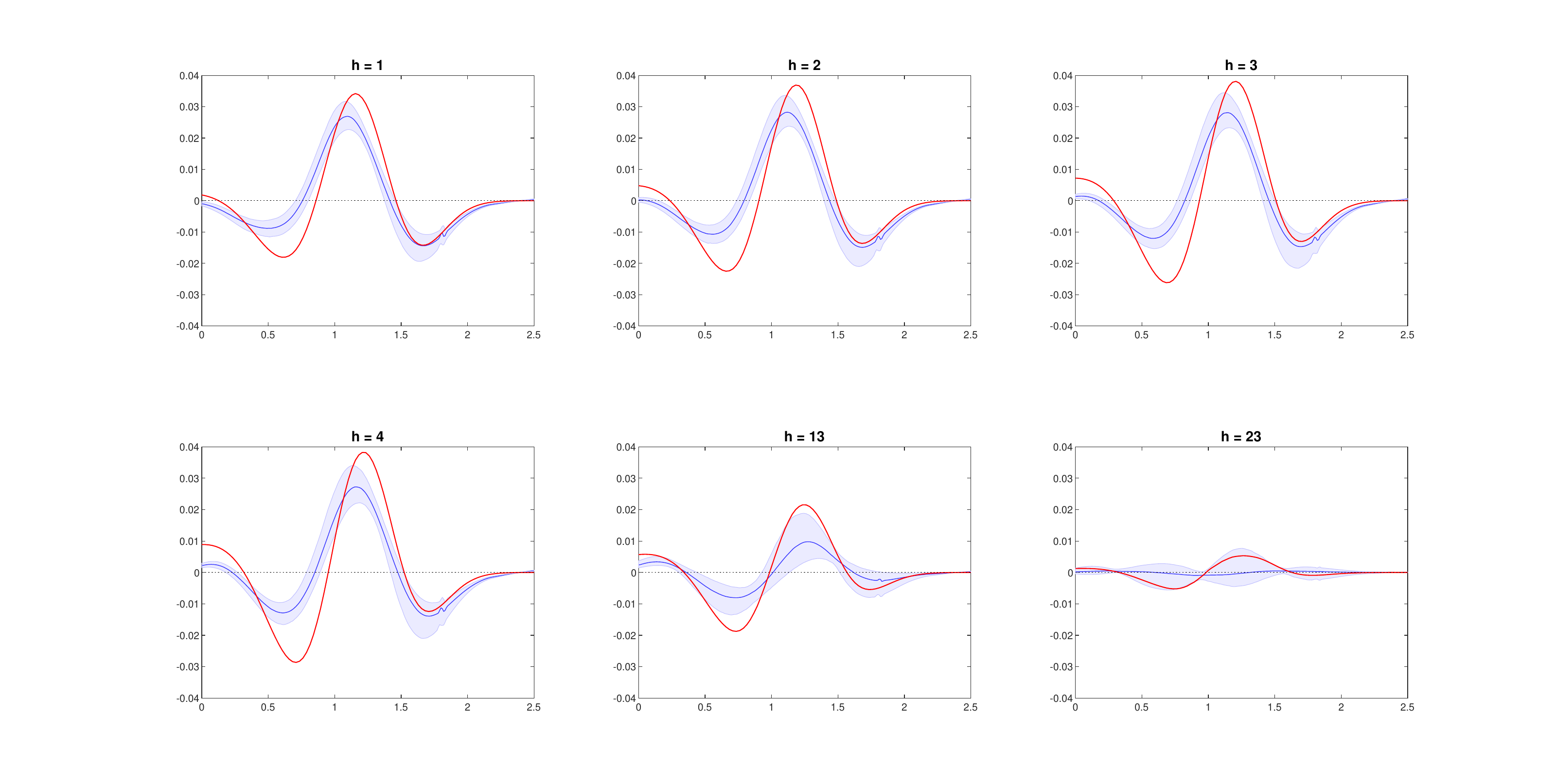}
\centering{}%
\begin{minipage}[t]{0.95\columnwidth}%
\begin{spacing}{0.69999999999999996}
{\footnotesize{}Notes: Red lines show the true responses of the asset
distribution to a one standard deviation productivity shock. The solid
blue lines represent the posterior median response, while dashed blue
lines delimit the 90\% credible bands. $h$ denotes the horizon at
which the response is measured. Notice that the horizons for which
responses are reported are the same as those in Figure 6 of \cite{chang2021heterogeneity}, the difference in the panels titles only
stems from different timing conventions. }
\end{spacing}
\end{minipage}
\end{figure}

The third experiment we perform is inspired by \cite{chang2021heterogeneity} and makes use of the \cite{krusell1998income} model as DGP. The log-linearized solution of the model proposed by \cite{krusell1998income} implies a VAR law of motion for the productivity level, the
capital stock, the employment level, and the centered moments of the distribution of assets among the employed. \cite{chang2021heterogeneity} generate 160 artificial observations from this VAR and use them to estimate their proposed model. In this section, we use the same artificial data kindly made available by the authors.\footnote{We obtained the data from Frank Schorfheide's website: web.sas.upenn.edu/schorf/working-papers/.} For inference, we set $p=1$, $K=7$,\footnote{Results with different truncation points $K$ are very similar and are reported in the appendix.} and specify the natural conjugate prior described in the previous section. Figure \ref{fig:Functional-IRFs--KS98} compares the functional IRFs implied by the estimated F-SVAR (blue) with the true responses implied by the DGP. While the inferred responses tend in some cases to underestimate the deviation of the shocked distribution from the steady state counterpart, Figure \ref{fig:Functional-IRFs--KS98} shows that the F-SVAR is still  able to reproduce the salient features of the dynamic responses. 

\subsection{Alternative Transformations of \texorpdfstring{$p_t(\cdot)$}{pt}\label{subsec:Alternative transformations}}
Although we have shown that the econometric strategy described in Section \ref{sec:FSVAR} works well in a variety of settings, and despite we have already discussed the advantages of making use of the LQD transformation, in this section we want to understand how the goodness of fit of the FPC approximation of the distribution is affected when different transformations $g(p_t(\cdot))$ are considered. Specifically we compare the use of the LQD with the cases in which FPCA is performed on the deviation from the mean (i) of density itself, i.e $g(p_t(\cdot))=p_t(\cdot)$, and (ii) of the logarithm of the density, i.e. $g(p_t(\cdot))=log(p_t(\cdot))$.
In order to do that, we implement a cross-validation exercise using samples of distributions generated by the DGPs described in Sections \ref{subsec:F-SVAR-DGP 1}, \ref{subsec:F-SVAR-DGP 2}, and \ref{subsec:F-SVAR-DGP KS}. For each DGP and different truncation points $K$, we extract FPCs from 80\%  of the distributions in the sample (randomly selected), and use those to approximate the remaining 20\% of the available distributions. More specifically, once we extract the basis $\boldsymbol\zeta\left(x\right)$ through FPCA from the training sub-sample, we estimate by OLS the $\alpha_t$ coefficients associated with the distributions in the validation sub-sample. The obtained approximations are then used to compute the Mean Integrated Squared Error over the validation sub-sample:

\begin{equation}
MISE=\frac{1}{T}\sum_{t=1}^{T}\int_{\Xi}\left(\hat{p}_{t}\left(\xi\right)-p_{t}\left(\xi\right)\right)^{2}d\xi
\end{equation}

where $\hat{p}_{t}\left(\xi\right)=g^{-1}\left(\bar{f}+\zeta\left(\xi\right)^{\prime}\hat{\alpha_{t}}\right)$.
We repeat the exercise 100 times and report the average MISE attained by each method in Table \ref{tab:MISE}.

The Table shows that performing FPCA directly on the distribution provides the most efficient approximation, i.e. it achieves the lowest MISE for a given number of FPCs. The nonlinear transformations operated by the logarithm and by the LQD, instead, deteriorate to some extent the goodness of fit of the approximation. Intuitively, FPCA is designed to provide an optimal approximation in the transformed space, but the goodness of fit inevitably decreases once the approximation is projected back to the density space. 

The Table, however, shows that the use of the LQD allows a substantially better approximation of the distribution when compared to the log-transformation. 
To summarize, although there is some loss in the goodness of fit provided by the implementation of FPCA on the LQD rather than on $p_t(\cdot)$ itself, we think that the advantages discussed in Section \ref{LQD sec}, and the superiority in terms of approximation errors when compared to the logarithm transformation, justify our preference for the use of LQD over alternative transformations. In addition, \cite{Petersen2016FunctionalDA} provide examples in which the goodness of the approximation obtained by performing FPCA on the LQD can even exceed that obtained by performing FPCA directly on $p_t(\cdot)$. This mainly happens when large horizontal variation is observed across the distributions in the sample. 

\begin{center}
\begin{table}[h]
\begin{centering}
\caption{Cross-validation exercise: MISE}
\label{tab:MISE}  
\par\end{centering}
\begin{centering}
\begin{tabular}{ccccccc}
 &  & \multicolumn{5}{c}{$K$}\tabularnewline
 &  & $1$ & $2$ & $3$ & $4$ & $5$\tabularnewline
\hline 
\multirow{3}{*}{\begin{turn}{90}
DGP1
\end{turn}} & $p_{t}\left(\cdot\right)$ & $1$ & $0.465$ & $0.249$ & $0.127$ & $0.069$\tabularnewline
 & $\log p_{t}\left(\cdot\right)$ & $1.879$ & $1.457$ & $0.996$ & $0.730$ & $0.594$\tabularnewline
 & LQD & $1.085$ & $0.644$ & $0.530$ & $0.370$ & $0.307$\tabularnewline
\hline 
\multirow{3}{*}{\begin{turn}{90}
DGP2
\end{turn}} & $p_{t}\left(\cdot\right)$ & $1$ & $0.102$ & $0.053$ & $0.032$ & $0.022$\tabularnewline
 & $\log p_{t}\left(\cdot\right)$ & $5.150$ & $1.982$ & $3.143$ & $1.671$ & $1.128$\tabularnewline
 & LQD & $2.334$ & $1.167$ & $0.678$ & $0.421$ & $0.337$\tabularnewline
\hline 
\multirow{3}{*}{\begin{turn}{90}
DGP3
\end{turn}} & $p_{t}\left(\cdot\right)$ & $1$ & $0.598$ & $0.486$ & $0.395$ & $0.324$\tabularnewline
 & $\log p_{t}\left(\cdot\right)$ & $1.908$ & $1.857$ & $1.467$ & $1.339$ & $1.260$\tabularnewline
 & LQD & $1.449$ & $1.199$ & $1.174$ & $1.066$ & $1.020$\tabularnewline
\hline 
\end{tabular}
\par\end{centering}
\centering{}Ratios relative to the MISE attained by the first approach
for $K=1$. 
\end{table}
\par\end{center}

\section{\large{The Distributional Implications of Uncertainty Shocks}\label{sec:JLN15}}

Having demonstrated the reliability of the adopted econometric procedure, the aim of this section is to apply these techniques to study the distributional consequences of uncertainty shocks. We start by discussing the dataset and how we identify an uncertainty shock before discussing the reaction of the macroeconomic aggregates to uncertainty shocks. Afterwards, we focus on how the cross-sectional distributions of earnings and consumption reacts to unexpected movements in economic uncertainty.  

\subsection{Data overview, structural identification and model specification}
We analyze the relationship between economic uncertainty, macro aggregates and the distributions of earnings and consumption by building on the linear VAR model put forth in \citet[][JLN]{jurado2015measuring}. In addition to the aggregate series discussed below, this amounts to including alternately the distribution of earning among employed people or the distribution of consumption as endogenous functional variable. We compile time series of earnings-to-GDP distributions using data constructed by \cite{chang2021heterogeneity} based on the Current Population Survey (CPS). 
When we focus on the consumption distribution, we employ the micro data from the Consumption Expediture Survey (CE) used by \cite{chang2024effects}. In this dataset, every observation is divided by the per capita consumption level, so that an observation equal to 1 represents a household consuming the national per capita level.\footnote{Both datasets can be downloaded from Frank Schorfheide's website: web.sas.upenn.edu/schorf/working-papers/.}
Since in \cite{chang2021heterogeneity}'s and \cite{chang2024effects}'s datasets the micro data are available at quarterly frequency, we convert the monthly SVAR(12) of JLN into a quarterly F-SVAR(4) model. 
The sample period for the F-VAR including the earnings distribution runs from 1989:Q1 to 2017:Q3. When we focus on the consumption distribution, the sample period goes from 1990:Q2 to 2016:Q4.

As endogenous aggregate series, we include $11$ series in the model. These are: real GDP (\texttt{rGDP}), real PCE (\texttt{rPCE}), the GDP deflator (\texttt{GDPdef}),
real wages (\texttt{rW}), real investments (\texttt{rINV}), labor productivity (\texttt{Lprod}), the unemployment rate (\texttt{unr}), the Federal Funds Rate (\texttt{ffr}), the S\&P500 index (\texttt{sp500}), the M2 growth rate (\texttt{M2}) and the macro-uncertainty measure constructed by JLN (\texttt{U}$_m$ \texttt{h3}).\footnote{All variables can be downloaded from the FRED MD database, except
for the uncertainty measure which is available at www.sydneyludvigson.com/macro-and-financial-uncertainty-indexes.

While real GDP, real PCE, the GDP deflator, real wages, real investments, labor productivity, the unemployment rate are available at quarterly frequency, the Federal Funds Rate, the S\&P500 index, the M2 growth rate and the macro-uncertainty measure are aggregated from the monthly series. } All variables are considered in logarithm, except for the interest rate and uncertainty measure. 

\begin{figure}[ht!]
\caption{\label{fig:ScreePlot-CPS} Share of variance explained by Functional PC}
\includegraphics[viewport=50bp 20bp 1920bp 430bp,scale=0.5]{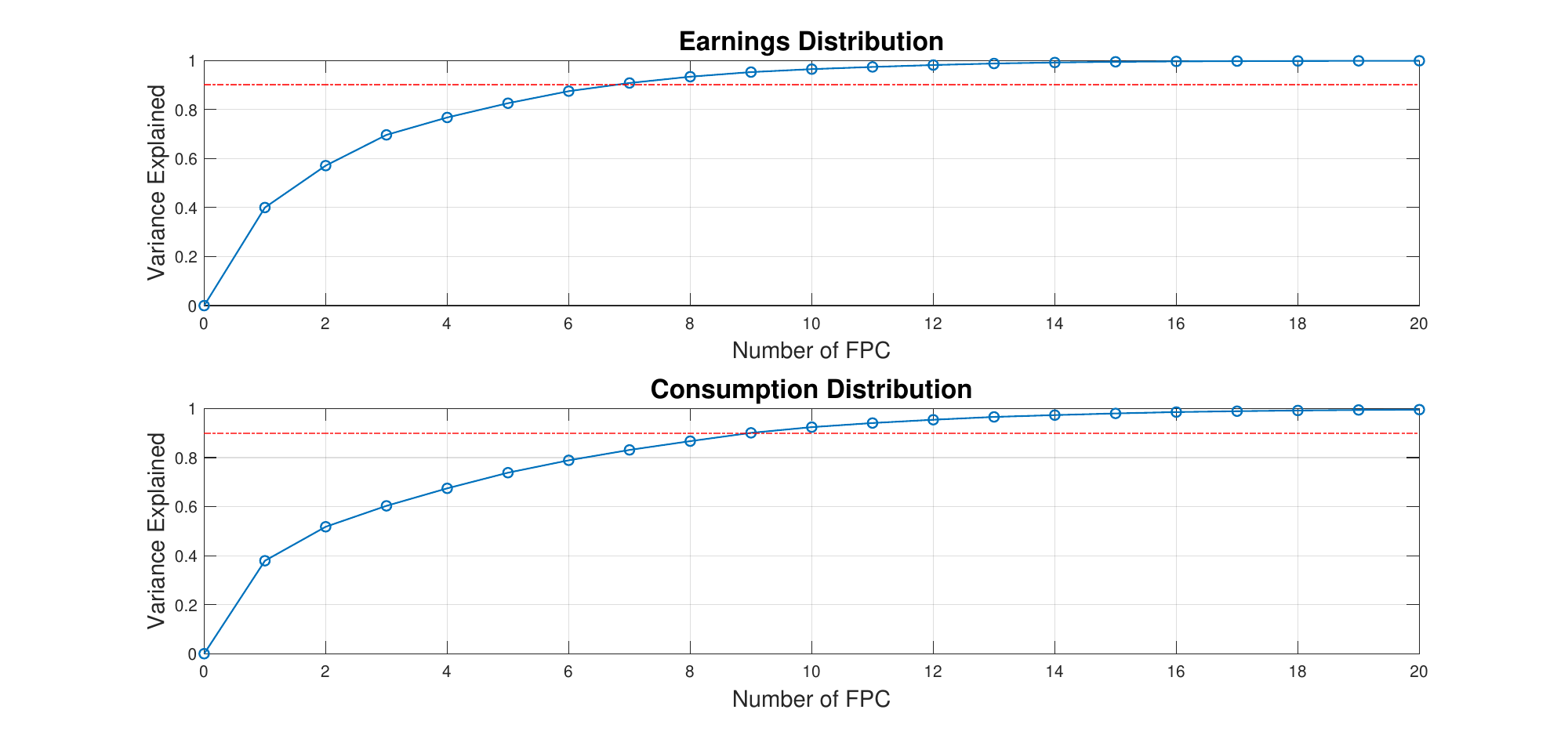}
\begin{minipage}[t]{1\columnwidth}%
\begin{spacing}{0.69999999999999996}
{\footnotesize{}Notes: The blue line denotes the cumulated share of in-sample variation explained by the Functional PC. The dashed red line denote the threshold we apply for selecting the truncation point $K$.}
\end{spacing}
\end{minipage}
\end{figure}

To these variables we append the $K$ factors $\alpha_{t}$ obtained through FPCA of the LQD associated to the earnings or the consumption data. The chosen value of $K$ is that for which the cumulative share of in-sample variance explained by the FPCs is closest to 90\%. The scree plot in Figure \ref{fig:ScreePlot-CPS} shows that, for earnings-to-GDP data, the first three FPCs are able to explain already 70\% of the in-sample variation observed in the demeaned LQD, while the subsequent four PCs explain an additional share that accounts for over 20\%. The first seven FPCs are therefore sufficient to explain more than 90\% of the total variation observed. For consumption data, on the other hand, the first three FPCs explain 60\% of the time-variation, and nine FPCs are enough to explain 90\% of the in-sample variance. The relatively small value of FPCs required to reflect the bulk of the variation observed in the function of interest is indeed one of the main advantages of using FPCA instead of alternative functional bases, such as splines, for which a larger number would be needed. Furthermore, in our case, the exact choice of $K$ does not appear to be crucial for the analysis, as long as it is not set to an excessively small number. In fact, we have experimented with different choices of $K$ and noticed that different values did not produce
any relevant differences in the results, as long as the  value was greater than $2$ for earning data and greater than $4$ for consumption data.\footnote{ Figures in Appendix \ref{sec:Uncertainty Shocks - Results with Different $K$} show functional IRFs obtained by setting $K=3$ or $K=15$ for earnings data and $K=5$ or $K=15$ for consumption data.}

As CPS and CE data are top-coded with censoring values changing several times along the sample period, we limit the support of the distribution to the smallest censoring values applied in the surveys, which is slightly above an earnings-to-DGP per-capita ratio equal to 2 for CPS data, and around 3.2 times the per capita level of consumption for CE data. Our analysis remains therefore silent about the dynamics concerning the higher end of the earnings and consumption distributions, the study of such dynamics is left for future research.

Following JLN, the macro uncertainty shock is identified by ordering
the uncertainty measure last among the endogenous variables in a Cholesky
identification scheme. The idea behind this strategy is to reflect
a structural source of fluctuation in the uncertainty measure that
remains after accounting for other contemporaneous developments affecting
the macroeconomy. 

\subsection{Dynamic reactions of the aggregate economy to uncertainty shocks}
\begin{figure}[ht!]
\caption{\label{fig:JLN-IRFs-} Impulse response of the macroeconomic aggregates to a one standard deviation  uncertainty shock}
\includegraphics[viewport=80bp 20bp 1920bp 500bp,scale=0.65]{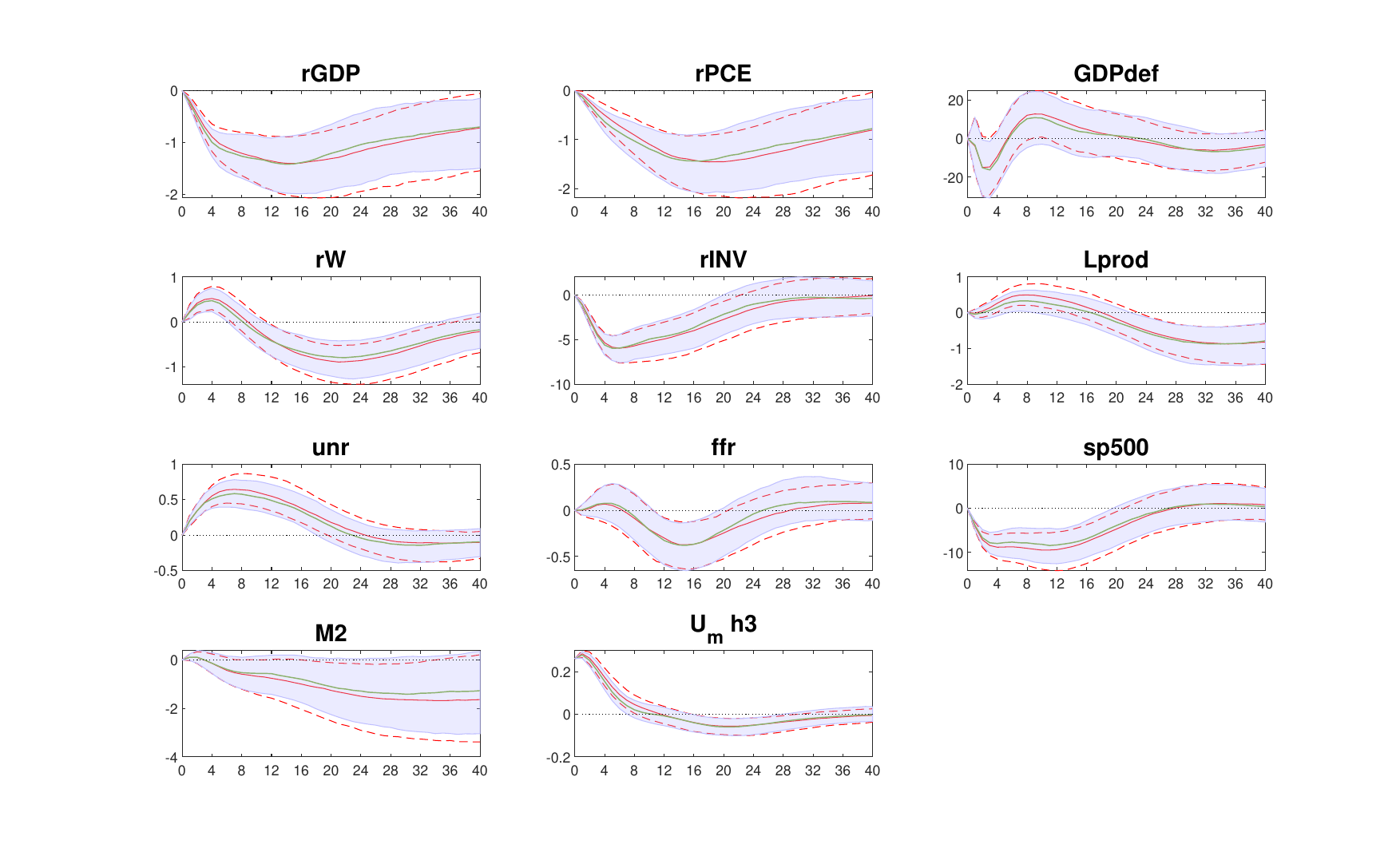}
\begin{minipage}[t]{1\columnwidth}%
\begin{spacing}{0.69999999999999996}
{\footnotesize{}Notes: The solid lines represent the posterior
median response, while dashed lines delimit the 68\% credible
bands. The horizon on the horizontal axis is expressed in quarters. Blue lines are generated by the F-SVAR, red lined are generated by the standard VAR.}
\end{spacing}
\end{minipage}
\end{figure}

We start our discussion by focusing on the responses of the aggregate macro series first. These are shown in  
Figure \ref{fig:JLN-IRFs-}. To understand the effects of adding the cross-sectional distribution of earnings or consumption to the VAR, we contrast the reactions of the F-SVAR (in blue) to the ones of a standard SVAR that features aggregate series only (in red).\footnote{In the Figure we report the IRFs generated by the F-VAR including the earnings distribution. Those generated by the F-VAR including the distribution of consumption are virtually identical.} All these responses are to a one standard deviation shock, and we report 68\% credible bands in both cases.

At a general level, the figure indicates that the responses are consistent with what has been found in the literature (\cite{bloom2009impact}, \cite{jurado2015measuring}, \cite{castelnuovo2019domestic}, \cite{carriero2023macro}).  Real output, consumption, investments and stock prices decline while the unemployment rate increases. As opposed to the earlier literature, we do not find evidence supporting an overshoot in real activity in the medium run.  Real Wages and labor productivity display a similar behavior: after a muted short-run reaction, both decline at medium to long response horizons.  Comparing the results between our F-SVAR and the SVAR points towards no discernible effects of including information in the form of the cross-section of earnings or consumption to the model.   

The fact that our model produces IRFs that are in line with the previous literature gives confidence about the ability of the model to identify the desired shock, and assures that the model, although more densely parameterized than a simple SVAR, is still capable of reproducing the established features
of the propagation of uncertainty shocks. It also suggests that the distributional dynamics related to uncertainty shocks, while relevant as we will see at the micro level, do not contain information that is not spanned by the set of variables considered by JLN.

\subsection{Dynamic reactions of the cross-sectional distributions}
We now turn to how the cross-sectional distribution of earnings-to-GDP reacts to uncertainty shocks. 

It is worth emphasizing again at this point that, for each impulse response horizon, we enforce the unit integration of the earnings distribution by construction. As a result, the distribution of earnings for which the changes in response to an uncertainty shock are shown below, only refers to employed people and does not contain information about variations of unemployment generated by the shock. The way we construct the distributional IRFs is therefore different from that in the empirical analysis of \cite{chang2021heterogeneity}, as they re-normalize the distribution at every horizon after the shock to integrate to $(1-\texttt{unr})$, implying that the resulting distributional IRFs in their paper also reflect changes in the ratio between employed people and total population. As a consequence of this difference, the functional IRFs we show throughout the paper always integrate to zero, while those shown by \cite{chang2021heterogeneity} in their empirical section integrate at each horizon to the change of the employment ratio generated by the shock. Despite accounting for the change of the mass of employed people has its merit, it requires artificially re-normalizing  the area under the distribution at every horizon. For this reason, we prefer to focus on the distribution of earnings among employed people and to let the distribution integrate to one (and its change to zero) at every horizon, as implied by the use of the LQD transformation. 

\begin{figure}[h]
\caption{\label{fig:Functional-IRFs--JLN_base} Functional responses of the earnings and consumption distributions to a one standard deviation uncertainty shock.}
\includegraphics[trim={2.2cm 0 0 0},clip, scale=0.55]{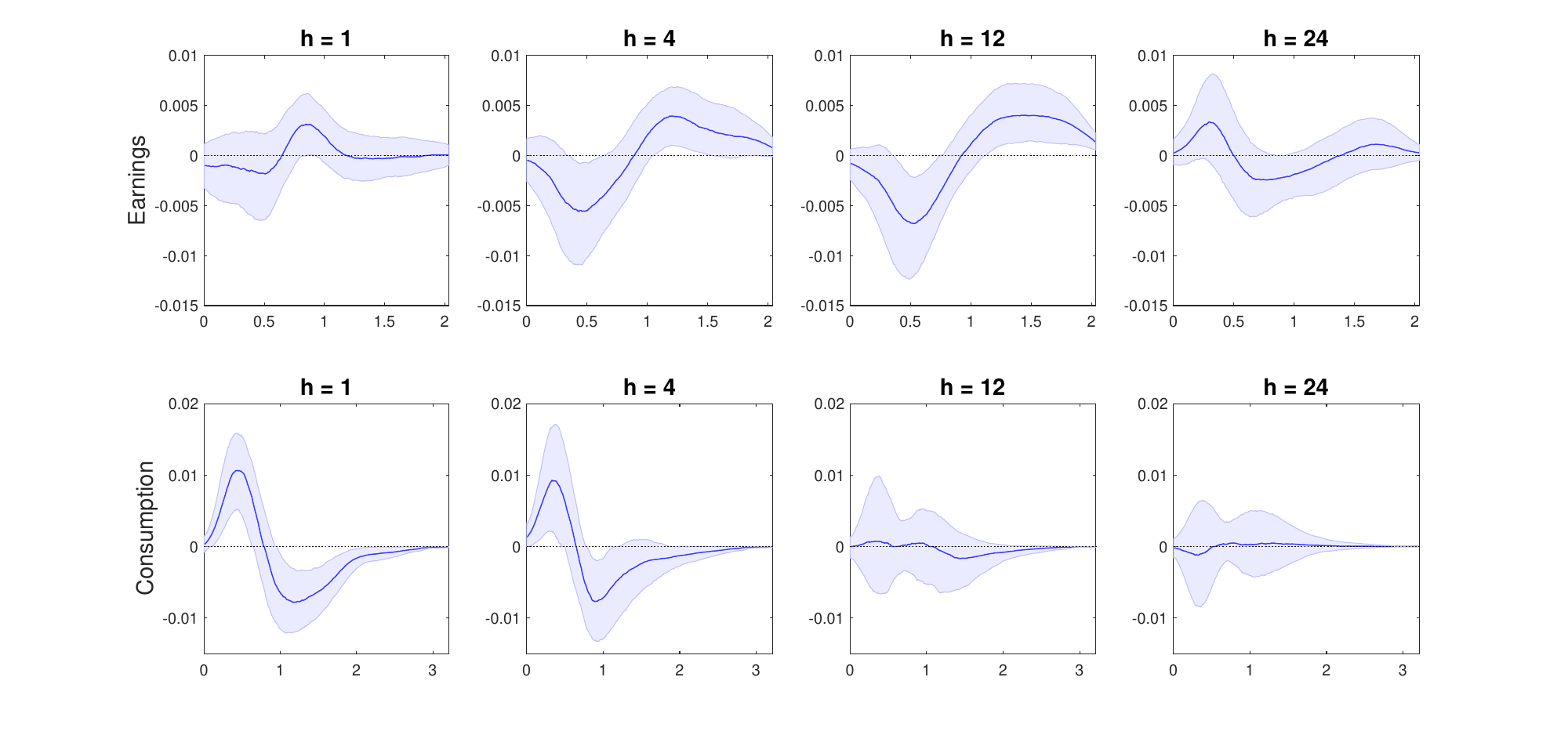}
\centering{}%
\begin{minipage}[t]{0.95\columnwidth}%
\begin{spacing}{0.69999999999999996}
{\footnotesize{}Notes: The first (second) row of the figure shows the difference between the earnings (consumption) distribution after a standard deviation shock and the one prevailing in the steady state. The solid blue lines represent the posterior
median response, while dashed blue lines delimit the 68\% credible
bands. $h$ denotes the horizon at which the response is measured.
In the first (second) row, the measure on the horizontal axis is the earnings-to-GDP per-capita (households consumption - to - per-capita consumption)
ratio.}
\end{spacing}
\end{minipage}
\end{figure}

Figure \ref{fig:Functional-IRFs--JLN_base} shows the change in the earnings-to-GDP and in the consumption distribution produced by the identified uncertainty shock at different horizons. In the first three years from the shock (i.e., for $h=1, 4, 12$), the proportion of employed people
receiving an income below the GDP per-capita level (i.e. earnings-to-DGP per-capita ratio equal to unity) decreases significantly, while the share of employed with salary between one and two times GDP
per-capita increases. At the same time, the mass in the lower part of the consumption distribution increases significantly, reflecting households cutting down their consumption from an average level to a low level.  

When we consider longer-run responses, however, the distributional consequences of uncertainty shocks change markedly. Specifically, the figure for $h=24$ suggests that the portion of employed people receiving low salaries increases considerably, while the share of people earning salaries at the GDP per-capita level drops. At the same horizon, the median response for the consumption distribution also shows a slight decrease of the mass in the lowest part of the distribution; the uncertainty around this response is however too large to support any conjecture.  

A first important consideration when interpreting these results is that the distribution of earnings only refers to employed people, therefore it is very likely that the immediate decrease of the mass in the left part of the distribution does not only reflect people moving from
the low-income category to a higher class.  It may well be that the decline is also due to a higher proportion of the people laid off formerly belonging to a low-income class. This is strongly supported by the opposite change seen in the distribution of consumption, where a significant share of households move from the central part to the lower tail.  

Another important consideration is that the functional IRFs in Figure \ref{fig:Functional-IRFs--JLN_base} must
be read in conjunction with the IRFs of aggregate variables, in particular
with the response of real wages and labour productivity. In fact, Figure \ref{fig:JLN-IRFs-} shows that the average real wage level actually increases at short horizons after the shock, but then it declines significantly, together with labour productivity, and reaches its trough between five and six years from the shock. This corroborates the existence of two stages in the propagation of uncertainty shocks, also evident from Figure \ref{fig:Functional-IRFs--JLN_base}. 

\begin{figure}[t!]
\caption{\label{QuantilesIRF}Responses of specific quantiles of the earnings and consumption distribution to a one standard deviation uncertainty shock.}
\includegraphics[viewport=100bp 0bp 70bp 445bp,scale=0.5]{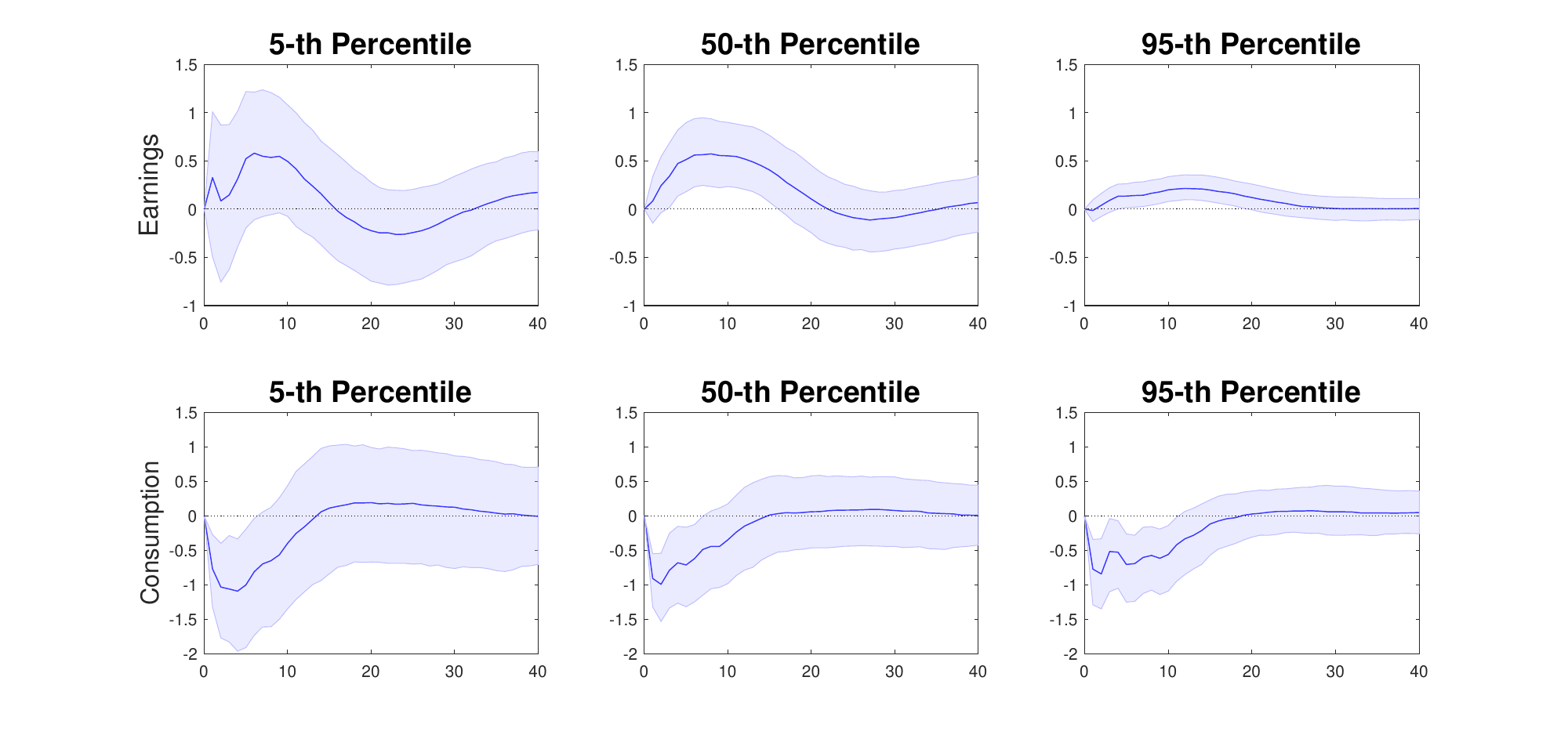}
\centering{}%
\begin{minipage}[t]{0.9\columnwidth}%
\begin{spacing}{0.69999999999999996}
{\footnotesize{}Notes: The first (second) row shows the percentage change of specific percentiles of the of the Earnings-to-GDP ratio (consumption) distribution. The solid blue lines represent the posterior
median response, while the dashed area represent the 68\% credible
bands. The horizon on the horizontal axis is expressed in quarters.}
\end{spacing}
\end{minipage}
\end{figure}

To gain a better understanding of how uncertainty impacts specific quantiles of the distributions, we now focus on the responses of the $5^{th}$, $50^{th}$ and $95^{th}$ quantiles of the earnings and consumption distribution to a one standard deviation uncertainty shock.  This is shown in Figure \ref{QuantilesIRF} and the corresponding IRFs are obtained by picking the relevant quantiles from the posterior of the functional responses of the distributions of interest. From the upper panels, it is clear that, while the entire distribution of earnings shift to the right, the median quantile is the one that displays the most significant percentage increase. This reflects  an increase of the share of employed receiving salaries close to the per-capita GDP level, suggesting that the distributional developments triggered by uncertainty tend to concern to a minor extent the people receiving very high income.
From the consumption side, the whole distribution shifts to the left, but the largest  declines (in percentage) is again observed in the left part of the distribution. 

\begin{figure}[t!]

\caption{\label{ClassesIRF}Responses of earnings and consumption classes to a one standard deviation uncertainty shock}

\includegraphics[viewport=190bp 10bp 800bp 440bp,scale=0.6]{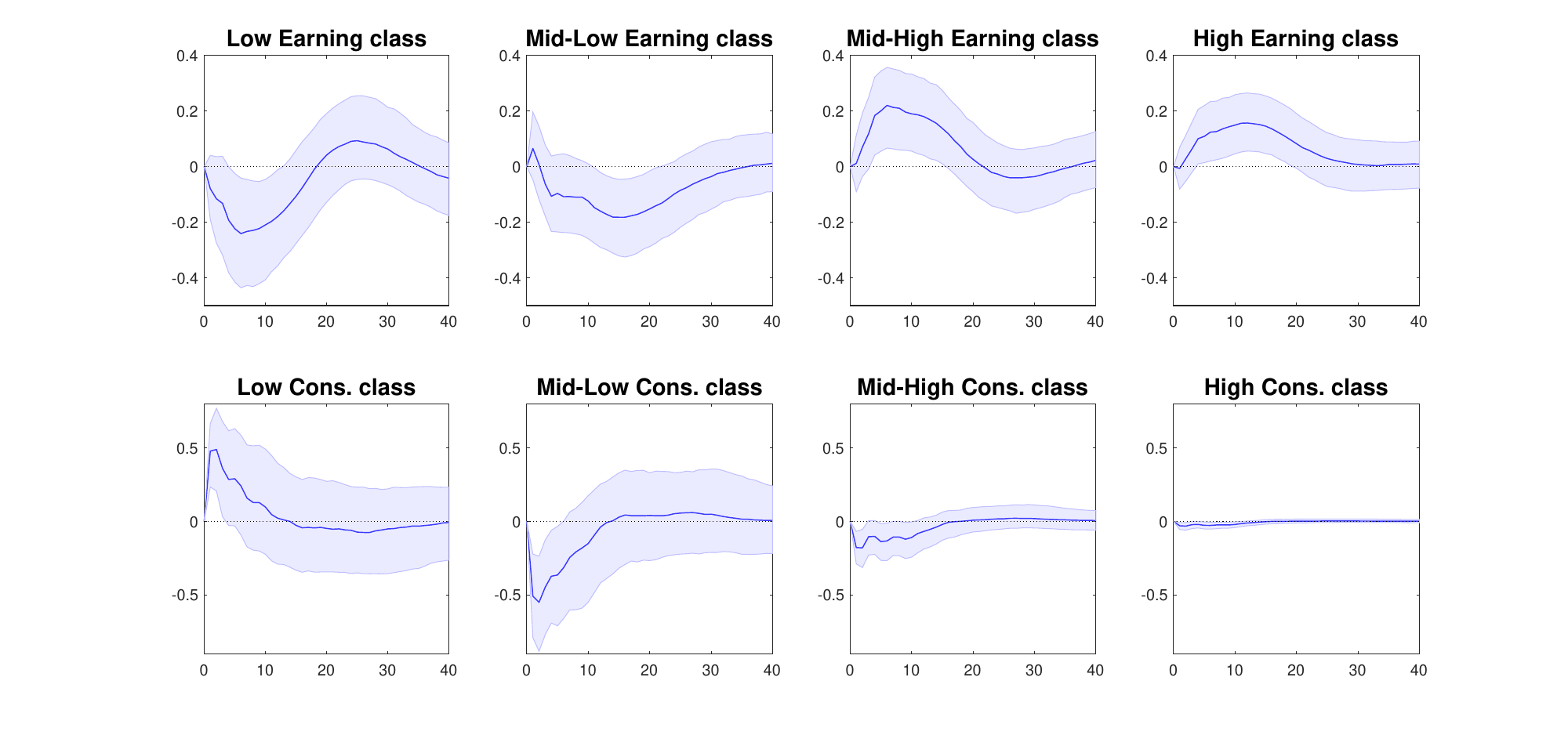}
\centering{}%
\begin{minipage}[t]{0.9\columnwidth}%
\begin{spacing}{0.69999999999999996}
{\footnotesize{}Notes: The first (second) row shows the response of the share of the employed (total) population belonging to specific earnings (consumption) classes. The solid blue lines represent the posterior
median response, while the dashed area represent the 68\% credible
bands. The horizon on the horizontal axis is expressed in quarters.}
\end{spacing}
\end{minipage}
\end{figure}

A related but distinct examination that the F-SVAR allows us to consider is the assessment of how the relative weight of each earnings and consumption class changes in response to the identified shock. In Figure \ref{ClassesIRF} we track the share of employed people belonging to four classes, each corresponds to one fourth of the distribution support. As for earnings, the figure suggests that the share of employed people belonging to the bottom income class decreases significantly in a first phase, while the relative weight of the classes whose earnings belong to the right half of the support increases. In a second phase, however, the share of low-income employed increases (although the credible bands also contain the zero line at larger horizons), at the cost of mid-low income class.

As for consumption, on the other hand, the uncertainty shock affects almost exclusively the relative weights of the low- and mid-low consumption classes. In fact, the figure shows that the fast increase in the proportion of households reporting low consumption levels is associated with an almost identical decrease in the proportion of households reporting consumption level in the second fourth of the support. Meanwhile, the relative weight of the consumption classes belonging to the upper half of the support remain almost unchanged. This makes it even clearer that the decrease of the mass in the left part of the earnings distribution is mainly due to an increase of the pool of unemployed, who are then compelled to cut down consumption, due to their inability to access consumption smoothing channels.

In summary, we conjecture that the propagation of uncertainty
shocks happens in two phases. In the short run, the rise in unemployment
concerns to a larger extent people at the lower end of the earnings
distribution, who therefore cut down their consumption, while the part of low-income workers that is not laid
off actually sees its relative earnings increased or kept constant. In the subsequent phase, when unemployment is finally reabsorbed, the size of the low-income group rises at the cost of middle-income employed people, who however do not seem to reduce their consumption level. This evidence is in line with the findings of \cite{choi2023impact}, who show that uncertainty shocks affect more adversely poor people, while it is in contrast with the results obtained by \cite{theophilopoulou2022impact} for the UK, where poor people income and consumption appeared to be sustained more vigorously by social benefits.

\begin{figure}[t!]
\caption{\label{fig:IRF-Gini-JLN}Responses of the Gini coefficient to a one standard deviation uncertainty shock}
\includegraphics[viewport=190bp 10bp 800bp 430bp,scale=0.5]{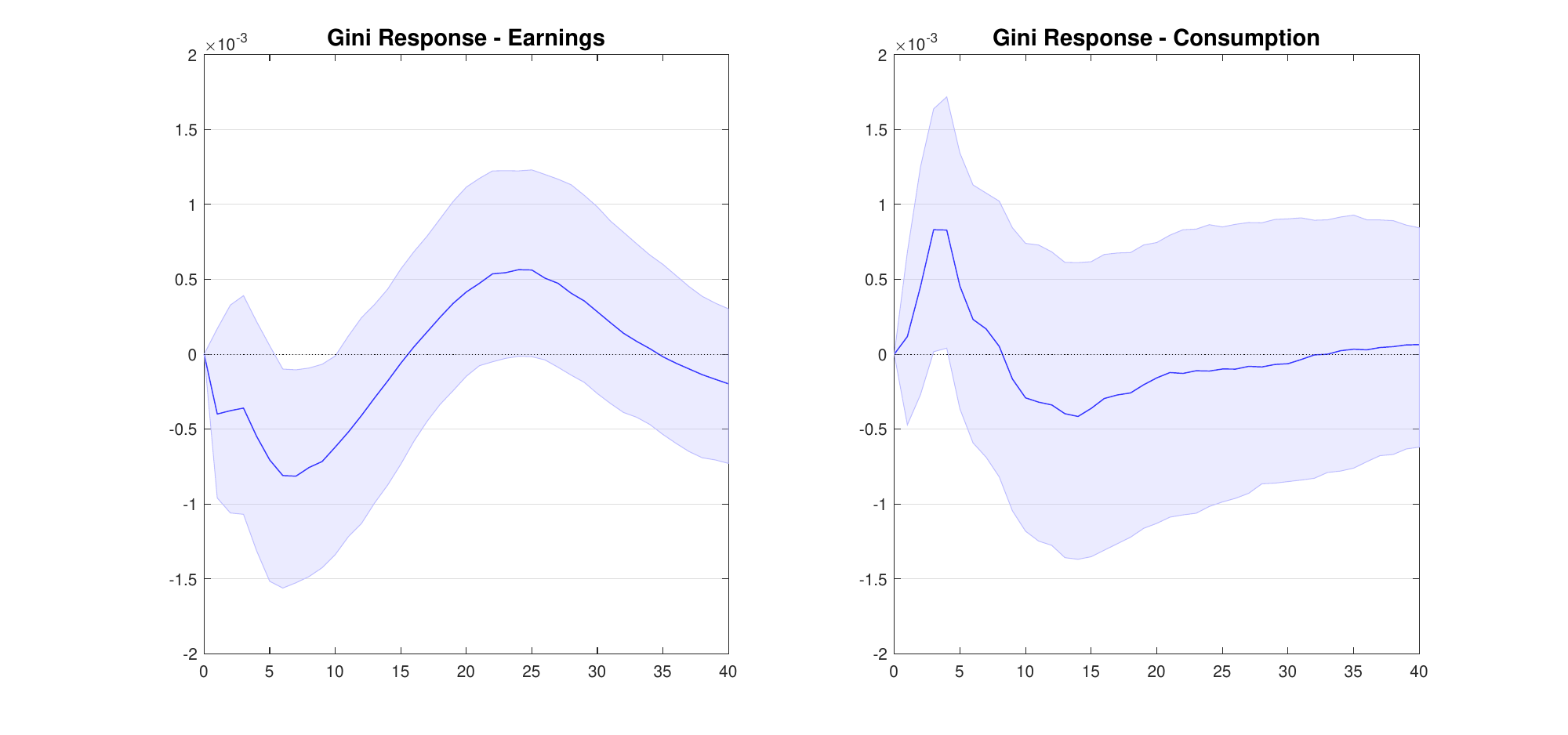}
\centering{}%
\begin{minipage}[t]{0.9\columnwidth}%
\begin{spacing}{0.69999999999999996}
{\footnotesize{}Notes: The solid blue lines represent the posterior
median response, while dashed blue lines delimit the 68\% credible
bands. The horizon on the horizontal axis is expressed in quarters.}
\end{spacing}
\end{minipage}
\end{figure}

It is interesting to see the reflection of the mechanisms just described
in the responses of the Gini coefficient, which measures the degree
of inequality in the distributions and can be computed from the functional
responses produced by the F-SVAR. Figure \ref{fig:IRF-Gini-JLN} shows
that the short-run developments generated by the uncertainty shock
reduces  the total level of income inequality among
employed people but increases the degree of consumption inequality significantly, while the subsequent phase of the propagation mechanism
causes a considerable rise in the Gini coefficient, which is reabsorbed
only eight years after the shock. This pattern is consistent with \cite{fischer2021regional} who show that an increase in uncertainty triggers an immediate decline in the Gini coefficient and a subsequent increase after around three years.

It is also worth noticing that the responses depicted in Figure \ref{fig:IRF-Gini-JLN} are very similar to those that would be obtained by a standard quarterly SVAR(4) in which the Gini coefficients computed on CPS an CEX data are appended to the vector of macro variables. These responses are reported in Figure \ref{fig:Gini Standard SVAR} and display the same two-phase propagation found by the F-SVAR. Clearly, the similarity of the result points to the robustness of our main findings, but does not imply that a standard SVAR would be sufficient to perform the analysis. The scope of the F-SVAR is much broader, as it gives the possibility to focus on any characteristic of interest of the earnings and consumption distributions, and to enforce  the restrictions that are inherent in the space of distributions automatically. On the other hand, any standard SVAR alternative would inevitably limit the analysis to the characteristics included in the vector of endogenous variables, and would not guarantee that responses of these characteristics will be admissible. For example, in a standard SVAR containing the Gini coefficient, the corresponding responses could take values below zero or larger than one; in a standard SVAR containing quantiles, the implied IRFs could be charachterized by quantile crossing or could take values outside of the domain, e.g. negative income or consumption levels.

This discussion demonstrates that simply considering how summary statistics, such as the Gini coefficient, react to uncertainty shocks paints an inevitably partial picture of the overall distributional dynamics, and could run into problems that undermine the reliability of the results. For this reason, we believe the F-SVAR represents a more suitable tool to answer the central empirical question of this paper. 

\subsection{Investigating the mechanism: Unemployment reactions across educational levels}
One important conjecture we made in the previous subsection is that the decline in the mass in the left tail of the earnings distribution is driven by changing employment levels across income classes.

To check that this is the case,  we study the responses to uncertainty shocks of unemployment rates computed for different categories of the work-force. Unfortunately, we are not aware of any dataset providing employment information for sub-groups of the population based on their previous income level. However, the Bureau of Labour Statistics (BLS) provides unemployment rates computed monthly for sub-groups based on educational attainments, which can serve as a proxy of income classes.\footnote{The average income differential between the categories is indeed substantial and reported periodically by the BLS. Figure \ref{fig:EarningsEdu} shows the evolution of the median weekly earnings level for the four categories in the time period for which it is available.} In particular, the BLS distinguishes between four educational levels: (i) less than a high school diploma, (ii) high school diploma but no college, (iii) some college but no degree or some associate degree, (iv) Bachelor's degree or higher.

Since the BLS series start in January 1992, using the quarterly dataset would imply too few observations. Hence, to understand how different education-specific unemployment rates react to an uncertainty shock we add the (logarithm) of these granular unemployment series in the monthly SVAR of \cite{jurado2015measuring}. The auxiliary VAR model  is estimated on  a sample going from 1992$:$M01 (the start of the BLS series) to 2019$:$M12 (to avoid having to deal with Covid-19-specific outliers in the data).

\begin{figure}[t!]
\caption{\label{fig:IRF_unempEdu}Dynamic responses of different unemployment categories to a one standard deviation uncertainty shock}
\includegraphics[viewport=0bp 0bp -30bp 300bp,scale=0.7]{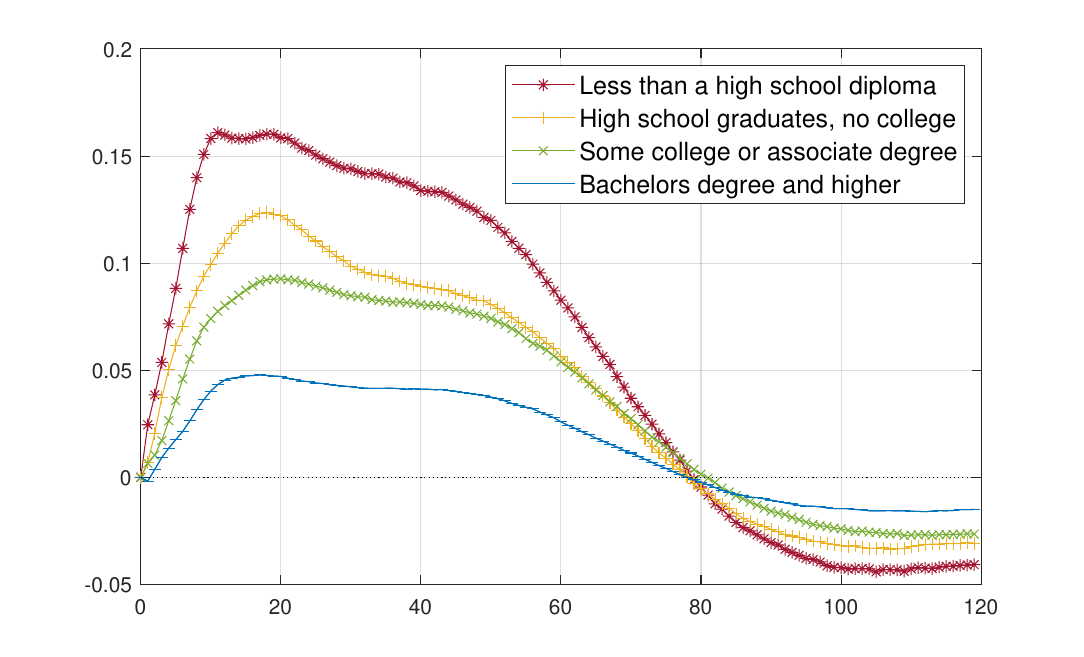}
\centering{}%
\begin{minipage}[t]{0.9\columnwidth}%
\begin{spacing}{0.69999999999999996}
{\footnotesize{}Notes: The four different lines depict the posterior median of different unemployment rates across education levels.}
\end{spacing}
\end{minipage}
\end{figure}

In Figure \ref{fig:IRF_unempEdu}, we report the (point-wise) posterior median IRFs obtained for the four sub-groups. The responses provide substantial evidence that workers with a lower education attainment, who tend to occupy the left part of the earnings distribution, are those that are lied off more often after an uncertainty shock. The figure tells a remarkably consistent story that employment reactions tend to become weaker with increasing educational attainment.  

\begin{figure}[ht!]
\caption{\textbf{\label{fig:IRFsd_unempEdu}Differences between IRFs of different categories}}
\includegraphics[viewport=110bp 0bp 80bp 430bp,scale=0.5]{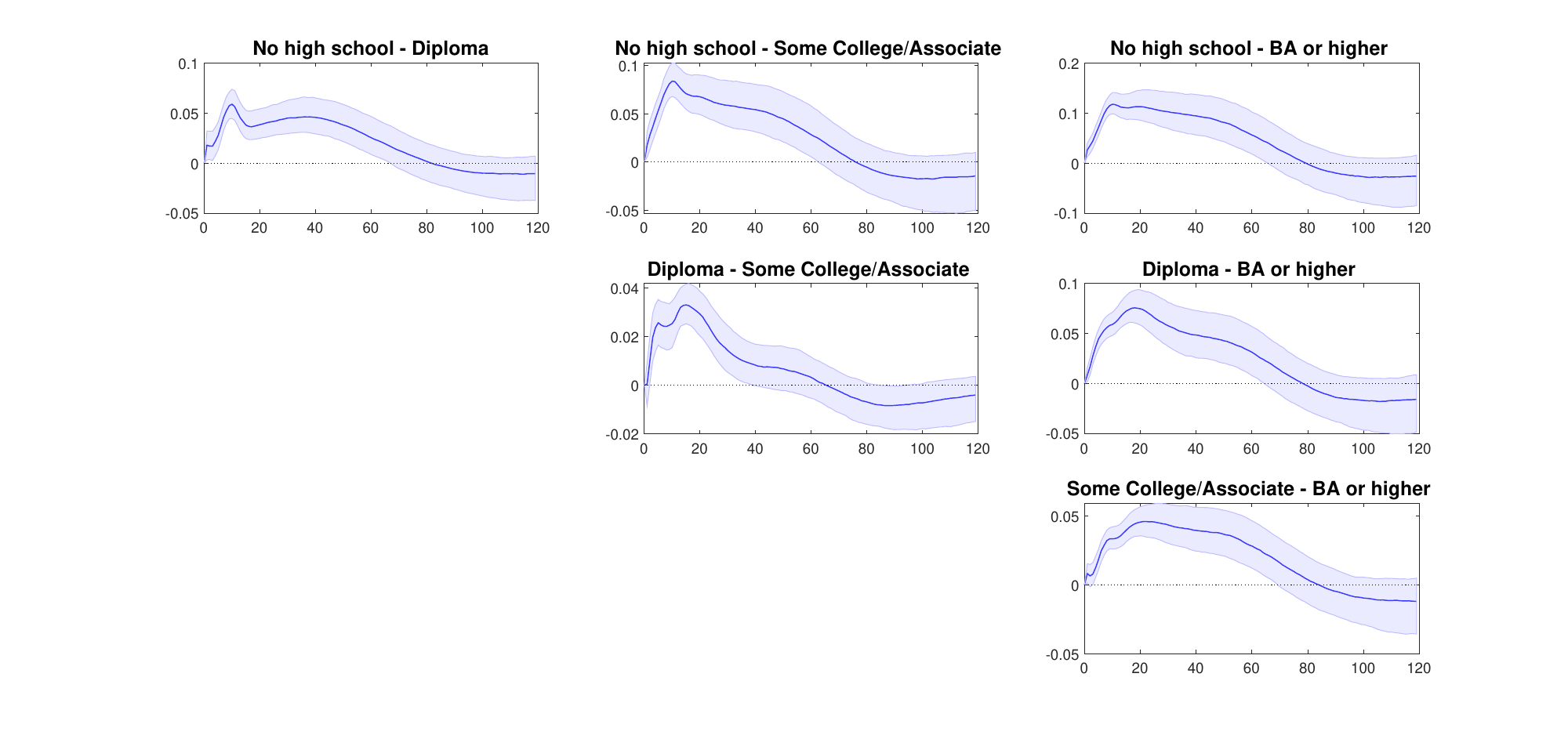}
\centering{}%
\begin{minipage}[t]{0.9\columnwidth}%
\begin{spacing}{0.69999999999999996}
{\footnotesize{}Notes: Solid lines show the median difference between IRFs of the unemployment rate computed for different categories of the population. Shaded areas represent 68\% credible bands. Categories are: (i) No high school: civilian population 25 years with less than a high school diploma; (ii) Diploma: civilian population 25 years graduated from high school but no college; (iii) Some College/Assoiate: civilian population 25 years who attended some college but did not graduate, or obtained some associate degree; (iv) BA or higher: civilian population 25 years with a Bachelor's degree or higher educational attainment.}
\end{spacing}
\end{minipage}
\end{figure}

Considering only posterior medians  neglects posterior uncertainty surrounding the estimates of the IRFs and one might ask whether differences across educational categories are statistically different form each other. To shed light on this, we report pairwise differences of the posterior distribution of impulse responses in  Figure \ref{fig:IRFsd_unempEdu}. Since these are the responses of the differences, we can assess whether a specific horizon $h$ IRF between two educational levels differs by simply checking whether the corresponding credible intervals include zero or not.

The pairwise comparisons in the figure reveal that the differences are, in fact, highly significant. Notice that these significant differences persist throughout the impulse response horizon, turning insignificant only after around 80 months.  This relatively simple analysis, therefore, supports our conjecture that the short-run decrease of the mass in the most-left part of the earnings distribution highlighted in Figure \ref{fig:Functional-IRFs--JLN_base} is, at least in part, due to a greater proportion of the people laid off belonging to a low-income class. If the insights collected from Euro-Area survey responses by \cite{coibion2024effect} hold also for US citizens, then the difference in the unemployment risk of different income classes can also explain the results we have obtained for the consumption distribution. In fact, according to their evidence, the households that cut down expenditure more aggressively in the presence of uncertainty are those whose income is more at risk in periods of high uncertainty, which we have shown include less specialized workers.

\section{Functional Local Projections\label{sec:FLP}}
Just as for standard scalar aggregate variables the IRFs can be estimated by Local Projections (LPs, see \cite{jorda2005estimation}), the response of distributions to the structural shock of interest can be estimated by Functional Local Projections (F-LPs). As a matter of fact, the reason why a F-VAR process was assumed for the joint dynamics of aggregate variables and the endogenous function $\widetilde{f}_t$ was to have a tractable model in terms of scalar variables $y_t$ and $\alpha_t$, which would allow us to infer the responses of the distribution $p_t(\cdot)$ to an uncertainty shock. However, this assumption is not essential, and nothing prevents the use of LPs to perform inference about the IRFs of $\alpha_t$, which we defined as $IRF_{\alpha,j,d,h}$, and to map them to distributional IRFs as described in Section \ref{sec: FIRFs}. This is the approach we follow in this section to show that all the results we have discussed so far are not determined by the modelling choice described in \ref{sec:FVAReq} or by the priors discussed in \ref{sec: inference}. 

Although the estimation of F-LPs is straightforward and can simply be done by OLS, two points are worth considering. First, as the steady state distribution, $p_{ss}(\cdot)$, cannot be implied by the model, we take it to be the one associated with the sample average $\bar{f}(\cdot)$, so that $p_{ss}(x)=g^{-1}\left(\bar{f}(c)\right)$. Second, unlike in standard LP applications, we are not interested in the IRF of one variable at a time. Instead, we are interested in the joint response of the whole vector $\alpha_t$, $IRF_{\alpha,j,d,h}$, which then determines the distributional IRF as: $p_{ss+h}(\cdot)-p_{ss}(\cdot)=g^{-1}\left(\boldsymbol\zeta(\cdot)'IRF_{\alpha,j,d,h}+\bar{f}\right)-p_{ss}(\cdot)$. As a result, when conducting frequentist inference, the variance of the estimator of $IRF_{\alpha,j,d,h}$ must account not only for the presence of serial correlation in the residuals arising in standard LP inference, but also for the cross correlation across residuals associated with the different $\alpha_{tk}$'s. \cite{newey1987simple} covariance estimator is therefore not enough, \cite{driscoll1998consistent}'s method is required in this context. 

Furthermore, since the shock identification strategy used in this paper only relies on timing assumptions, it is easy to incorporate them in the estimation of the LP for $\alpha_t$, as explained by \cite{plagborg2021local}. More specifically, 
continuing to assume that the uncertainty shock is ordered last in Cholesky-type identification scheme, $IRF_{\alpha,N,1,h}$ is given by the $\beta_{1,U}^h$ coefficient in the following regression:
\begin{equation}
\alpha_{t+h}=a^{h}+B_{1}^{h}\left[\begin{array}{cc}
y_{t,\setminus{U}}^{\prime}, & \alpha_{t}^{\prime}\end{array}\right]^{\prime}+\beta_{1,U}^{h}\texttt{U}_{m,t}+\sum_{l=1}^{p}B_{l+1}^{h}\left[\begin{array}{cc}
y_{t-l}^{\prime}, & \alpha_{t-l}^{\prime}\end{array}\right]^{\prime}+e_{h,t},
\label{eq:FLP}
\end{equation}
where we have defined as $y_{t,\setminus{U}}$ the vector of all aggregate variables except \texttt{U}$_{m,t}$. 

Since in this section we perform inference employing frequentist methods, we cannot make use of priors to shrink irrelevant parameters toward zero. \footnote{Although performing Bayesian inference on F-LP coefficients is possible following the approach of \cite{ferreira2023bayesian}, eliciting priors for such parameters and accounting for the autocorrelation of projection residuals are not standard tasks. For these reasons, we choose to apply standard frequentist methods and to leave the  Bayesian treatment of F-LPs for future research.} Furthermore, as there are 10 variables in the vector $y_{t,\setminus{U}}$, the number of parameters to estimate in (\ref{eq:FLP}) can become excessively large when we consider many lags, $p$, and many FPCs, $K$. For this reason, in this section we set $p=1$ and $K=3$ when focusing on the earnings distribution, and $p=1$ and $K=5$ when focusing on the consumption distribution. 

To demonstrate that F-LPs are a viable alternative to the F-SVAR, we repeated all the three artificial data experiments described in Section \ref{sec:Simulation} performing inferences about $IRF_{\alpha,N,1,h}$ via LP. In the interest of space, the resulting figures are reported in Appendix \ref{sec:F-LP Simulations}. The results show that also F-LPs do a good job in estimating the responses of interest. Moreover, the figures in Appendix \ref{sec:F-LP Simulations} also suggest that the well known variance-bias trade-off that characterizes the comparison between standard LPs and standard SVARs carries over to the comparison between both approaches. In fact, while at short horizons the uncertainty around F-LP-based estimates is very limited, at long horizons the variance of the estimator grows substantially. This does not hold for the F-SVAR in which the credible bands never inflate excessively. Such comparison is definitely worth more research and we leave a thorough assessment for future work. 

\begin{figure}[t!]
\caption{\label{fig:Agg_LP_JLN}Impulse response of the macroeconomic aggregates to a one standard deviation uncertainty shock}
\includegraphics[viewport=190bp 10bp 800bp 430bp,scale=0.6]{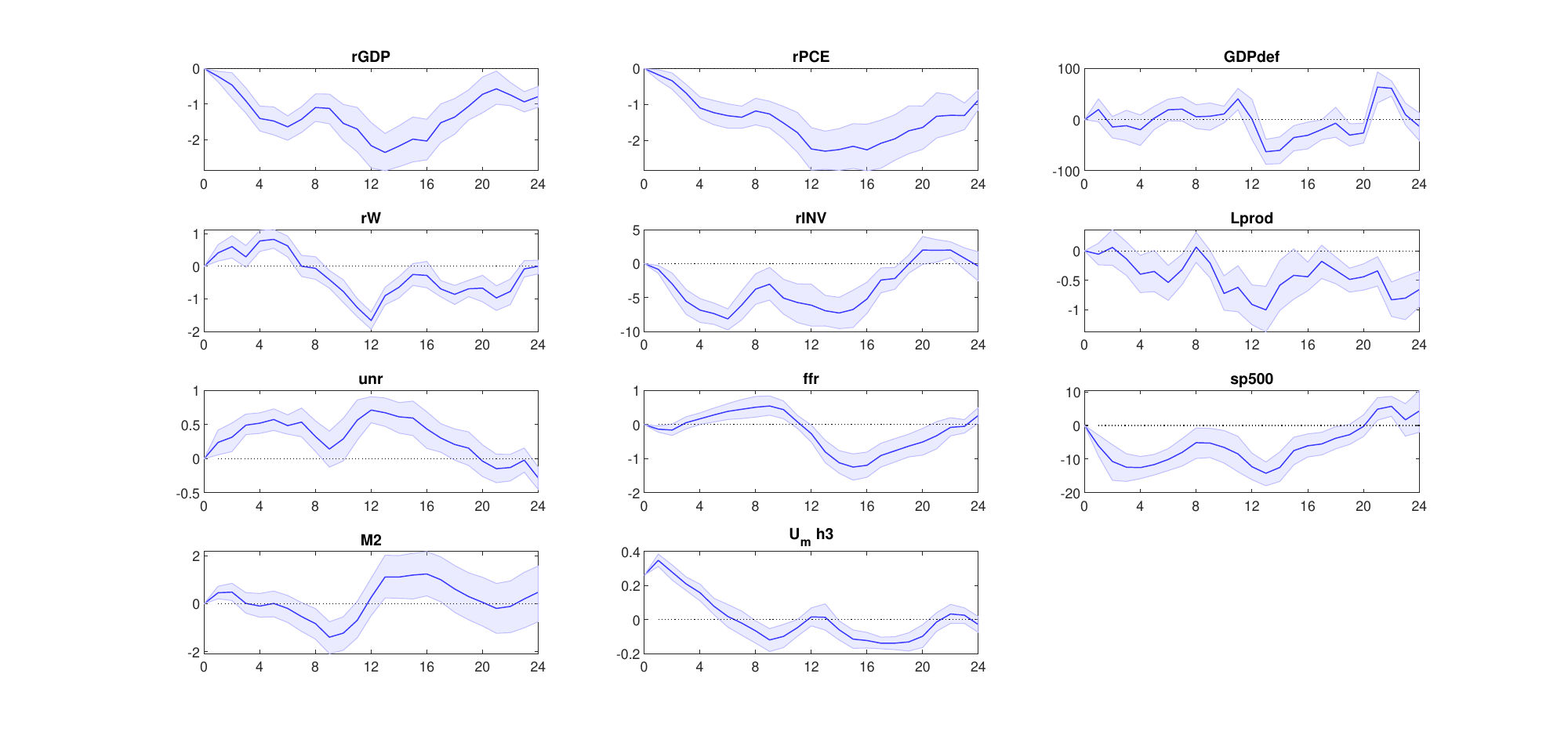}
\centering{}%
\begin{minipage}[t]{0.9\columnwidth}%
\begin{spacing}{0.69999999999999996}
{\footnotesize{}Notes: The solid blue lines represent the posterior
median response, while dashed blue lines delimit the 68\% credible
bands. The horizon on the horizontal axis is expressed in quarters.}
\end{spacing}
\end{minipage}
\end{figure}

Additionally, in order to show that the regression in (\ref{eq:FLP}) actually allows us to estimate the response to the same shock studied in the previous section, in Figure \ref{fig:Agg_LP_JLN} we report the IRFs of the macro aggregates estimated by LP. Apart from erratic behaviours peculiar to LP, the responses are very similar to those estimated through the F-SVAR and shown in Figure \ref{fig:JLN-IRFs-}.

\begin{figure}[t!]
\caption{Distributional LP\label{fig:FLP_JLN}}
\includegraphics[viewport=190bp 10bp 800bp 430bp,scale=0.6]{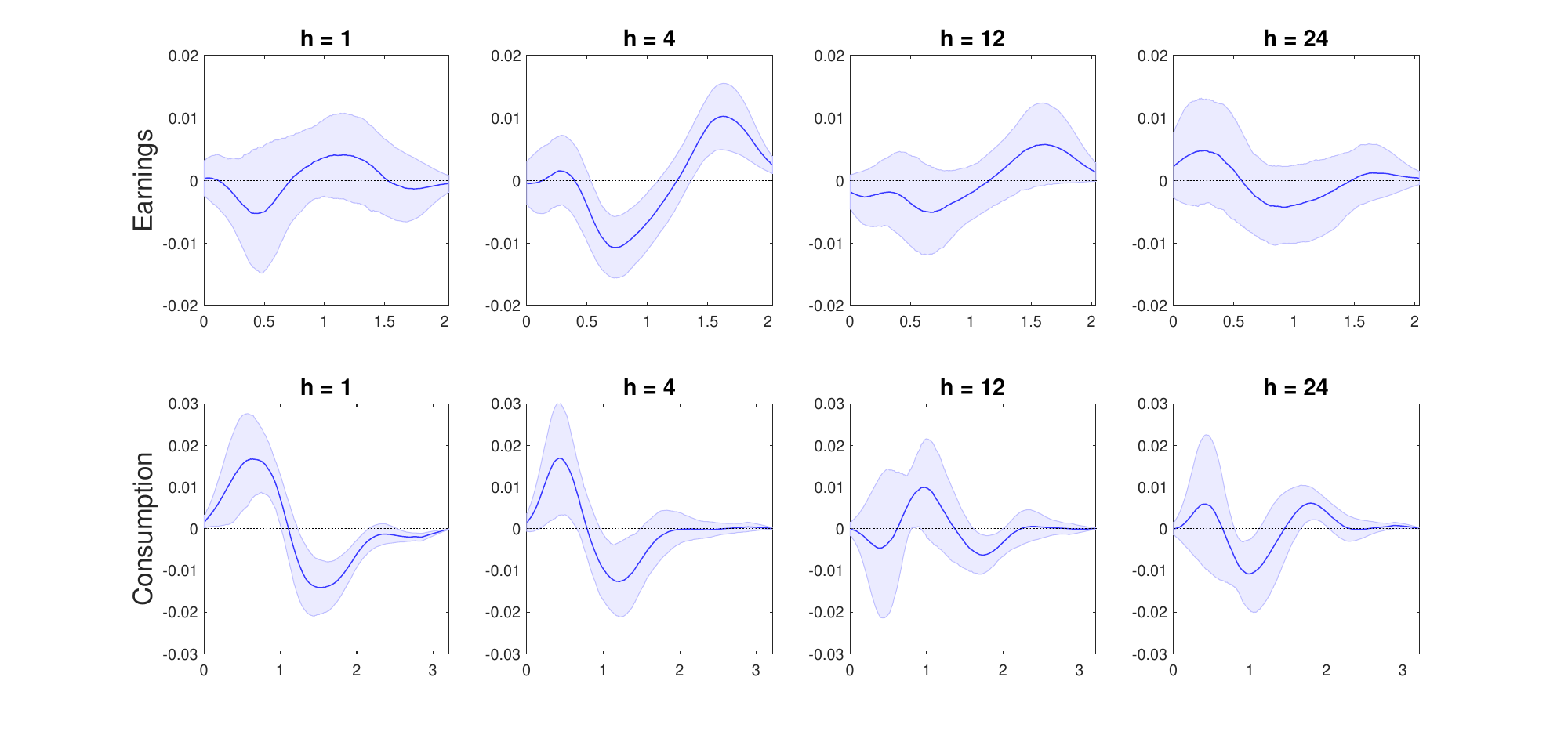}
\centering{}%
\begin{minipage}[t]{0.9\columnwidth}%
\begin{spacing}{0.69999999999999996}
{\footnotesize{}Notes: The first (second) row of the figure shows the difference between the earnings (consumption) distribution after a standard deviation shock and the one prevailing in the steady state. The solid blue lines represent the posterior
median response, while dashed blue lines delimit the 68\% credible
bands. $h$ denotes the horizon at which the response is measured.
In the first (second) row, the measure on the horizontal axis is the earnings-to-GDP per-capita (households consumption - to - per-capita consumption)
ratio.}
\end{spacing}
\end{minipage}
\end{figure}

Confident that the LP in (\ref{eq:FLP}) is able to retrieve the responses to the uncertainty shock of interest, Figure \ref{fig:FLP_JLN} depicts the distributional IRFs estimated by F-LPs. Although there are some differences with respect to Figure \ref{fig:Functional-IRFs--JLN_base}, the general features highlighted by the two methods are very similar, especially for horizons up to one year. 

\section{Conclusion\label{sec:Summary-and-concluding}}
Since the distributional consequences of uncertainty shocks have remained relatively unexplored by the literature, we introduced a Functional VAR model to study the propagation of such shocks on the earnings and consumption distributions. 

After showing that the adopted econometric technique is able to reproduce distributional developments in response of structural shocks in a variety of challenging simulated experiments, we used the Functional VAR methodology to extend the popular analysis of \cite{jurado2015measuring} and to take into account the dynamics of the earnings and consumption distributions. In particular, the Functional VAR we employ is based on a Functional Principal Component Analysis of the Log Quantile Density associated with the distribution of earnings and consumption, which is in turn estimated starting from available survey data using kernel methods. Furthermore, we also showed that the same type of analysis can be easily carried out estimating Functional LP. 

Our results show that the effects of uncertainty shocks unfold in two phases. In a first phase, the economic contraction triggered by the shock is coupled with an increase of unemployment and a drop in stock prices. At this stage, the fraction of employed people earning less than the GDP per capita level decreases, and the share of higher-income workers increases. In a second phase, the unemployment level is gradually reabsorbed and labour productivity decreases. At the same time, the fraction of employed low-income people increases, while the employed medium-income class is reduced in size. As a consequence, the uncertainty shock reduces significantly the total level of earnings inequality among employed people in the short run, as measured by the Gini coefficient, while in the medium term it causes a considerable rise in it,

A potential explanation for this pattern is that, in the short run, a larger proportion of less specialized low-income workers are laid off due to the economic contraction, while those that are not laid off see their earnings kept constant. In the longer run, however, the pool of unemployed people is reabsorbed, but labour productivity declines due to the previous foregone investments. The lower labor productivity can in turn explain the increase in the share of low-income employed relative to the steady state distribution, at the expense of the middle class. 

\newpage
\clearpage
\small{\setstretch{0.85}
\addcontentsline{toc}{section}{References}
\doublespacing
\bibliographystyle{cit_econometrica.bst}
\bibliography{lit}}
\newpage

\appendix
\thispagestyle{empty}\null

\begin{center}
{\Huge{}\vspace{5cm}
Online Appendix}{\Huge\par}
\par\end{center}

\vspace{5cm}

\noindent 

\pagebreak{}

\pagenumbering{arabic}

\renewcommand{\thefootnote}{A\arabic{footnote}}
\renewcommand{\thepage}{A\arabic{page}}
\renewcommand{\thetable}{A\arabic{table}}
\renewcommand{\thefigure}{A\arabic{figure}}

\setcounter{footnote}{0} 
\setcounter{section}{0}
\setcounter{table}{0}
\setcounter{figure}{0}

\section{F-SVAR DGP\label{sec:FSVAR-DGP 1 & 2}}

The reduced form parameters of the VARs used to simulate the data
in Section \ref{subsec:F-SVAR-DGP 1} and \ref{subsec:F-SVAR-DGP 2} are: 

{\footnotesize{}
\[
\Pi_{0}=\left[\begin{array}{c}
0\\
0\\
0\\
0\\
0
\end{array}\right];
\]
}

{\footnotesize{}
\[
\Pi_{1}=\left[\begin{array}{ccccccc}
0.85 & -0.15 & 0.15  & 0.15  & -0.25 \\
-0.2 &  0.85 & -0.15 & -0.25  & -0.2 \\
0.15 & -0.15 & 0.85  & -0.15 & 0 \\
0.1 &  0.15  & -0.2 & 0.85  & -0.2 \\
-0.25 &  0.15  & 0.15  & 0.15  & 0.85
\end{array}\right];
\]
}

{\footnotesize{}
\[
\Pi_{2}=\left[\begin{array}{ccccccc}
-0.3 & 0.1 & 0.15 & -0.15 & -0.1 \\
-0.1 & -0.3 & 0.1 & 0.15 & 0.15 \\
-0.05 & 0.1 & -0.3 & 0.05 & -0.1 \\
0.15 & -0.1 & -0.05 & -0.3 & -0.05\\
0.15 & -0.15 & 0.1 & -0.1 & -0.3 
\end{array}\right];
\]
}

{\footnotesize{}
\[
\Pi_{3}=\left[\begin{array}{ccccccc}
0.15 & 0 & 0 & 0 & 0\\
0 & 0.15 & 0 & 0 & 0\\
0 & 0 & 0.15 & 0 & 0\\
0 & 0 & 0 & 0.15 & 0\\
0 & 0 & 0 & 0 & 0.15
\end{array}\right];
\]
}

{\footnotesize{}
\[
\Pi_{4}=\left[\begin{array}{ccccccc}
0.05 & 0 & 0 & 0 & 0\\
0 & 0.05 & 0 & 0 & 0\\
0 & 0 & 0.05 & 0 & 0\\
0 & 0 & 0 & 0.05 & 0\\
0 & 0 & 0 & 0 & 0.05
\end{array}\right];
\]
}

\noindent The $Omega$ matrix is formed sampling 100 random matrices $R$ with elements equal to the product of two standard normal draws multiplied by $0.1$, and setting $Omega=R*R'$. the matrix $A_{0}^{-1}$ is then set to be the lower-tringular
Cholesky factor of $\Omega$.

\newpage
\section{\normalsize{F-SVAR DGP 1 - Results with Different \texorpdfstring{$K$}{K}}\label{sec:FSVAR-DGP 1 - Results with Different $K$}}

\begin{figure}[ht!]
\caption{\label{fig:Functional-IRFs-FSVAR_dgpk1}\textbf{Functional IRFs - F-SVAR
DGP 1 - $K=1$}}
\begin{centering}
\includegraphics[bb=1000bp 0bp 0bp 450bp,scale=0.5]{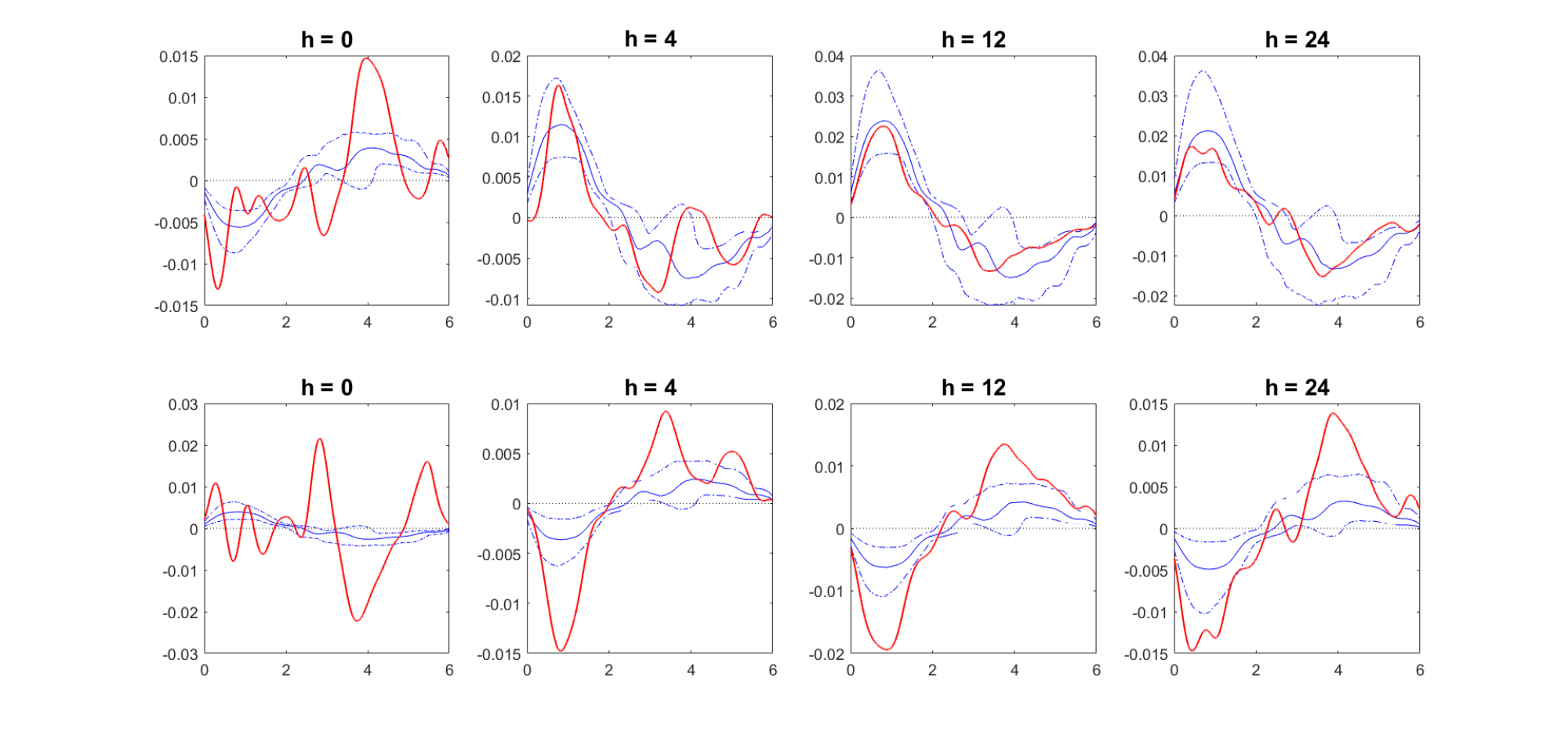}
\par\end{centering}
\centering{}%
\begin{minipage}[t]{0.9\columnwidth}%
\begin{spacing}{0.7}
{\footnotesize{}Notes: Red lines show the true responses of $p_{t}\left(\xi\right)$
to one standard deviation $\varepsilon_{1}$ and $\varepsilon_{2}$.
The solid blue lines represent the posterior median response, while
dashed blue lines delimit the 90\% credible bands. $h$ denotes the
horizon at which the response is measured. }
\end{spacing}
\end{minipage}
\end{figure}

\begin{figure}[ht!]
\caption{\label{fig:Functional-IRFs-FSVAR_dgpk2}\textbf{Functional IRFs - F-SVAR
DGP 1 - $K=2$}}
\begin{centering}
\includegraphics[bb=1000bp 0bp 0bp 450bp,scale=0.5]{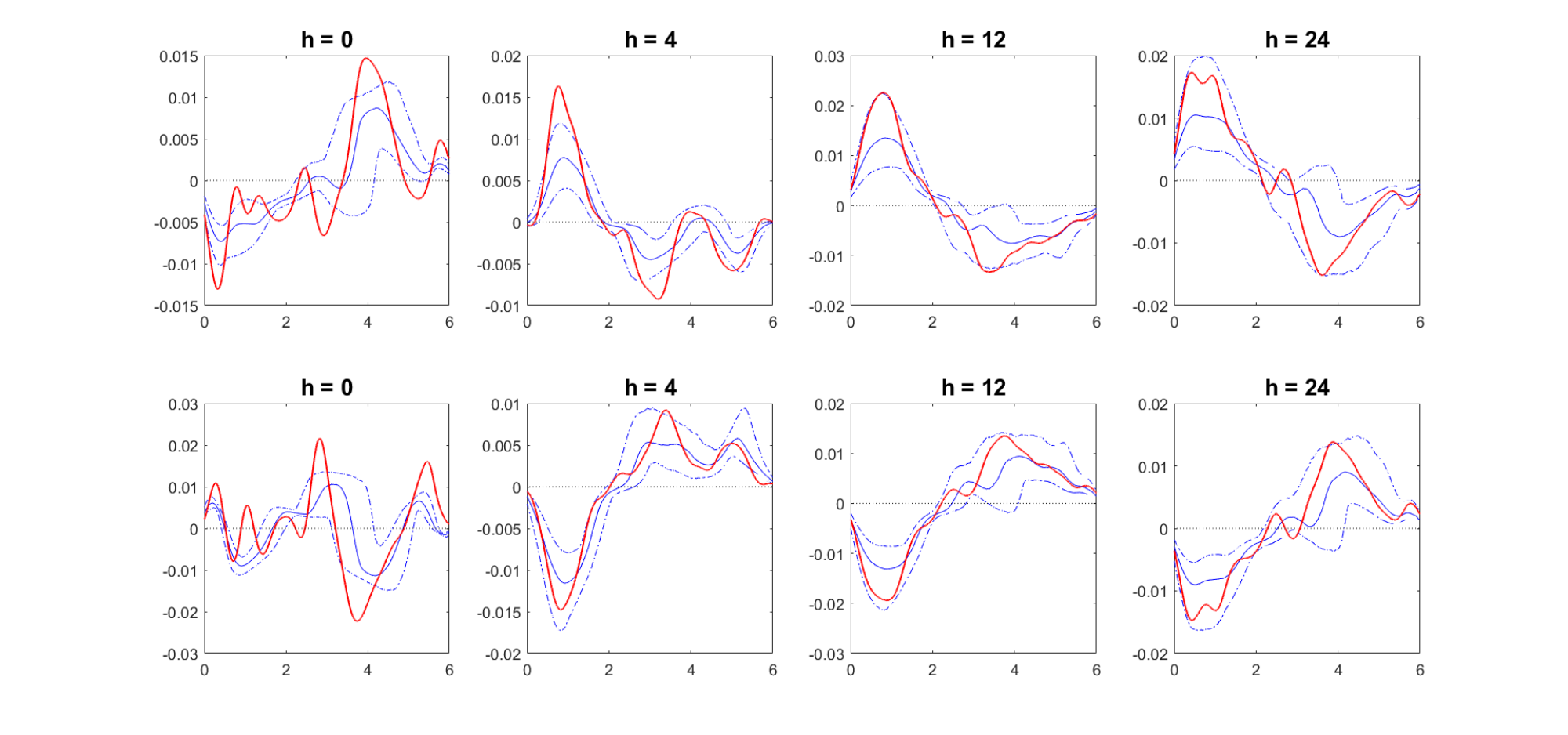}
\par\end{centering}
\centering{}%
\begin{minipage}[t]{0.9\columnwidth}%
\begin{spacing}{0.7}
{\footnotesize{}Notes: Red lines show the true responses of $p_{t}\left(\xi\right)$
to one standard deviation $\varepsilon_{1}$ and $\varepsilon_{2}$.
The solid blue lines represent the posterior median response, while
dashed blue lines delimit the 90\% credible bands. $h$ denotes the
horizon at which the response is measured. }
\end{spacing}
\end{minipage}
\end{figure}

\begin{figure}[ht!]
\caption{\label{fig:Functional-IRFs-FSVAR_dgpk3}\textbf{Functional IRFs - F-SVAR
DGP 1 - $K=3$}}
\begin{centering}
\includegraphics[bb=1000bp 0bp 0bp 450bp,scale=0.5]{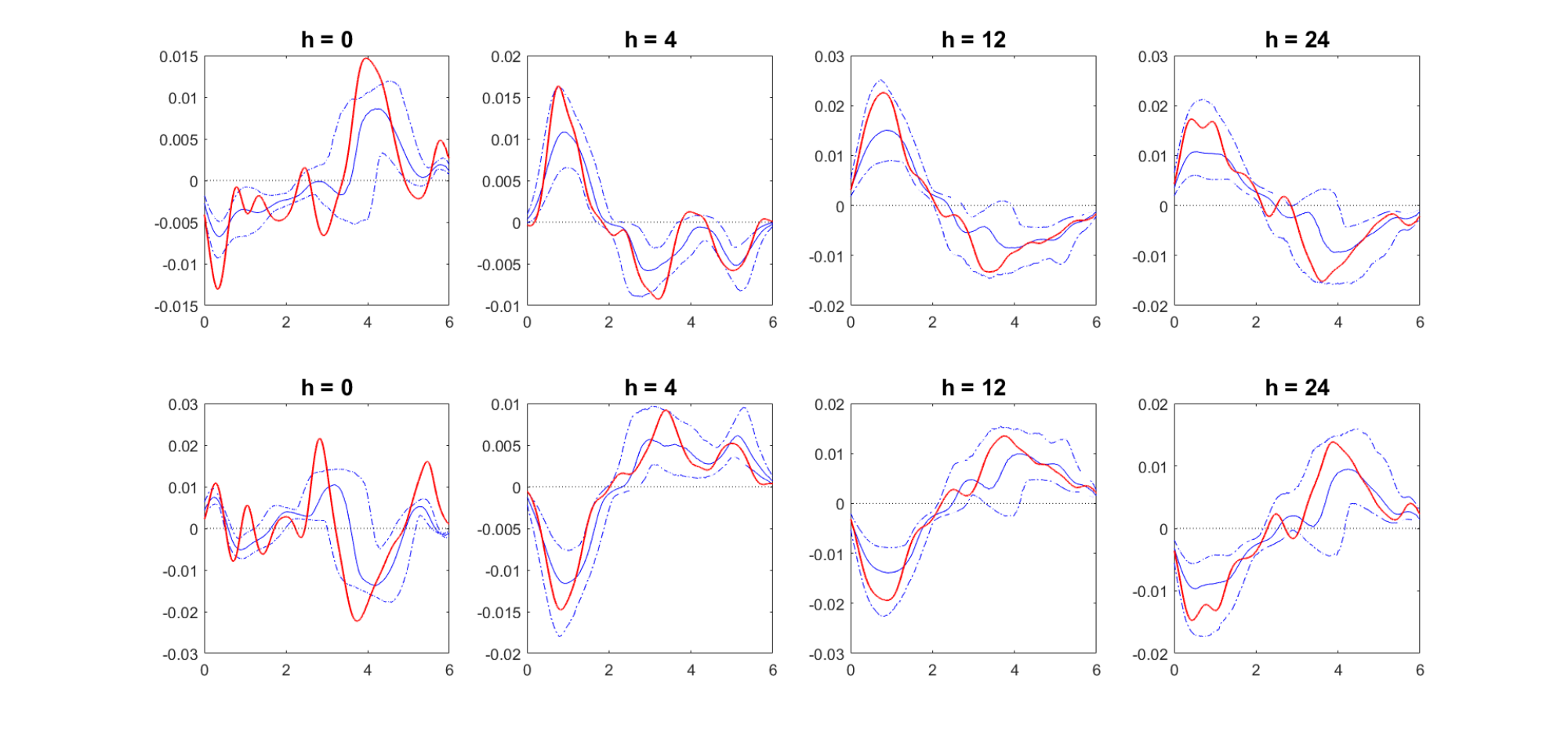}
\par\end{centering}
\centering{}%
\begin{minipage}[t]{0.9\columnwidth}%
\begin{spacing}{0.7}
{\footnotesize{}Notes: Red lines show the true responses of $p_{t}\left(\xi\right)$
to one standard deviation $\varepsilon_{1}$ and $\varepsilon_{2}$.
The solid blue lines represent the posterior median response, while
dashed blue lines delimit the 90\% credible bands. $h$ denotes the
horizon at which the response is measured. }
\end{spacing}
\end{minipage}
\end{figure}

\begin{figure}[ht!]
\caption{\label{fig:Functional-IRFs-FSVAR_dgpk15}\textbf{Functional IRFs - F-SVAR
DGP 1 - $K=15$}}
\begin{centering}
\includegraphics[bb=1000bp 0bp 0bp 450bp,scale=0.5]{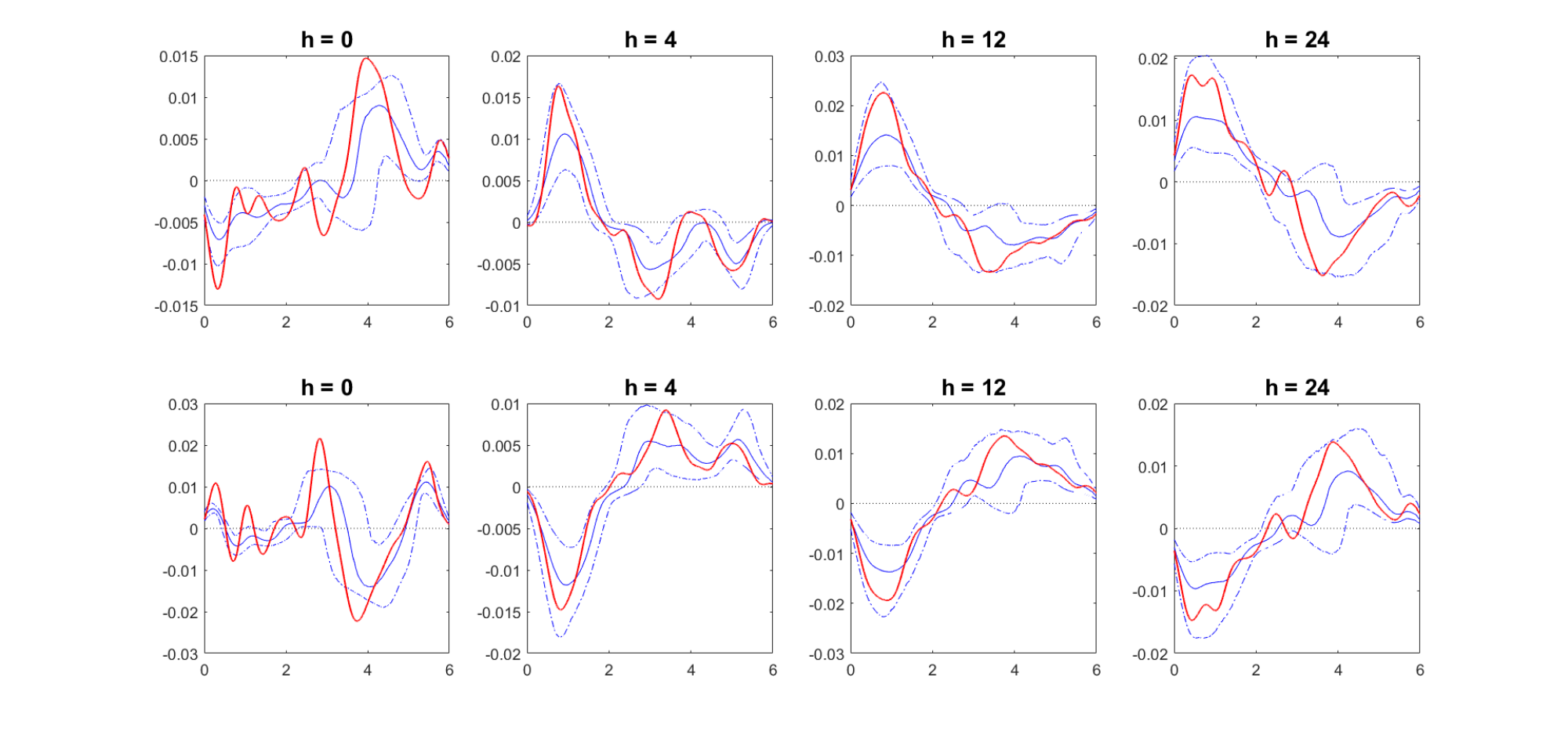}
\par\end{centering}
\centering{}%
\begin{minipage}[t]{0.9\columnwidth}%
\begin{spacing}{0.7}
{\footnotesize{}Notes: Red lines show the true responses of $p_{t}\left(\xi\right)$
to one standard deviation $\varepsilon_{1}$ and $\varepsilon_{2}$.
The solid blue lines represent the posterior median response, while
dashed blue lines delimit the 90\% credible bands. $h$ denotes the
horizon at which the response is measured. }
\end{spacing}
\end{minipage}
\end{figure}

\clearpage
\section{\normalsize{F-SVAR DGP 2 - Results with Different \texorpdfstring{$K$}{K}}\label{sec:FSVAR-DGP 2 - Results with Different $K$}}

\begin{figure}[ht!]
\caption{\label{fig:Functional-IRFs-FSVAR_dgp2k3}\textbf{Functional IRFs - F-SVAR
DGP 2 - $K=3$}}
\begin{centering}
\includegraphics[bb=40bp 20bp 1030bp 450bp,scale=0.5]{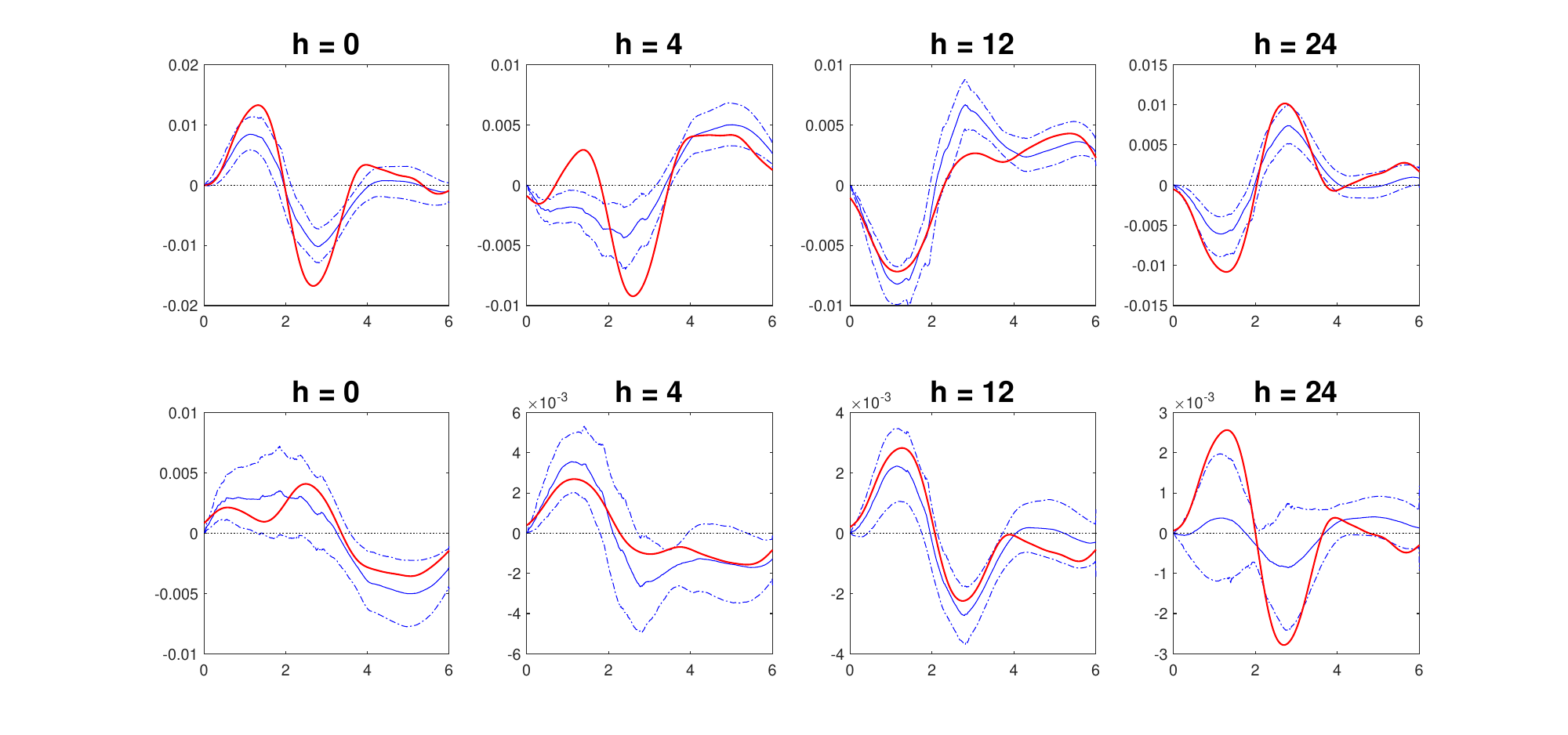}
\par\end{centering}
\centering{}%
\begin{minipage}[t]{0.9\columnwidth}%
\begin{spacing}{0.7}
{\footnotesize{}Notes: Red lines show the true responses of $p_{t}\left(\xi\right)$
to one standard deviation $\varepsilon_{1}$ and $\varepsilon_{2}$.
The solid blue lines represent the posterior median response, while
dashed blue lines delimit the 90\% credible bands. $h$ denotes the
horizon at which the response is measured. }
\end{spacing}
\end{minipage}
\end{figure}

\begin{figure}[ht!]
\caption{\label{fig:Functional-IRFs-FSVAR_dgp2k15}\textbf{Functional IRFs - F-SVAR
DGP 2 - $K=15$}}
\begin{centering}
\includegraphics[bb=40bp 20bp 1030bp 450bp,scale=0.5]{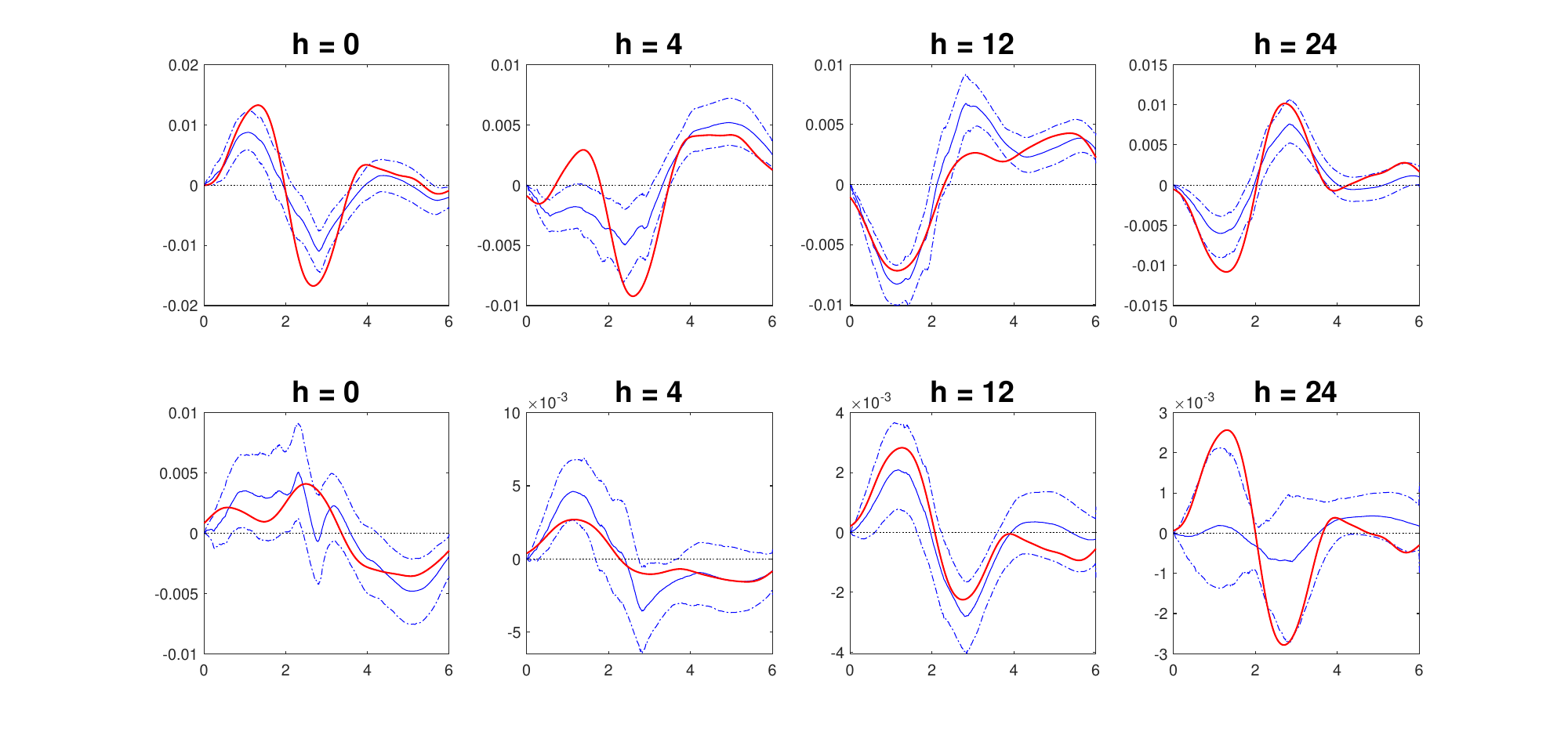}
\par\end{centering}
\centering{}%
\begin{minipage}[t]{0.9\columnwidth}%
\begin{spacing}{0.7}
{\footnotesize{}Notes: Red lines show the true responses of $p_{t}\left(\xi\right)$
to one standard deviation $\varepsilon_{1}$ and $\varepsilon_{2}$.
The solid blue lines represent the posterior median response, while
dashed blue lines delimit the 90\% credible bands. $h$ denotes the
horizon at which the response is measured. }
\end{spacing}
\end{minipage}
\end{figure}

\clearpage
\section{\normalsize{Krusell and Smith (1998) DGP - Results with Different \texorpdfstring{$K$}{K}}\label{sec:Functional-IRFs--KS98 - Results with Different $K$}}

\begin{figure}[ht!]
\caption{\textbf{\label{fig:Functional-IRFs--KS98-K3}Functional IRFs - Krusell
and Smith (1998) DGP - $K=3$}}

\includegraphics[viewport=110bp 0bp 80bp 430bp,scale=0.5]{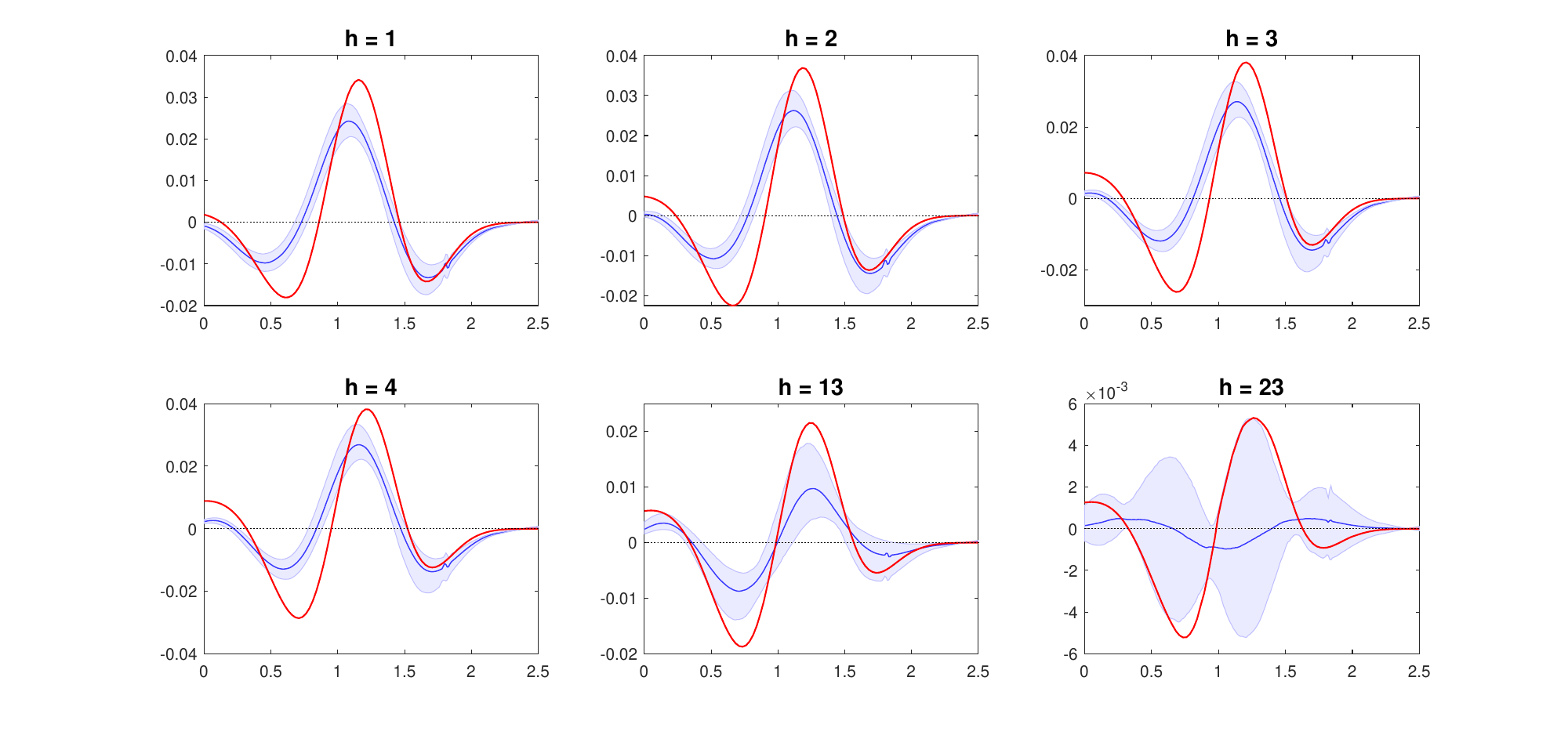}
\centering{}%
\begin{minipage}[t]{0.9\columnwidth}%
\begin{spacing}{0.69999999999999996}
{\footnotesize{}Notes: Red lines show the true responses of the asset
distribution to a one standard deviation productivity shock. The solid
blue lines represent the posterior median response, while dashed blue
lines delimit the 90\% credible bands. $h$ denotes the horizon at
which the response is measured. Notice that the horizons for which
responses are reported are the same as those in Figure 6 of \cite{chang2021heterogeneity}, the difference in the panels titles only
stems from different timing conventions. }
\end{spacing}
\end{minipage}
\end{figure}

\begin{figure}[ht!]
\caption{\textbf{\label{fig:Functional-IRFs--KS98 - K15}Functional IRFs - Krusell
and Smith (1998) DGP -$K=15$}}

\includegraphics[viewport=80bp 0bp 70bp 430bp,scale=0.5]{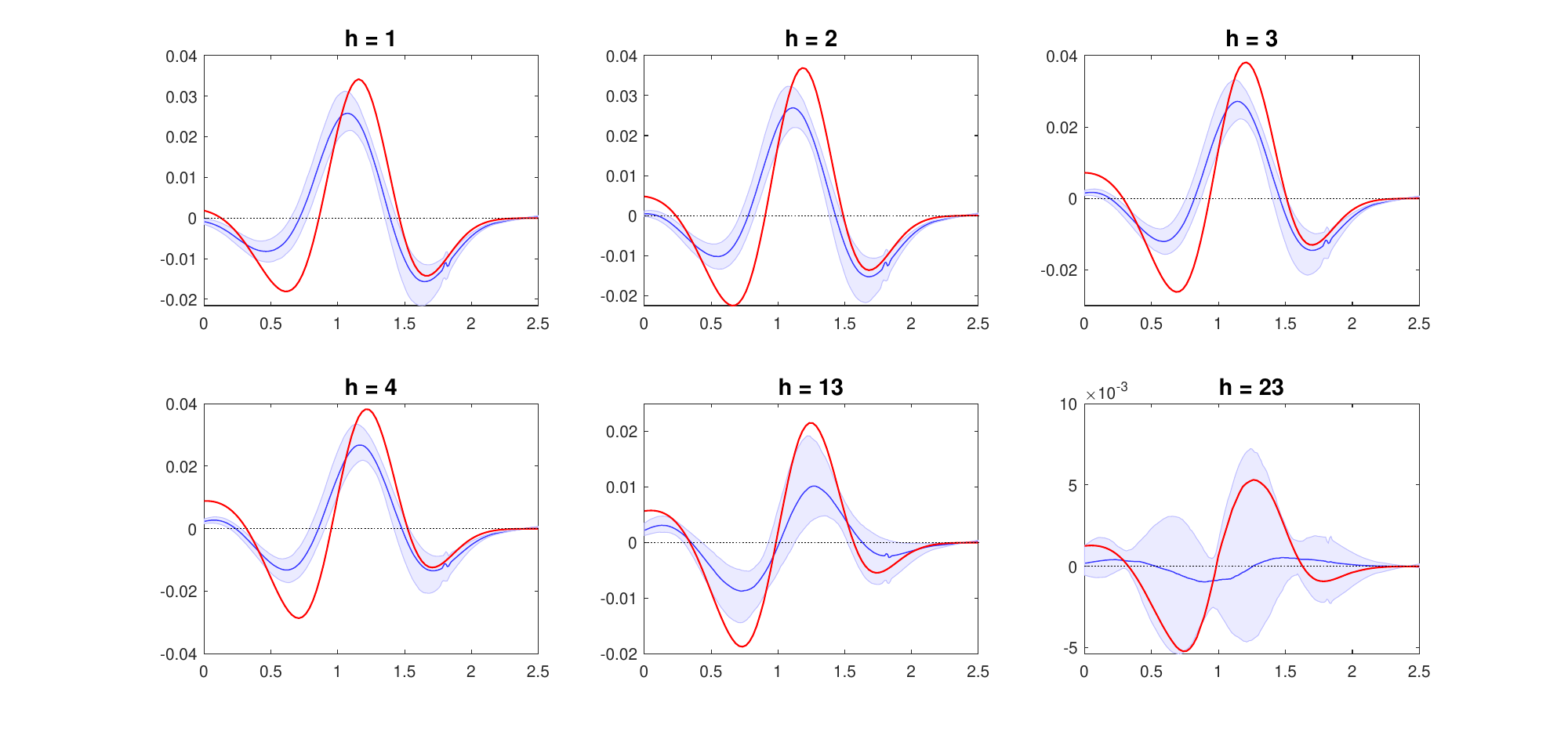}
\centering{}%
\begin{minipage}[t]{0.95\columnwidth}%
\begin{spacing}{0.69999999999999996}
{\footnotesize{}Notes: Red lines show the true responses of the asset
distribution to a one standard deviation productivity shock. The solid
blue lines represent the posterior median response, while dashed blue
lines delimit the 90\% credible bands. $h$ denotes the horizon at
which the response is measured. Notice that the horizons for which
responses are reported are the same as those in Figure 6 of \cite{chang2021heterogeneity}, the difference in the panels titles only
stems from different timing conventions. }
\end{spacing}
\end{minipage}
\end{figure}

\clearpage
\section{\large{Uncertainty Shocks - Results with Different \texorpdfstring{$K$}{K}}\label{sec:Uncertainty Shocks - Results with Different $K$}}

\begin{figure}[ht!]
\caption{\label{fig:Functional-IRFs--JLN}Functional IRFs - Effects of Uncertainty
Shocks on Earings and Consumptio Distribution - $K=3$ and $K=5$ respectively}

\includegraphics[bb=40bp 20bp 1030bp 450bp,scale=0.5]{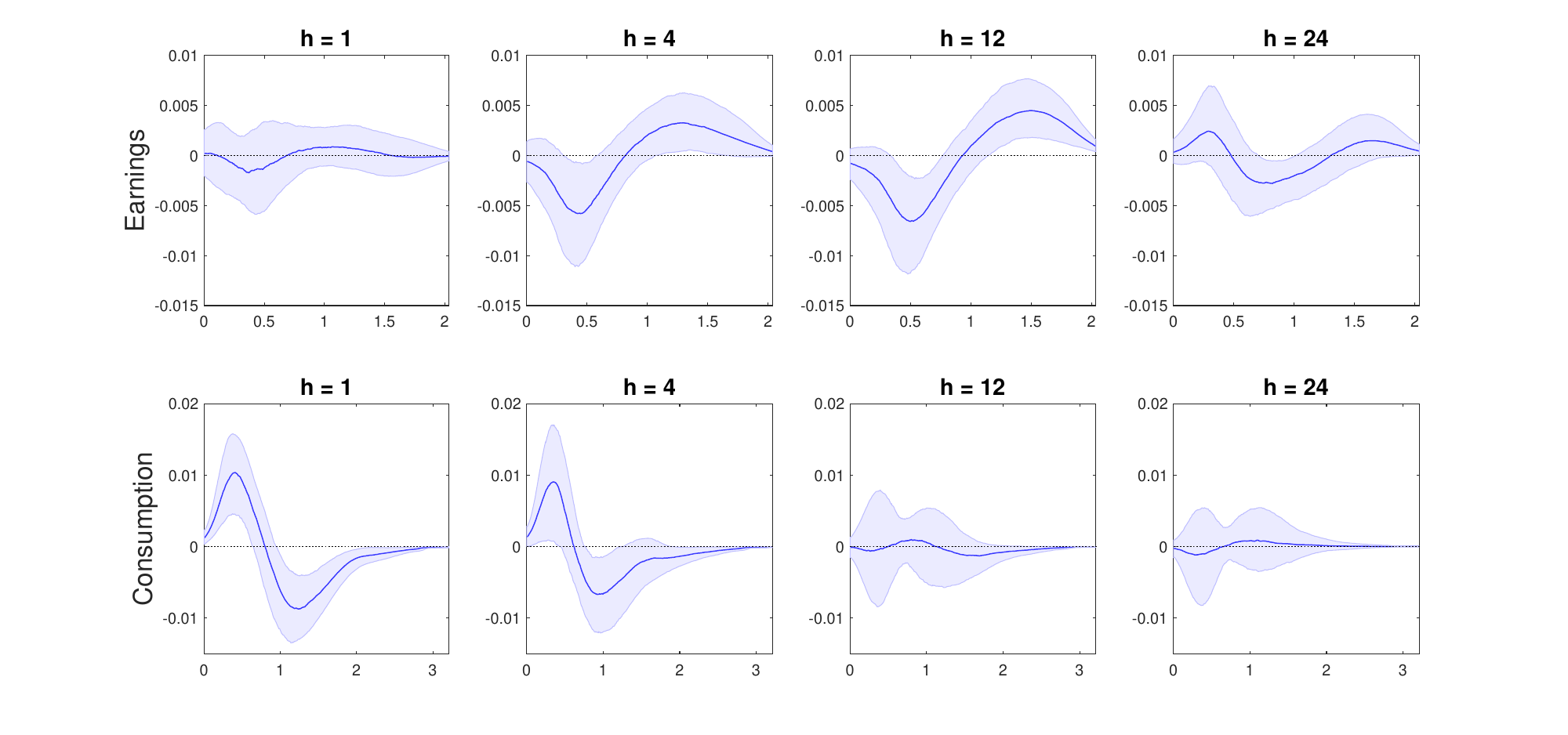}
\centering{}%
\begin{minipage}[t]{1\columnwidth}%
\begin{spacing}{0.69999999999999996}
{\footnotesize{}Notes: The solid blue lines represent the posterior
median response, while dashed blue lines delimit the 68\% credible
bands. $h$ denotes the horizon at which the response is measured.}
\end{spacing}
\end{minipage}
\end{figure}

\begin{figure}[ht!]
\caption{\label{fig:Functional-IRFs--JLN2}Functional IRFs - Effects of Uncertainty
Shocks on Earnings and Consumption Distribution - $K=15$}

\includegraphics[bb=40bp 20bp 1030bp 450bp,scale=0.5]{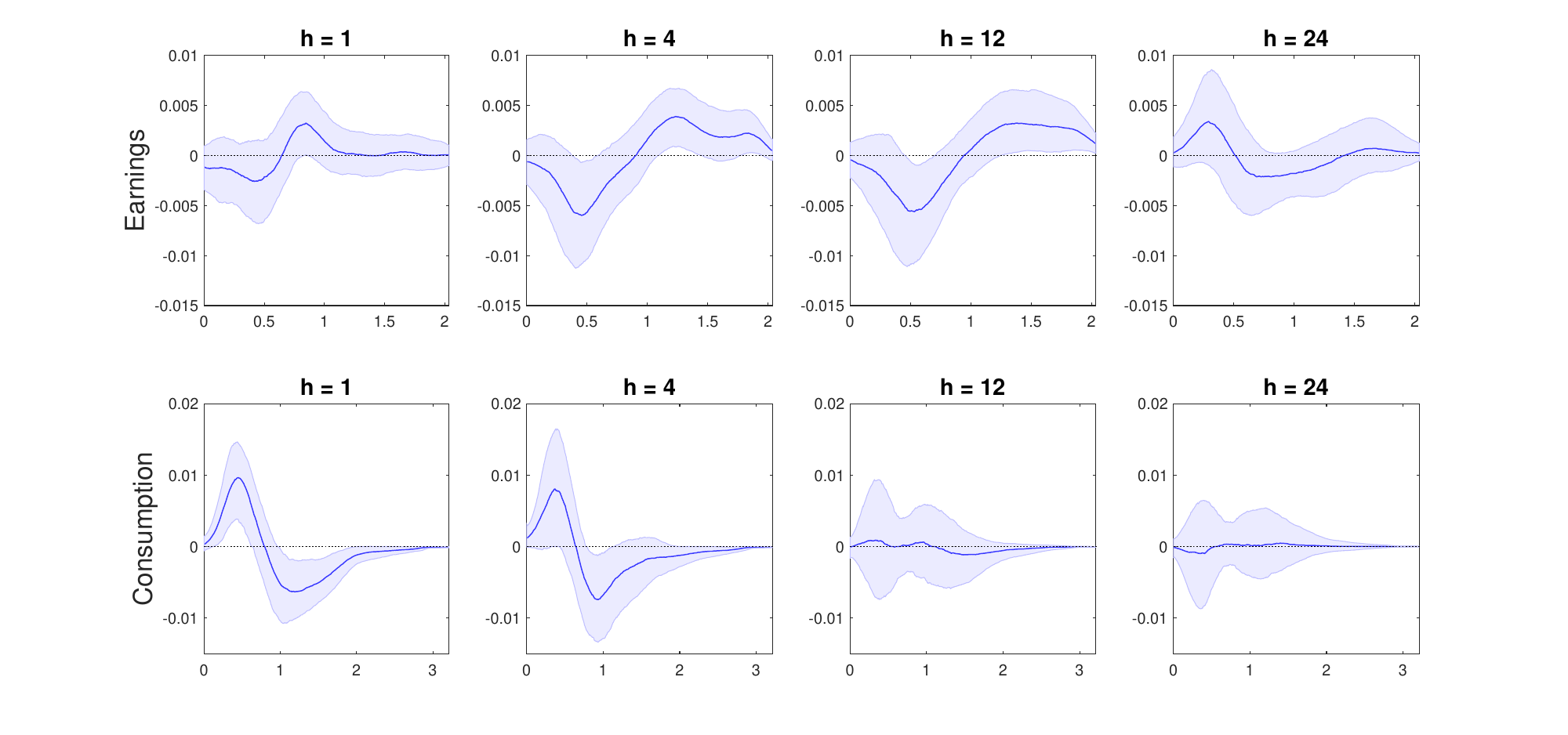}
\centering{}%
\begin{minipage}[t]{1\columnwidth}%
\begin{spacing}{0.69999999999999996}
{\footnotesize{}Notes: The solid blue lines represent the posterior
median response, while dashed blue lines delimit the 68\% credible
bands. $h$ denotes the horizon at which the response is measured.
The measure on the horizontal axis is the earnings-to-DGP per-capita
ratio.}
\end{spacing}
\end{minipage}
\end{figure}

\clearpage
\section{\large{Standard SVAR Results}\label{sec:Uncertainty Shocks - Gini Standard SVAR}}

\begin{figure}[ht!]
\caption{\label{fig:Gini Standard SVAR}}

\includegraphics[viewport=100bp 0bp 900bp 450bp,scale=0.50]{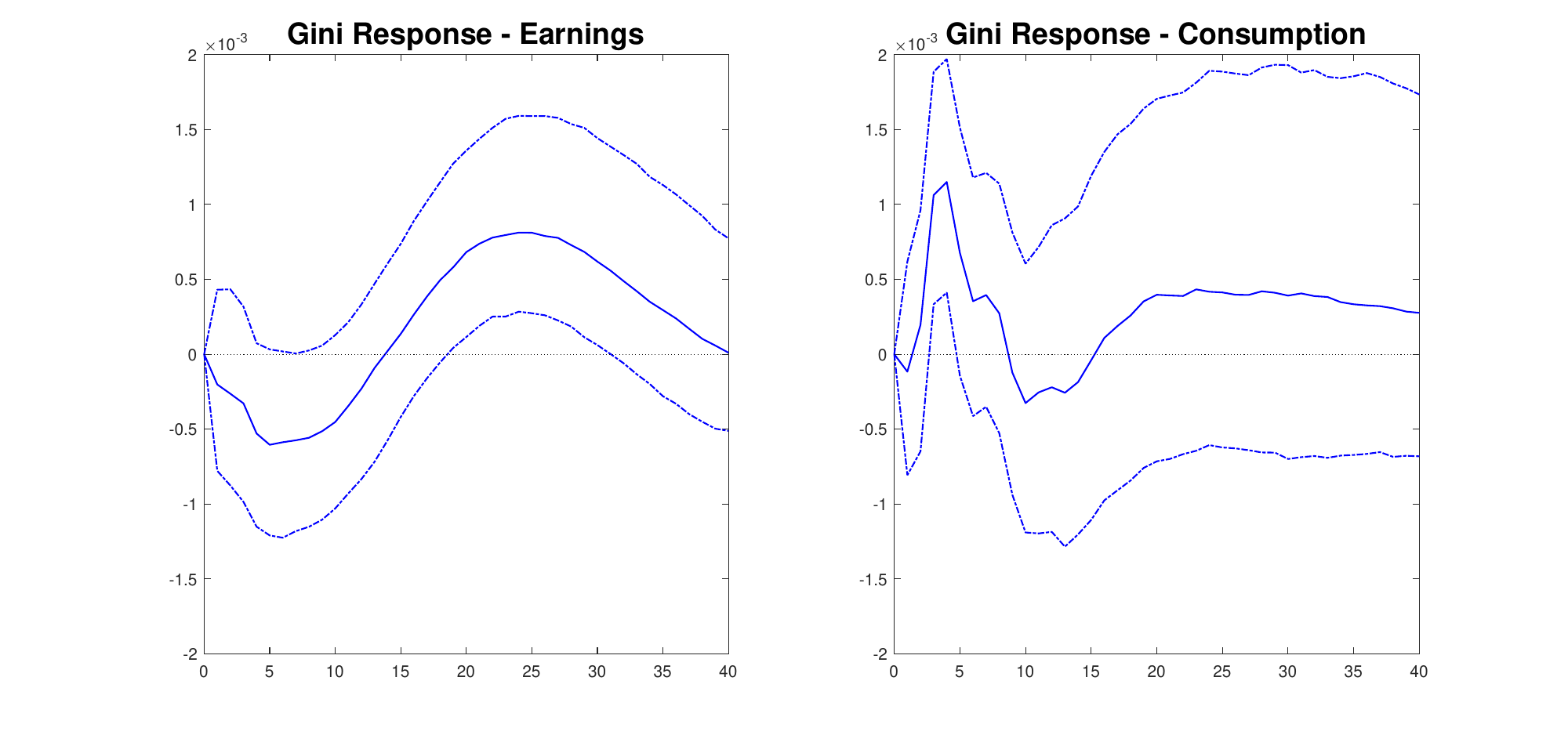}
\centering{}%
\begin{minipage}[t]{1\columnwidth}%
\begin{spacing}{0.69999999999999996}
{\footnotesize{}Notes: The solid red lines represent the posterior
median response to a one-standard deviation shock, while dashed red lines delimit the 68\% credible
bands. The horizon is measured in quarters.}
\end{spacing}
\end{minipage}
\end{figure}

\clearpage
\section{\normalsize{IRFs Unemployment by Educational Attainment}\label{sec:IRFsd_unempEdu}}

\begin{figure}[ht!]
\caption{\textbf{\label{fig:EarningsEdu}Median weekly earnings}}

\includegraphics[viewport=110bp 0bp 80bp 430bp,scale=0.5]{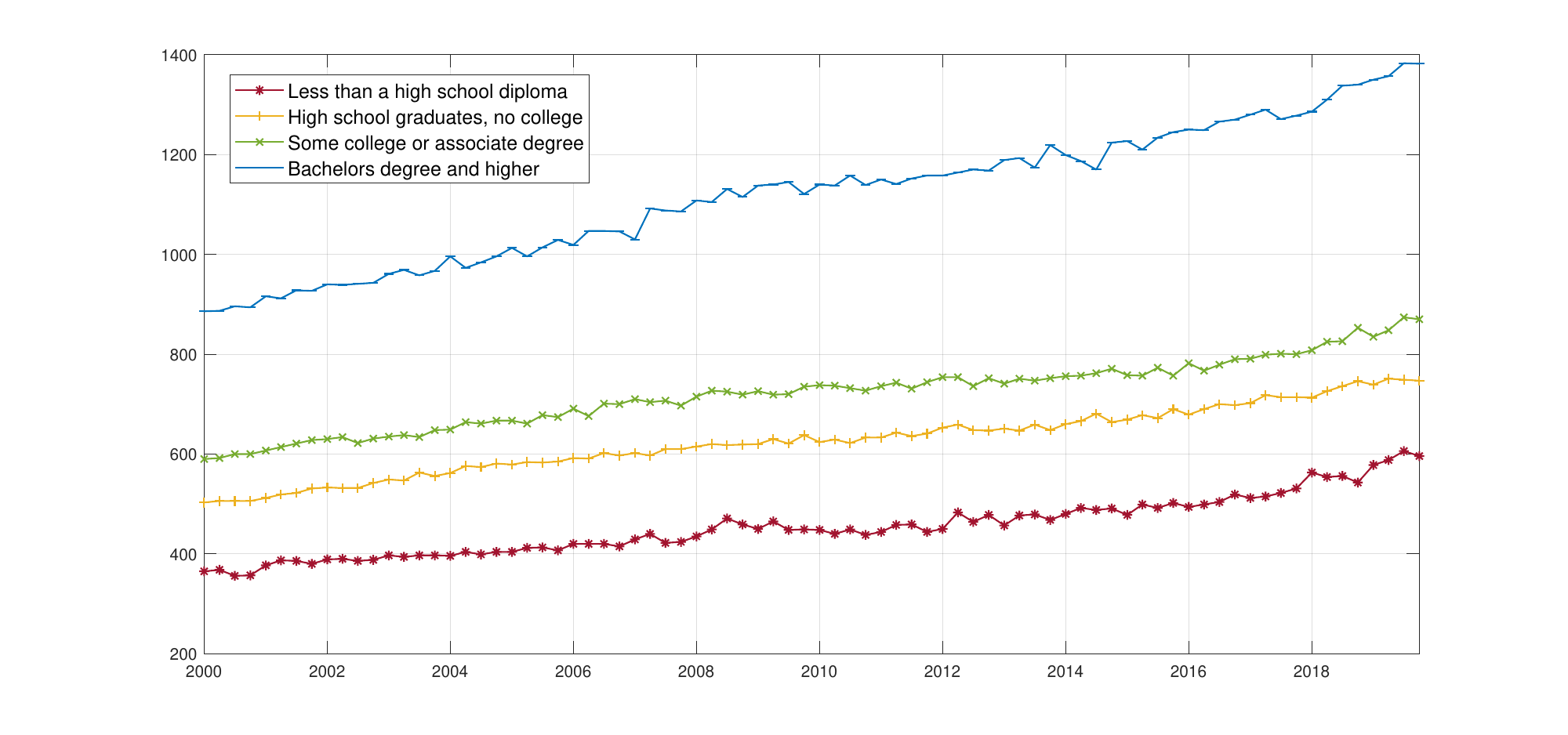}
\centering{}%
\begin{minipage}[t]{0.9\columnwidth}%
\begin{spacing}{0.69999999999999996}
{\footnotesize{}Notes: The lines show the median weekly earnings of the four categories.}
\end{spacing}

\end{minipage}
\end{figure}

\clearpage
\section{F-LP: Artificial Data Experiments\label{sec:F-LP Simulations}}

\subsection{\normalsize{DGP 1}\label{sec:F-LP Simulations1}}

\begin{figure}[ht!]
\caption{\label{fig:F-LP_dgp 1}\textbf{F-LP: DGP 1}}
\begin{centering}
\includegraphics[bb=30bp 20bp 1030bp 450bp,scale=0.5]{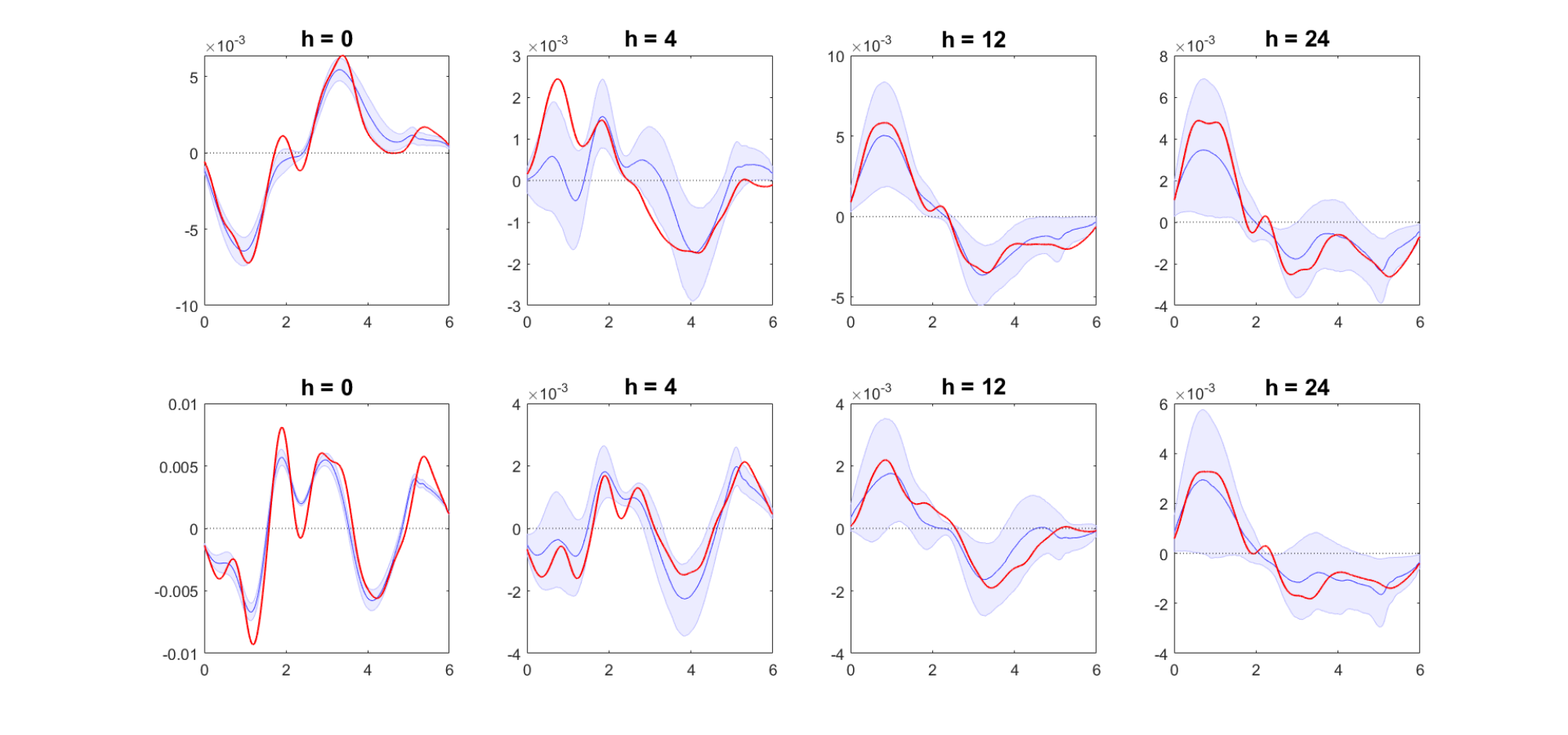}
\par\end{centering}
\centering{}%
\begin{minipage}[t]{0.9\columnwidth}%
\begin{spacing}{0.7}
{\footnotesize{}Notes: Red lines show the true responses of $p_{t}\left(\xi\right)$
to one standard deviation shocks to $\varepsilon_{1}$ (upper panels) and $\varepsilon_{2}$ (lower panels).
The solid blue lines represent the posterior median response, while
dashed blue lines delimit the 90\% credible bands. $h$ denotes the
horizon at which the response is measured. }
\end{spacing}
\end{minipage}
\end{figure}

\begin{figure}[ht!]
\caption{\label{fig:F-LP_vsVARdgp 1}\textbf{F-LP vs F-SVAR: DGP 1}}
\begin{centering}
\includegraphics[bb=30bp 20bp 1030bp 450bp,scale=0.5]{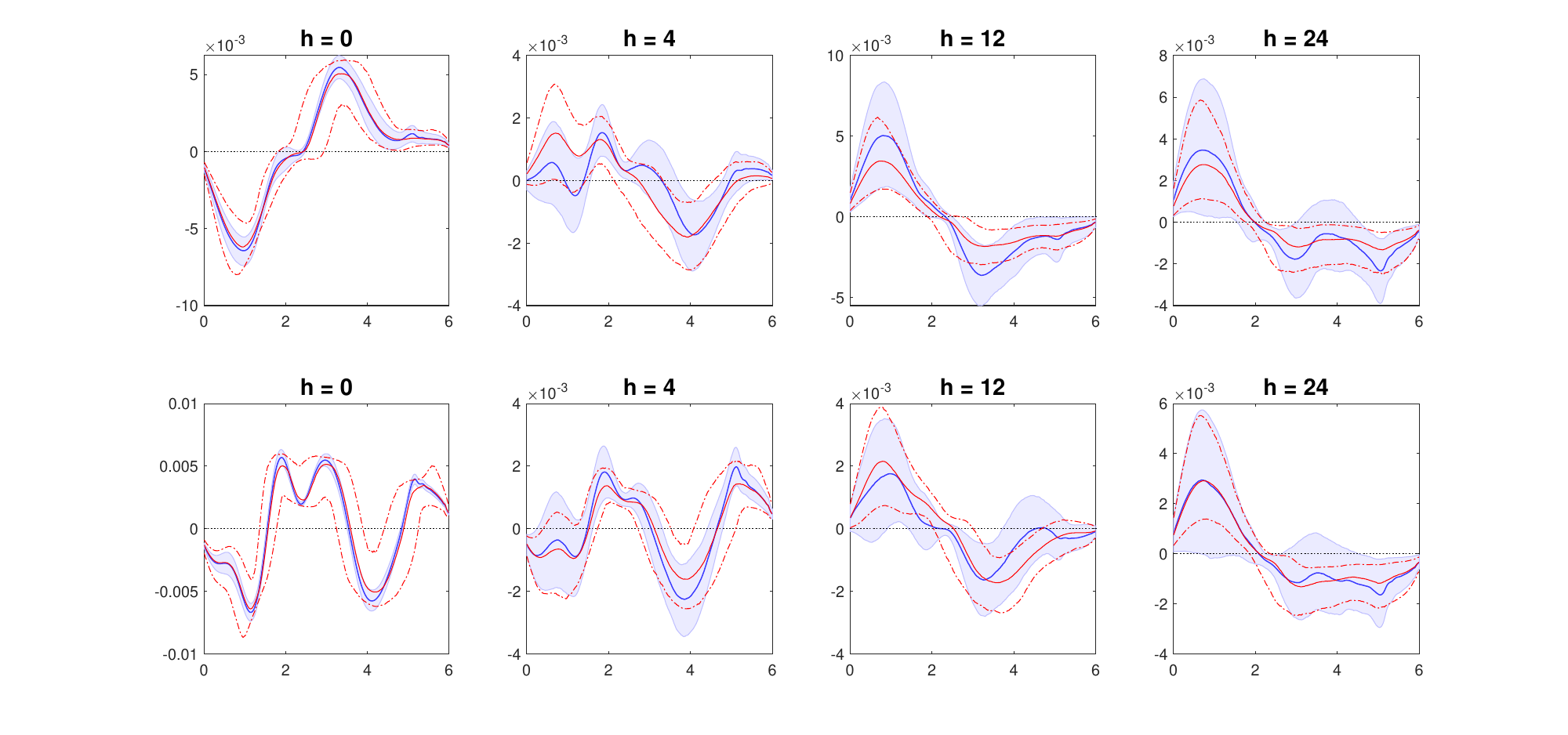}
\par\end{centering}
\centering{}%
\begin{minipage}[t]{0.9\columnwidth}%
\begin{spacing}{0.7}
{\footnotesize{}Notes: The solid red line shows and the dashed red lines show the posterior median and the 90\% credible bands of the responses inferred by the F-SVAR.
The solid blue lines represent the estimated F-LP, while the light blue area delimits the associated 90\% confidence interval. }
\end{spacing}
\end{minipage}
\end{figure}

\newpage

\subsection{\normalsize{DGP 2}\label{sec:F-LP Simulations2}}

\begin{figure}[ht!]
\caption{\label{fig:F-LP_dgp 2}\textbf{F-LP: DGP 2}}
\begin{centering}
\includegraphics[bb=30bp 20bp 1030bp 450bp,scale=0.5]{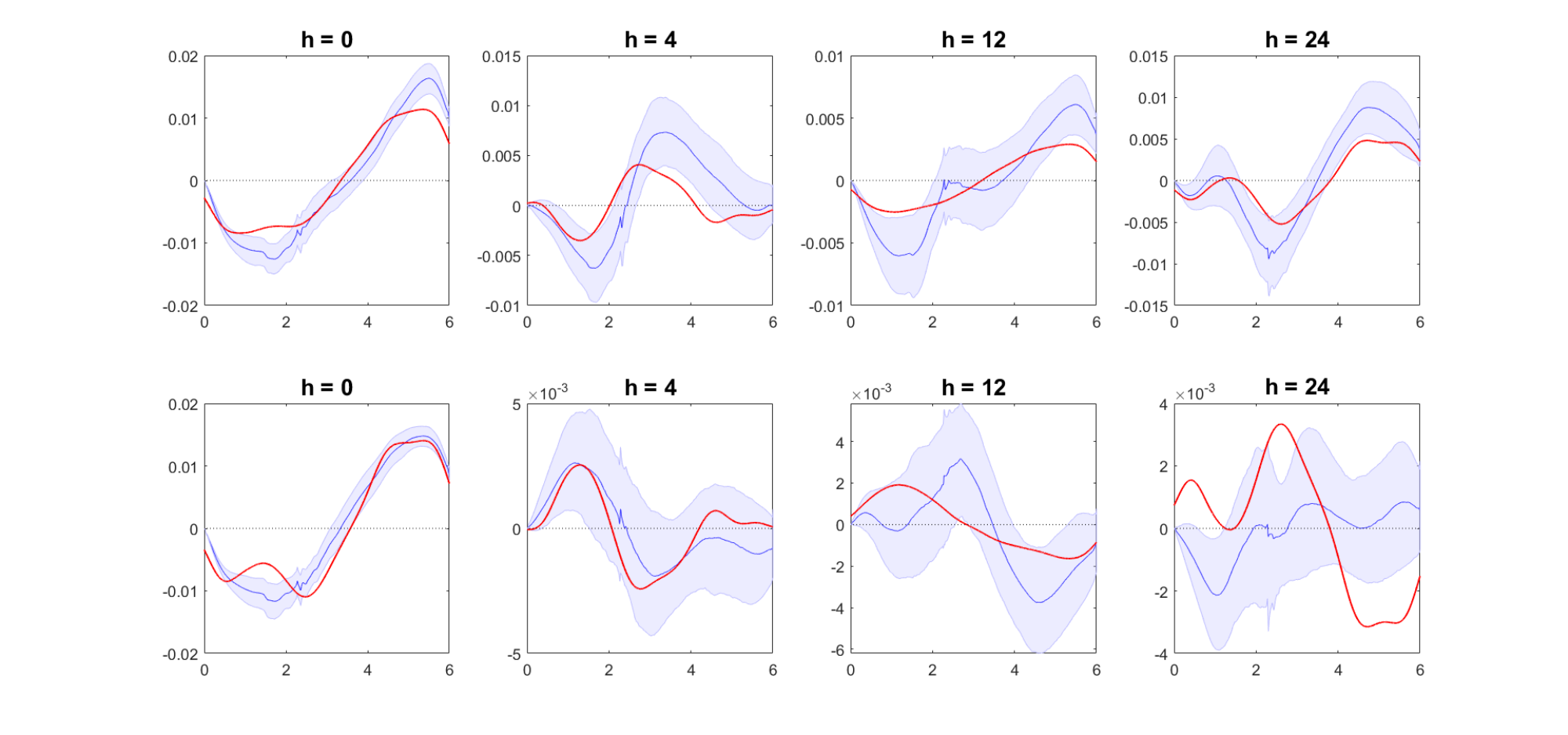}
\par\end{centering}
\centering{}%
\begin{minipage}[t]{0.9\columnwidth}%
\begin{spacing}{0.7}
{\footnotesize{}Notes: Red lines show the true responses of $p_{t}\left(\xi\right)$
to one standard deviation $\varepsilon_{1}$ (upper panels) and $\varepsilon_{2}$ (lower panels).
The solid blue lines represent the posterior median response, while
dashed blue lines delimit the 90\% credible bands. $h$ denotes the
horizon at which the response is measured. }
\end{spacing}
\end{minipage}
\end{figure}

\begin{figure}[ht!]
\caption{\label{fig:F-LP_vsVARdgp 2}\textbf{F-LP vs F-SVAR: DGP 2}}
\begin{centering}
\includegraphics[bb=30bp 20bp 1030bp 450bp,scale=0.5]{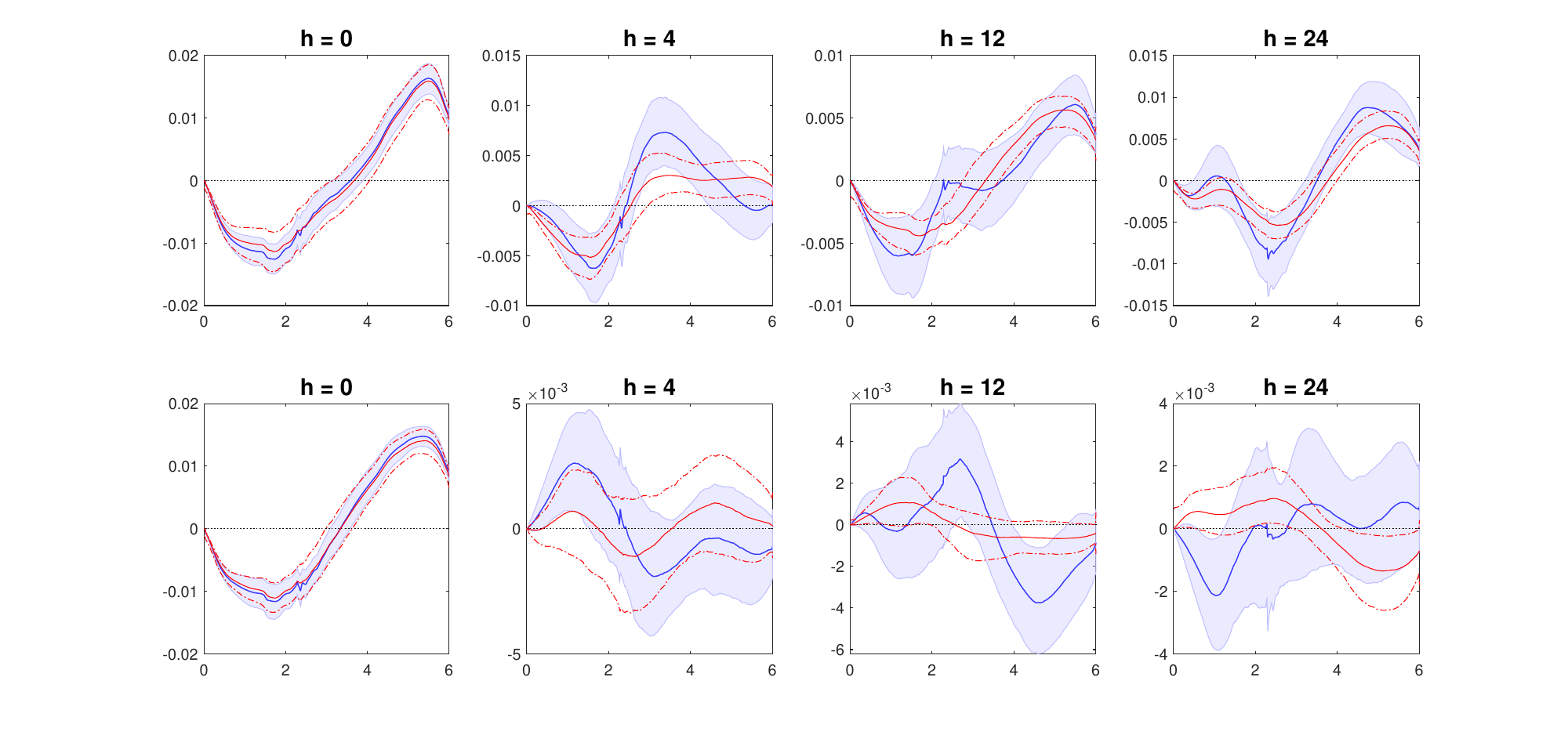}
\par\end{centering}
\centering{}%
\begin{minipage}[t]{0.9\columnwidth}%
\begin{spacing}{0.7}
{\footnotesize{}Notes: The solid red line shows and the dashed red lines show the posterior median and the 90\% credible bands of the responses inferred by the F-SVAR.
The solid blue lines represent the estimated F-LP, while the light blue area delimits the associated 90\% confidence interval. }
\end{spacing}
\end{minipage}
\end{figure}

\newpage
\subsection{\normalsize{Krusell and Smith (1998) DGP}\label{sec:F-LP SimulationsKS}}

\begin{figure}[ht!]
\caption{\label{fig:F-LP_KS}\textbf{F-LP: Krusell and Smith (1998) DGP}}
\begin{centering}
\includegraphics[bb=30bp 20bp 1030bp 450bp,scale=0.5]{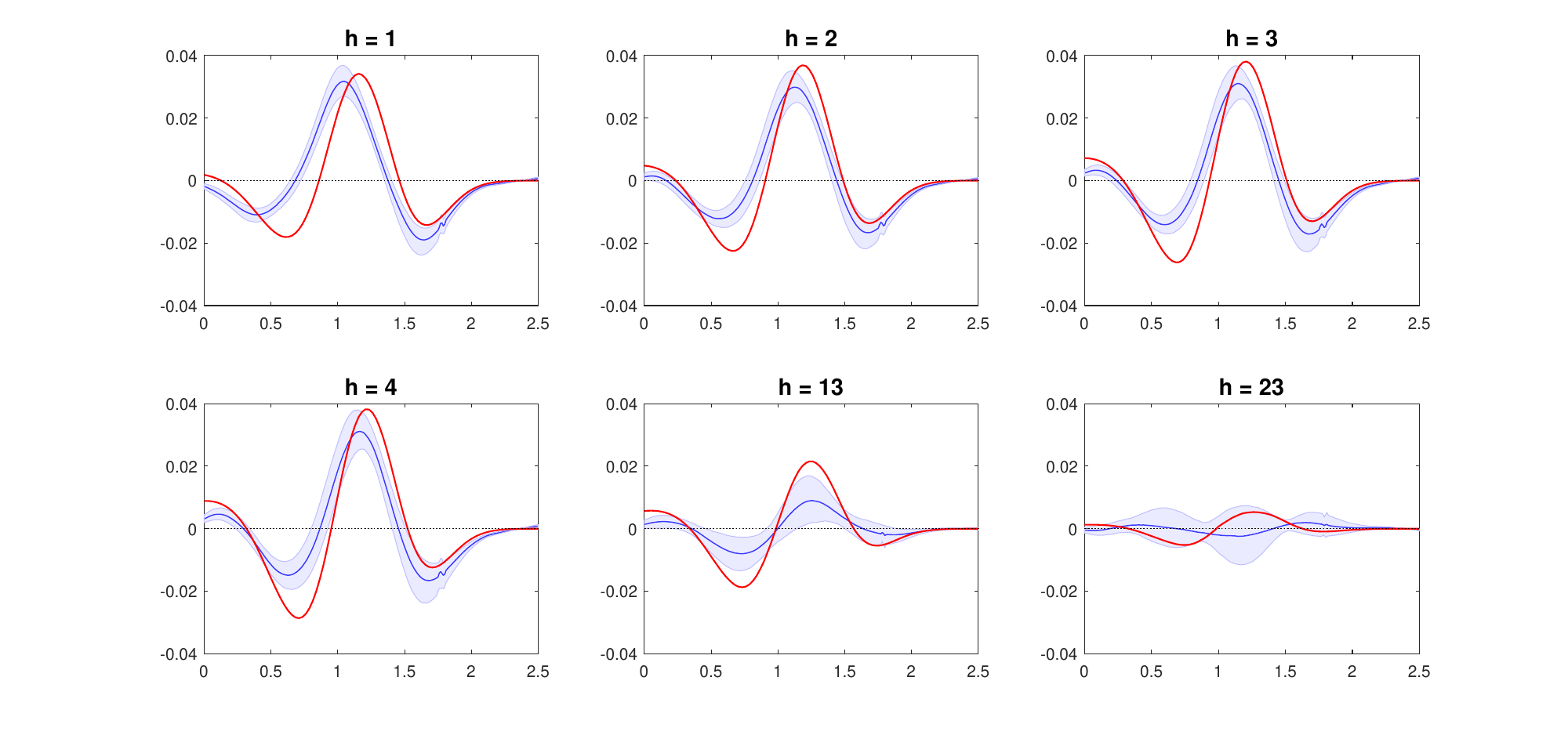}
\par\end{centering}
\centering{}%
\begin{minipage}[t]{0.9\columnwidth}%
\begin{spacing}{0.7}
{\footnotesize{}Notes: Red lines show the true responses of the asset
distribution to a one standard deviation productivity shock. The solid
blue lines represent the posterior median response, while dashed blue
lines delimit the 90\% credible bands. $h$ denotes the horizon at
which the response is measured. Notice that the horizons for which
responses are reported are the same as those in Figure 6 of \cite{chang2021heterogeneity}, the difference in the panels titles only
stems from different timing conventions. }
\end{spacing}
\end{minipage}
\end{figure}

\end{document}